\documentclass[twocolumn, numberedappendix]{aastex62}

\usepackage{amstext}
\usepackage{amsmath}
\usepackage{amssymb}
\usepackage{graphicx}
\usepackage{upgreek}

\newcommand{\beq}{\begin{equation}}
\newcommand{\eeq}{\end{equation}}

\graphicspath{{./}{figures/}}

\begin{document}

%\title{Circumstellar Properties, Light Curves, and Population Statistics of Engine Driven Explosions Following Compact Object -- Massive Star Coalescence }
\title{Explosions Driven by the Coalescence of a Compact Object with the Core of a Massive-Star Companion Inside a Common Envelope: Circumstellar Properties, Light Curves, and Population Statistics}

\correspondingauthor{Sophie Lund Schr\o der}
\email{sophie.schroder@nbi.ku.dk}

\author[0000-0003-1735-8263]{Sophie Lund Schr\o der}
\affiliation{DARK, Niels Bohr Institute, University of Copenhagen, Blegdamsvej 17, 2100 Copenhagen, Denmark}
\affiliation{Harvard-Smithsonian Center for Astrophysics, 60 Garden Street, Cambridge, MA, 02138, USA}

\author[0000-0002-1417-8024]{Morgan MacLeod}
\altaffiliation{NASA Einstein Fellow}
\affiliation{Harvard-Smithsonian Center for Astrophysics, 60 Garden Street, Cambridge, MA, 02138, USA}
%\email{morgan.macleod@cfa.harvard.edu}

\author[0000-0003-4330-287X]{Abraham Loeb}
\affiliation{Harvard-Smithsonian Center for Astrophysics, 60 Garden Street, Cambridge, MA, 02138, USA}

\author[0000-0003-1817-3586]{Alejandro Vigna-G\'{o}mez}
\affiliation{Birmingham Institute for Gravitational Wave Astronomy and School of Physics and Astronomy, University of Birmingham,Birmingham, B15 2TT, United Kingdom} 
\affiliation{Monash Centre for Astrophysics, School of Physics and Astronomy, Monash University, Clayton, Victoria 3800, Australia}
\affiliation{The ARC Center of Excellence for Gravitational Wave Discovery -- OzGrav}

\author[0000-0002-6134-8946]{Ilya Mandel}
\affiliation{Monash Centre for Astrophysics, School of Physics and Astronomy, Monash University, Clayton, Victoria 3800, Australia}
\affiliation{The ARC Center of Excellence for Gravitational Wave Discovery -- OzGrav}
\affiliation{Birmingham Institute for Gravitational Wave Astronomy and School of Physics and Astronomy, University of Birmingham,Birmingham, B15 2TT, United Kingdom}

\begin{abstract}
We model explosions driven by the coalescence of a black hole or neutron star with the core of its massive-star companion. Upon entering a common envelope phase, a compact object may spiral all the way to the core.  The concurrent release of energy is likely to be deposited into the surrounding common envelope, powering a merger-driven explosion. We use hydrodynamic models of binary coalescence to model the common envelope density distribution at the time of coalescence. We find toroidal profiles of material, concentrated in the binary's equatorial plane and extending to many times the massive star's original radius. We use the spherically-averaged properties of this circumstellar material (CSM) to estimate the emergent light curves that result from the interaction between the blast wave and the CSM. We find that typical merger-driven explosions are brightened by up to three magnitudes by CSM interaction.  From population synthesis models we discover that the brightest merger-driven explosions, $M_V \sim -18$ to $-19$, are those involving black holes because they have the most massive and extended CSM. Black hole coalescence events are also common; they represent about 50\% of all merger-driven explosions and approximately 0.3\% of the core-collapse rate.   Merger-driven explosions offer a window into the highly-uncertain physics of common envelope interactions in binary systems by probing the properties of systems that merge rather than eject their envelopes. 
\end{abstract}


\section{Introduction} \label{sec:intro}

Binary and multiple systems are ubiquitous among massive stars. Of these systems, a large fraction are at separations so close that they will interact over the stars' lifetimes \citep{2012Sci...337..444S,2014ApJ...782....7D,2017ApJS..230...15M}. As these multiple-star systems evolve to leave behind compact object stellar remnants, the stage is set for interactions between the evolving stars and the stellar remnants. In some cases, a phase of escalating, unstable mass transfer from a massive star donor onto a compact-object companion leads to a common envelope phase \citep{1976IAUS...73...75P}, in which the compact object is immersed within the envelope of the massive star and spirals closer to the massive star's core \citep[e.g.][]{1978ApJ...222..269T,1993PASP..105.1373I,1995ApJ...445..367T,2000ARA&A..38..113T,2000ApJ...532..540A,2013A&ARv..21...59I,2017PASA...34....1D}. 

Common envelope interactions can lead to either the ejection of the shared, gaseous envelope and a surviving, binary pair or to the merger of the donor star core with the companion. The distinction between these cases for a given binary remains highly uncertain, but is of significant interest given its importance \citep[e.g.][]{2002ApJ...572..407B,2004ApJ...601L.179K,2007PhR...442...75K,2008ApJS..174..223B,2015ApJ...801...32A,2017ApJ...846..170T,2018MNRAS.481.4009V}, for example, to estimating rates of compact object mergers and associated gravitational-wave transients \citep{2016PhRvL.116f1102A,2016PhRvL.116x1103A, 2017PhRvL.118v1101A,2017PhRvL.119p1101A,2017ApJ...851L..35A,2018arXiv181112940T,2018arXiv181112907T}. In the case of massive stars interacting with lower-mass, compact object companions, the theoretical expectation is that only donor stars with the most-extended, weakly bound hydrogen envelopes are susceptible to ejection, while the remainder of systems are likely to merge \citep{2016A&A...596A..58K}. 

What happens when a compact object merges with the helium core of a massive star? At least two possible outcomes have been suggested. Perhaps a neutron star embedded in a stellar envelope could burn stably, forming a Thorne-Zytkow object \citep{1977ApJ...212..832T,1995MNRAS.274..485P,2007ASPC..367..541P,2014MNRAS.443L..94L,2018MNRAS.475L..49M}. Alternatively, it is possible for the angular momentum of the merger to lead to the formation of a disk around the compact object (be it a neutron star or a black hole), from which material forms a rapidly accreting neutrino-cooled disk \citep{1991ApJ...376..234H,1993ApJ...411L..33C,1996ApJ...459..322C,1996ApJ...460..801F,1998ApJ...502L...9F,1999ApJ...518..356P,1999ApJ...524..262M,1999ApJ...526..152F,2001ApJ...550..410M,2001ApJ...550..357Z,2006ApJ...641..961L,2007ApJ...657..383C,2011MNRAS.415..944B,2012ApJ...752L...2C,2014ApJ...793L..36F,2019ApJ...871..117S}. Accretion of the surrounding core material liberates on the order of $\eta M_\odot c^2 \sim \eta 10^{54}$~erg, where $\eta$ is an efficiency factor of order 0.1 \citep{2002apa..book.....F}. Much of this energy emerges in neutrinos, which stream freely away from the accretion object \citep{1993ApJ...411L..33C,1996ApJ...460..801F,2001ApJ...550..357Z}. A fraction, however, emerges as Poynting flux or mechanical energy (either a disk wind or collimated outflow) and can feed back on the surroundings, powering a blast wave with energy similar to a supernova \citep{1998ApJ...502L...9F,2001ApJ...550..357Z,2012ApJ...752L...2C,2013ApJ...772...30D,2014ApJ...793L..36F,2019ApJ...871..117S,2019MNRAS.482.4233G,2019arXiv190201187S,2019MNRAS.484.4972S}.

\citet{2012ApJ...752L...2C} observed that, in the case that the merger was initiated by a preceding common envelope phase, the blast wave necessarily interacts with the dense surrounding medium of the common envelope ejecta \citep{2018MNRAS.475.1198S, 2019arXiv190201187S,2019MNRAS.482.4233G,2019MNRAS.484.4972S}. The resulting transient is described as a "common envelope jets supernova" by \citet{2019MNRAS.484.4972S}. One of the key points that \citet{2012ApJ...752L...2C} and \citet{2019MNRAS.484.4972S} mention is that the distribution of common envelope ejecta is crucial in shaping the observed light curve \citep[see][for a similar discussion in the context of rapidly-fading supernovae]{2018MNRAS.475.3152K}. Here, we pursue this line of examination. 

Light curves from supernova explosions of single stars within a dense circumstellar material (CSM) have been considered as the origin of the variations in type II supernova for many years \citep{1997ARep...41..672C,2001MNRAS.326.1448C,1994MNRAS.268..173C,2004MNRAS.352.1213C,2007ApJ...671L..17S,2011ApJ...729L...6C,2013MNRAS.428.1020M,2013ApJ...763...42O,2015MNRAS.449.4304D,2017ApJ...838...28M}, and even a population of objects that appear to transition from type I to type II \citep{2017ApJ...835..140M}. Specifically type IIn, with their narrow line features, are known to be interacting with a slow moving CSM \citep{2012ApJ...744...10K,2013A&A...555A..10T}. But the origin of the material around the pre-SN star is not yet clear. Ideas include late stage stellar winds or small outbursts of gas \citep{2012MNRAS.423L..92Q,2014ApJ...780...96S,2017MNRAS.470.1642F} 
or formations of a disk-like torus \citep{2018MNRAS.477...74A,2018ApJ...856...29M}. With this paper we add to the calculations by \citet{2012ApJ...752L...2C} the light curves expected from an engine-driven explosion inside a merger ejecta profile. We show how the atypical CSM distribution leads to light curves powered in part by CSM interaction, resembling type IIn and IIL and lacking a plateau phase \citep[e.g.][]{2017ApJ...851..138D,2018ApJ...867....4M,2018PASA...35...49E}.

 We briefly review the engine-driven explosion model in \ref{sec:explosion}. We then self-consistently model the circum-merger density distribution from the common envelope phase using a three-dimensional hydrodynamic simulation in Section \ref{sec:merger}. To explore the impact of this CSM on the resultant engine-driven explosions, we produce a number of spherically-symmetric radiative transfer models of blast waves interacting with the (spherically-averaged) common-envelope ejecta profiles in Section \ref{sec:lightcurves}. Next, we map the expected populations of compact object--core mergers and their resultant transients in Section \ref{sec:pop}. Finally, in Section \ref{sec:discussion} we study the imprint of CSM on merger-driven explosions, compare to known supernovae, and discuss possible identification strategies. In Section \ref{sec:conclusion} we conclude.

\section{Merger-driven explosions}\label{sec:explosion}

\subsection{Inspiral, Merger, and Central Engine}

Following its inspiral through the common envelope, a compact object can tidally disrupt and merge with the helium core of the massive star  \citep{2001ApJ...550..357Z,2012ApJ...752L...2C,2018MNRAS.475.1198S,2019MNRAS.484.4972S}. The disrupted material forms an accretion disk surrounding the now-central compact object. The local densities of more than $10^2$~g~cm$^{-3}$ imply that accretion occuring on a dynamical timescale is very rapid of the order of $10^{-3}$ to $10^{-1}M_\odot$~s$^{-1}$ \citep{1998ApJ...502L...9F,2001ApJ...550..357Z}, making these conditions very similar to those of a classical collapsar scenario of rapid accretion onto a newly formed black hole \citep{1999ApJ...524..262M,2001ApJ...550..410M,2018arXiv181000098S}. Neutrinos mediate the bulk of the accretion luminosity at these accretion rates \citep{1999ApJ...524..262M}, so accretion can occur onto either neutron stars or black holes under these conditions \citep{1996ApJ...460..801F}. At higher-still accretion rates, neutrino energy deposition can overturn the accretion flow \citep[e.g.][]{2012ApJ...746..106P}. This rapid accretion, over approximately the dynamical time of the core \citep{2001ApJ...550..357Z}, $10^{3}$~s, will lead a neutron star companion object to quickly collapse to a black hole. 

These conditions of hypercritical accretion from the core onto an embedded black hole set the stage for an explosion  powered by this central, accreting engine. In the context of collapsing helium stars stripped of their hydrogen envelopes, the result is a long gamma ray burst (GRB), in which a relativistic jet, perhaps powered by the coupling between the magnetic field in the accretion disk coupling to the rotational energy of the black hole (via the \citet{1977MNRAS.179..433B} process \citep[e.g.][]{2008MNRAS.385L..28B}) tunnels out of the helium star and is accelerated to high Lorentz factor \citep{1999ApJ...524..262M,2001ApJ...550..410M}. 

In the context of a helium core surrounded by an extended hydrogen envelope, the beamed power of the jet is not expected to be able to tunnel out of the envelope under most circumstances. This is because the jet head must displace the ambient stellar material and, therefore, expands at a rate which balances the ram pressure of this interaction with the momentum flux of the jet \citep[e.g.][]{ 2003MNRAS.345..575M}. Following the estimates of \citet{2012MNRAS.419L...1Q}'s Section 2.2, a typical jet whose working surface expands at a few percent the speed of light (as it displaces the surrounding stellar envelope)  reaches only approximately $0.03c \times 10^3{\rm s} \sim 10^{12}$~cm before the core-accretion event ends and the jet shuts off. 
Lacking the pressure to continue driving its expansion, the jet is choked by the surrounding gas distribution, and expands laterally, distributing its power more isotropically into the envelope material and powering an outburst \citep[e.g.][]{2014ApJ...788L...8M,2016PhRvD..93h3003S}.  We note that \citet{2012MNRAS.419L...1Q,2012ApJ...752...32W,2013ApJ...772...30D} discuss an alternate scenario in which the jet might escape if long-lived accretion from the envelope material persists over a duration of hundreds of days. The typical energies range from $10^{50}-10^{52}$~erg \citep{1998ApJ...502L...9F,2001ApJ...550..357Z}, as we discuss in Section \ref{sec:explosion_model}.  

An alternative process of stable burning has been suggested when the engulfed object is a neutron star. In these cases a Thorne-Zytkow Object is said to form -- a star with a neutron-star core, which might stably burn hydrogen for up to $10^5$~yr \citep{1977ApJ...212..832T,1995MNRAS.274..485P,2007ASPC..367..541P}, potentially showing unique surface features \citep{2014MNRAS.443L..94L}.  If the core of the helium star is near its original density when neutron star enters it, the flow convergence rate is high enough that it seems very difficult to avoid a neutrino-cooled accretion state \citep{1993ApJ...411L..33C,1998ApJ...502L...9F}. However, it is interesting to note that because the neutron star enters the core from the outside in (rather than inside-out as in a collapsar) there is a possibility that feedback from lower accretion-rate common envelope inspiral \citep[e.g.][]{1996ApJ...460..801F} would prevent the object from ever reaching the hypercritical, neutrino-cooled accreting branch \citep[see, for example, the trajectories of infall and accretion in][]{2015ApJ...798L..19M,2018ApJ...857...38H}. Thus, more work is needed to conclusively distinguish between these alternatives. 

\subsection{Model Adopted: Engine Mass and Energetics}\label{sec:explosion_model}

Before merger the donor star consists of a core of mass $M_{\rm core}$ and an envelope of mass $M_{\rm env}$. The total mass of the star is  $M_\ast = M_{\rm core} + M_{\rm env}$.
The compact-object companion that ends up in the center of the donor's core has mass $M_2 = q  M_\ast$. 
When the two stars merge, $M_2$ accretes a fraction, $f_{\rm acc}$,  of $M_{\rm core}$. This either causes it to grow if it is already a black hole, or, if it is a neutron star, collapse to a black hole, then grow to a final mass, 
\begin{equation}
M_{\rm BH, final} = M_2 + M_{\rm acc} =   M_2 + f_{\rm acc}  M_{\rm core} .
\label{eq:m_core}
\end{equation}
The released energy from the accretion is, therefore, $\Delta E_{\rm acc} = \eta M_{\rm acc} c^2 $, where $\eta \sim 0.1$. 
 Much of this energy is radiated by neutrino emission. However, some energy emerges mechanically, from a magnetohydrodynamic disk wind \citep[e.g.][]{2018ApJ...867..130F}. The mechanical power is an uncertain fraction of the accretion energy, such that $\Delta E_{\rm mec} =  f_{\rm mec}  \Delta E_{\rm acc}$. \citet{2013ApJ...772...30D} estimate $f_{\rm mec}\sim 10^{-3}$ given an inflow-outflow model in which the accreting material decreases as a power law with radius, implying 
\beq
\Delta E_{\rm mec} \sim 3\times10^{50} \left( \frac{f_{\rm mec} }{10^{-3}} \right) \left( \frac{\eta }{0.1} \right) \left( \frac{M_{\rm acc} }{2M_\odot} \right)~{\rm erg}.
\eeq
The energy that emerges in a jet via the \citet{1977MNRAS.179..433B} process can be estimated as \citep{2001ApJ...550..357Z}, 
\beq
\Delta E_{\rm BZ} \sim 10^{52}a^2  \left( \frac{M_{\rm BH} }{3M_\odot} \right)^2  \left( \frac{B }{10^{15}~{\rm G}} \right)^2 \left( \frac{t_{\rm acc} }{10^{2}~{\rm s}} \right)~{\rm erg},
\eeq
 where $B$ is the magnetic field in the disk, $t_{\rm acc}$ is the timescale of accretion, and $a$ is the black hole's dimensionless spin. Because $M_{\rm acc} \gtrsim M_{\rm BH}$, values of order unity for $a$ are expected regardless of the initial spin state. 
In what follows, we will assume that the compact object accretes the entire core mass ($f_{\rm acc}=1.0$) and we explore explosion energies between $3\times10^{50}$~erg and $10^{52}$~erg. We further assume that regardless of the precise mechanism or energetics, the energy is shared roughly spherically with the hydrogen envelope \citep{2012ApJ...752L...2C,2018MNRAS.475.1198S,2019arXiv190201187S}.

\section{Circumstellar Material Expelled During Binary Coalescence}\label{sec:merger}
%\subsection{What is this simulation (q,HSE star, 3D) - reference morgan paper}
%\subsection{show 3D gas pictures, plane + vertical}
%\subsection{extracting 1D rho-profile}
As a basis for our analysis, we model binary coalescence and the circumbinary ejecta that results from the merger of two example mass ratio binaries. Here we describe our numerical method, the unstable mass exchange that leads the binary to merge, and the CSM mass distribution that this runaway mass transfer creates. The distribution of CSM at large scales, $r\sim 10^{15}$~cm, is of particular importance for the outburst light curves. We therefore focus our numerical modeling on the early phases of runaway, unstable mass exchange that expels this largest-scale CSM.  

\subsection{Hydrodynamic Models of Binary Coalescence}\label{sec:hydro}
 Our models are simulated within the Eulerian hydrodynamic code Athena++ (Stone, J.M., in preparation), and are based on the methodology described in \citet{2018ApJ...863....5M,2018ApJ...868..136M,2018arXiv181207594M}. We use a spherical polar coordinate system surrounding the donor star in a binary pair. We model the interaction of this donor star with a softened point mass representing an unresolved companion object. The domain extends over the full three-dimensional solid angle from 0.1 to 100 times the donor-star's original radius. 
 
 The donor star is modeled by polytropic envelope, with structural index $\Gamma=1.35$. The donor has a core mass of 25\% its total mass. The gas in the simulation domain follows an ideal-gas equation of state, with index $\gamma=1.35$. These choices are intended to approximately represent a convective, isentropic envelope of a massive star in which radiation pressure is important in the equation of state \citep[in which case $\Gamma=\gamma \rightarrow 4/3$, e.g.][]{2017ApJ...838...56M,2017ApJ...845..173M}. 
 
 We initialize the calculation at a separation slightly smaller than the analytic Roche limit separation, where the donor star overflows its Roche lobe \citep{1983ApJ...268..368E}, and halt the calculation when the companion star has plunged to 10\% of the donor's original radius -- the inner boundary of our computational domain. The binary is initialized in a circular orbit and the donor star is initially rotating as a  solid body  with rotational frequency matching the orbital frequency at the Roche limit separation. 
 
The calculations themselves are carried out in dimensionless units where the donor's mass, radius, and gravitational constant are all set to unity. They may, therefore, be rescaled to a physical binary of any mass or size. Below, we report on two models which have secondary to donor star mass ratios of $q\equiv M_2/M_\ast=0.1$ and $q=0.3$. As we will see in Section \ref{sec:pop}, $q\sim 0.1$ is a very common mass ratio, while $q=0.3$ is near the upper end of the range of events that result in mergers.

%\begin{itemize}
%\item Athena++
%\item previous papers full method
%\item physical system/coordinate/unit description
%\item particulars of systems simulated
%\end{itemize}

\subsection{Unstable Mass Transfer Leading to Binary Merger}

\begin{figure}[tbp]
\begin{center}
\includegraphics[width=0.44\textwidth]{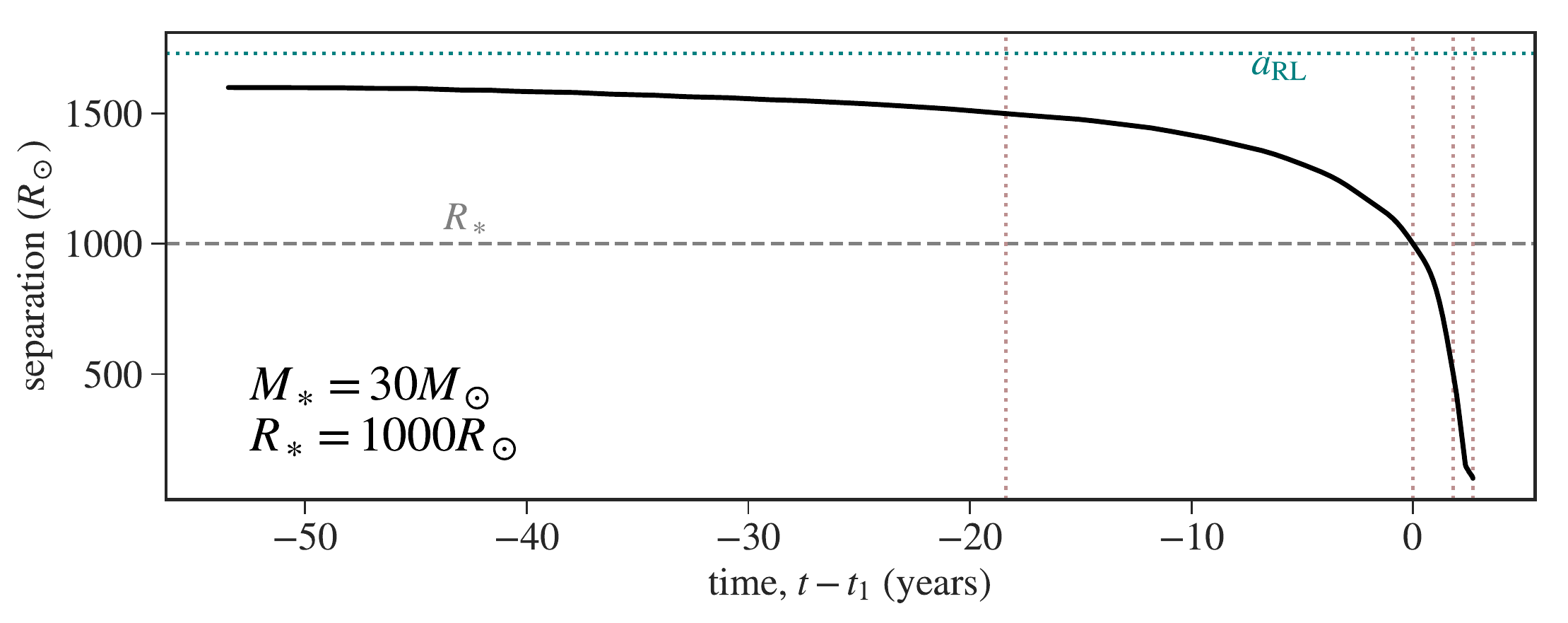}
\includegraphics[width=0.49\textwidth]{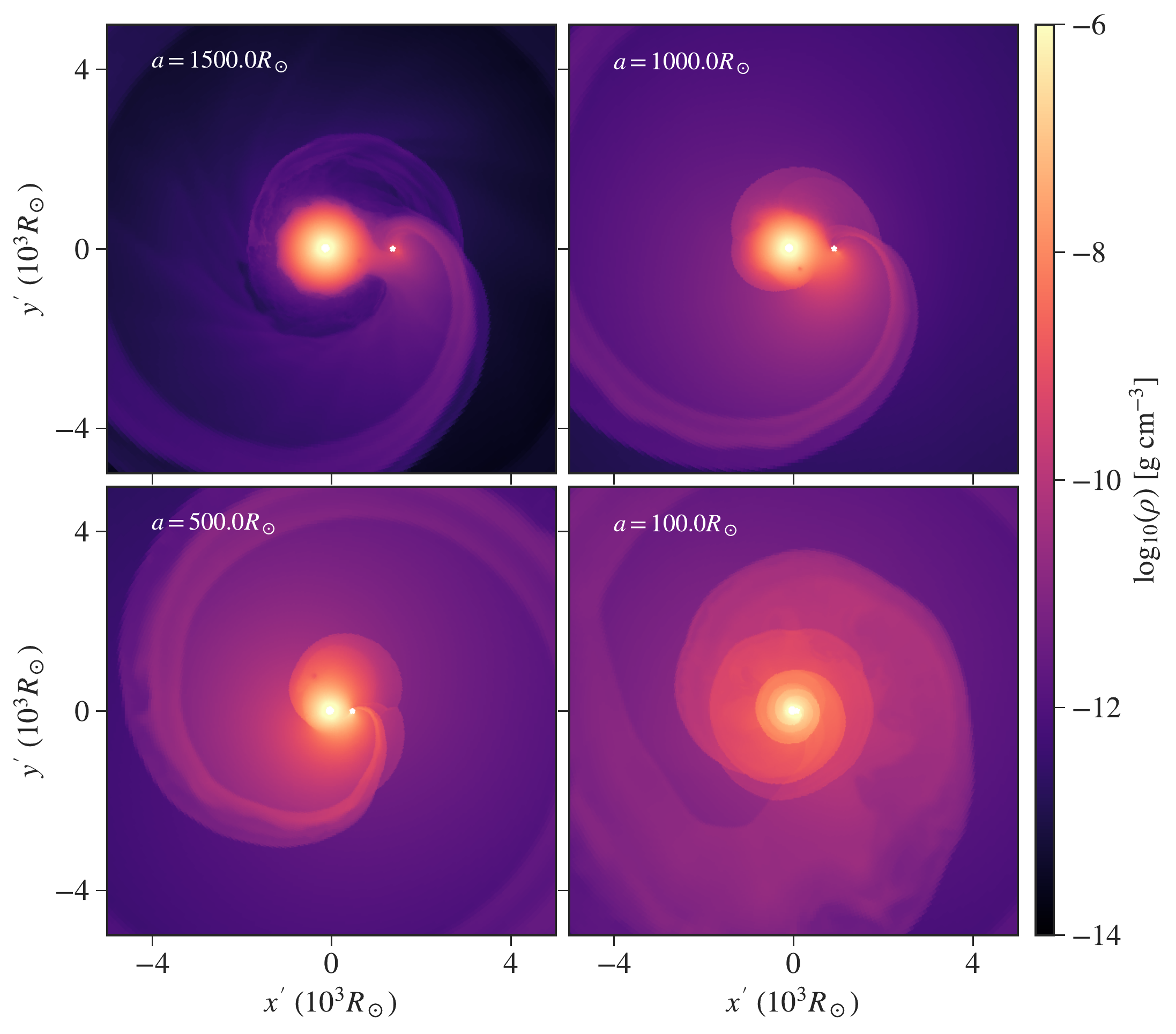}
\caption{Runaway Roche lobe overflow leading to binary coalescence for binary mass ratio $q=0.1$. The upper panel shows binary separation as a function of time during our simulation. The lower panel shows four snapshots (marked with vertical lines in the upper panel) of the gas density distribution in the orbital plane. We show these slices through the orbital plane in a rotated $x'-y'$ coordinate system so that the companion star lies along the $+x$-axis. The dimensionless simulations are rescaled to donor star mass of $30 M_\odot$ and radius of $1000R_\odot$ for specificity in these images. Following Roche lobe overflow, mass is pulled from the donor star and expelled from the binary system dragging the binary to tighter separations.  }
\label{fig:ts}
\end{center}
\end{figure}

Mass transfer is unstable in our model binary system in that it runs away to ever increasing rates and drives the binary toward merger. This process begins with Roche lobe overflow of the donor star into the vicinity of its companion. In general, mass transfer proceeds unstably when the loss of material from the donor causes the donor star to increasingly overflow its Roche lobe -- either because it grows in radius, or because its Roche lobe shrinks. In binary systems such conditions are often realized in binary pairs where a more massive donor star transfers mass onto a less massive accretor, causing the binary separation to shrink.

In Figure \ref{fig:ts}, we show the binary system separation as a function of time in our model system, and snapshots of the gas density distribution in the orbital plane. In these figures, we have rescaled our dimensionless simulations to a fiducial donor star mass of $30 M_\odot$ and radius of $1000R_\odot$. In the upper panel, time is zeroed at the time at which the companion plunges within the original donor-star's radius, $t_1$, where $a(t_1)=R_\ast$. Over the preceding 50 years, the binary separation continuously shrinks, at first gradually, but with increasing rapidity \citep{2018ApJ...863....5M}. After the companion object plunges within the donor's envelope it spirals to the inner boundary of our computational domain (at 10\% the donor's radius) within about 5 orbital cycles, or two years, approximately the orbital period at the donor star's surface. 

Mass loss from the donor star at the expense of orbital energy drives this rapid decrease in binary separation and the pair's coalescence. Turning our attention to the lower panels of Figure \ref{fig:ts}, we note that as the donor star overflows its Roche lobe, material is pulled, primarily from the vicinity of the $L_1$ Lagrange point, toward the companion object. As the orbital separation decreases, from $1500R_\odot$ to $1000R_\odot$ to $500R_\odot$, the breadth and intensity of this mass transfer stream increase dramatically. \citet{2018ApJ...863....5M} studied the dynamics of this runaway, unstable Roche lobe overflow in detail, and found that the mass loss rate from the donor increases by orders of magnitude over this period. However, \citet{2018ApJ...863....5M} also show that the analytic model of \citet{1972AcA....22...73P} coupled to a point-mass binary orbit evolution model captures the key features of these stages once the specific angular momentum of the ejecta has been measured \citep[e.g.][]{1963ApJ...138..471H}. 

These high mass exchange rates quickly exceed the Eddington limit mass accretion rate that material can accrete onto a compact object companion, and much of the material pulled from the donor is lost to the circumbinary environment \citep[as seen in the snapshots of Figure \ref{fig:ts}, though in the case of the simulation this is because accretion onto the companion object is not modeled, see][]{2018ApJ...868..136M}. Much of this mass loss occurs near the $L_2$ Lagrange point, near the lower-mass companion object \citep{1979ApJ...229..223S,2016MNRAS.455.4351P,2016MNRAS.461.2527P,2017MNRAS.471.3200M,2018ApJ...868..136M}. By the final panel, where the separation is $100R_\odot$, the core of the original donor and companion object are mutually immersed in a significantly extended common envelope that originated from the donor star \citep{1976IAUS...73...75P}. Once immersed, some binary systems deposit enough energy into their environments to expel this envelope. Others do not, and the companion object merges with the core -- powering the sort of engine-driven explosions discussed in Section \ref{sec:explosion}. 

Because orbital tightening and coalescence of the binary system is a direct result of angular momentum loss to ejected material, the amount of ejecta relates directly to the binary properties. First using semi-analytic scalings \citep{2017ApJ...835..282M}, then hydrodynamic simulation results \citep{2018ApJ...868..136M}, we have found that the expelled mass at the onset of coalescence (defined as mass at $r>R_\ast$ at $t=t_1$) is always on the order of 25\% the mass of the merging companion object \citep[Section 4.2 of][]{2018ApJ...868..136M}. In the calculation shown in Figure \ref{fig:ts}, which has a mass ratio $q=0.1$, at $t_1$ the ejecta mass (measured as the mass at radius greater than the donor's original radius) is 16\% the companion's mass, or approximately $0.49M_\odot$. At the termination of our calculation, when the separation has decreased by a further factor of ten, the ejecta mass has increased to roughly 150\% of the companion object's mass, or $4.45M_\odot$. By comparison, our calculation with $q=0.3$ expels nearly identical percentages of mass relative to the more massive black hole:  $1.44M_\odot$ at a separation equal to the donor's radius and $13.9M_\odot$ in our final snapshot (separation 10\% the donor's radius).  In the following, we analyze the distribution of this material in the circumstellar environment.

\subsection{Resultant Circumstellar Distribution}

\begin{figure*}[tbp]
\begin{center}
\includegraphics[height=8cm]{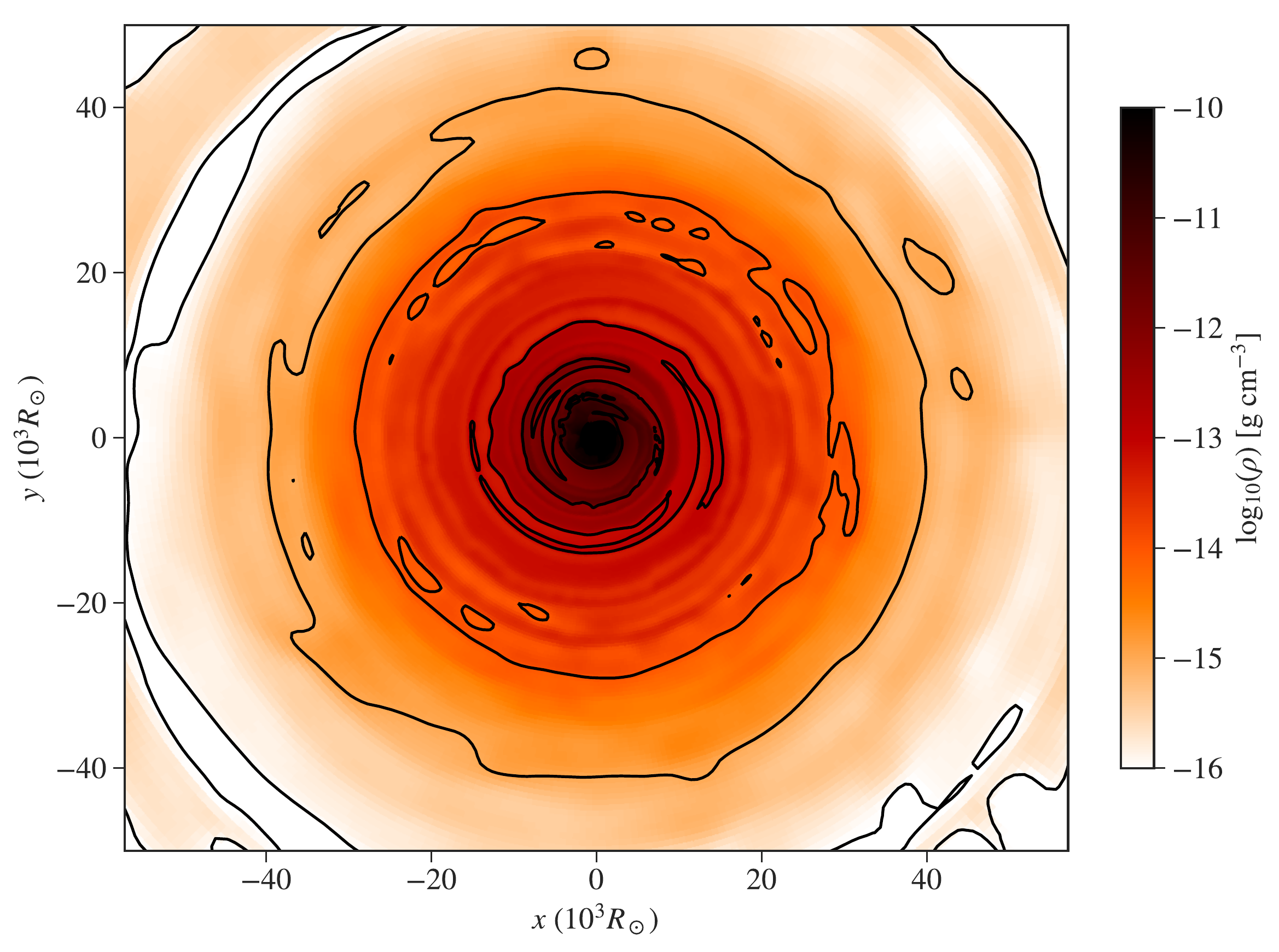}
\hspace{0.3cm}
\includegraphics[height=8cm]{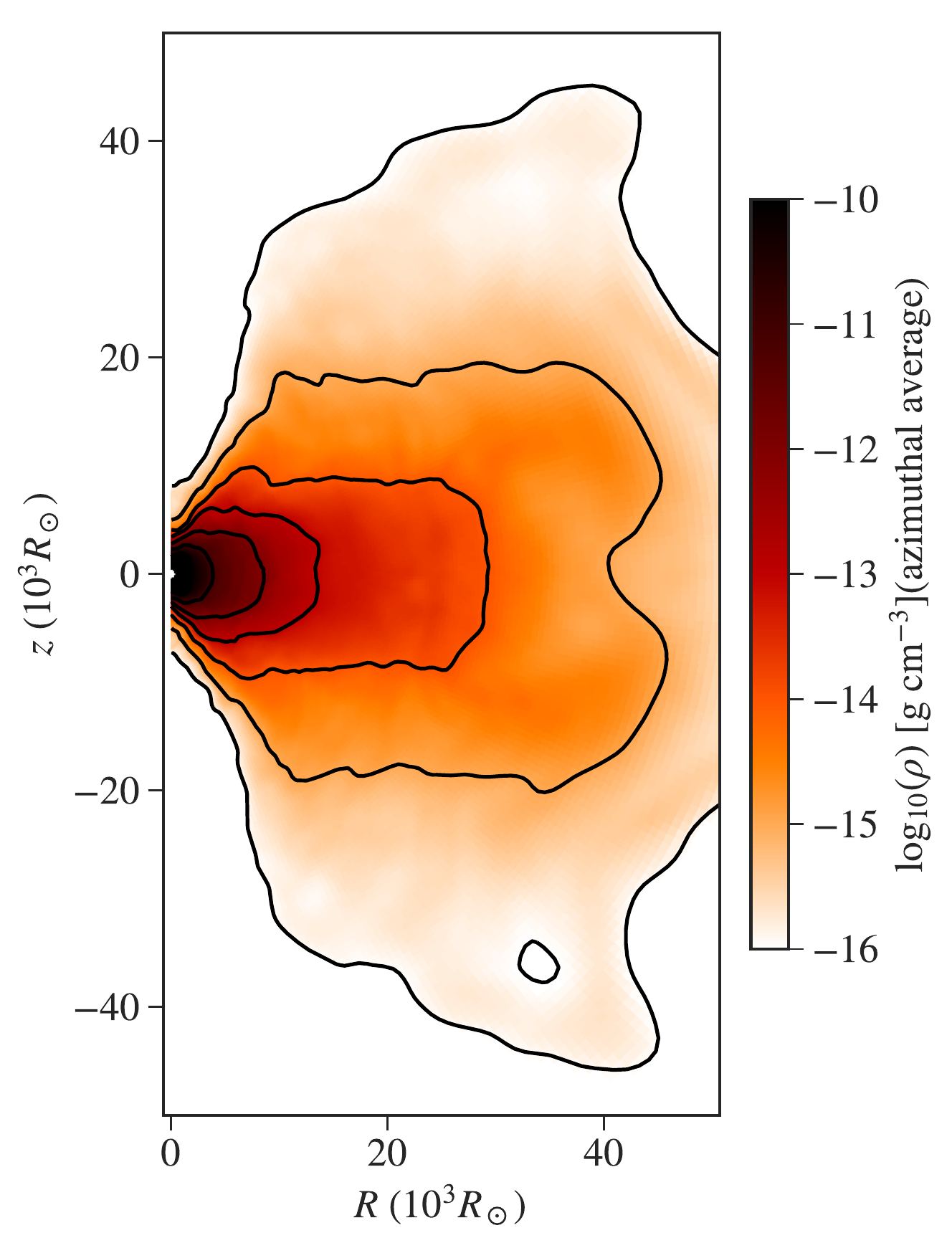}
\caption{Three dimensional distribution of ejecta near the time of merger. As in Figure \ref{fig:ts}, we have scaled to a fiducial donor mass of $30 M_\odot$ and radius of $1000R_\odot$.  When the compact object merges with the donor's core, it is surrounded by an extensive, thick torus of debris expelled by the merger itself. }
\label{fig:largescale}
\end{center}
\end{figure*}

Next, we analyze the three-dimensional distribution of debris expelled by the merger episode. To do so, we analyze the final snapshot of our hydrodynamic simulation, when the separation has tightened to one-tenth the donor's original radius, or $100R_\odot$ in our fiducial, $30M_\odot$, $1000R_\odot$ model. 
Because the binary separation is tightening extremely rapidly at this phase, material ejected subsequently in the merger does not affect the largest-scale gas distribution prior to the compact object's merger with the core. 
 This, therefore, is the CSM that any explosive outburst will interact with as it expands, particularly when we consider the crucial scales of interest of $10^{14} $ to $10^{15}$~cm that lie near the photosphere of the explosive transient.

In Figure \ref{fig:largescale}, we show the large-scale density distribution out to 50 times the initial donor radius ($5\times10^4R_\odot$, or approximately $3.5\times10^{15}$~cm). The panels show a slice through the orbital, $x-y$ plane, and an aziumthal average, plotted in $z-R$, perpendicular to the orbital plane. Figure \ref{fig:largescale} shows that a thick, extended circumbinary torus of expelled material from the donor's envelope has formed around the merging pair of stars. This torus is roughly azimuthally symmetric, but has distinct structure in polar angle, with relatively evacuated poles and dense equator, representative of the fact that material is flung away from the merging binary in the equatorial plane. \citet{2016MNRAS.455.4351P,2016MNRAS.461.2527P, 2017ApJ...850...59P} analyzed the thermodynamics of similar outflows and show that heating, arising  continuously from internal shocks, and radiative diffusion and cooling very likely regulate the torus scale height. Thus, the precise scale height observed in Figure \ref{fig:largescale}, modeled under the simplification of an ideal-gas equation of state, would likely be modified by the inclusion of more detailed physics. 

\citet{2018ApJ...868..136M} analyzed the kinematics of this torus material and found that the radial velocities of the most extended material are low relative to the escape velocity of the original donor star (roughly 100~km~s$^{-1}$ for our fiducial model). Thus, the majority of these (earlier) ejecta around bound to the merging binary. Some of the material at smaller radii (the later ejecta) is moving more rapidly, at velocities similar to the escape velocity. It therefore collides with the earlier, slow moving ejecta \citep{2018ApJ...868..136M}. Qualitatively, these velocities are similar to other sources of stellar mass loss like winds or non-terminal outbursts, in that they are similar to the giant star's escape velocity and are much less than the later explosion's blast wave velocity. 

Though the axisymmetric torus structure discussed above is clearly structured in polar angle, for the sake of computational efficiency, we model the interaction of the explosive blast wave with a one-dimensional (spherically-symmetric) density distribution derived from these models. In future work, it may be extremely interesting to relax this simplification.  To derive one-dimensional profiles, we spherically-average our model results about the donor star's core. 

\begin{figure}[tbp]
\begin{center}
\includegraphics[width=0.48\textwidth]{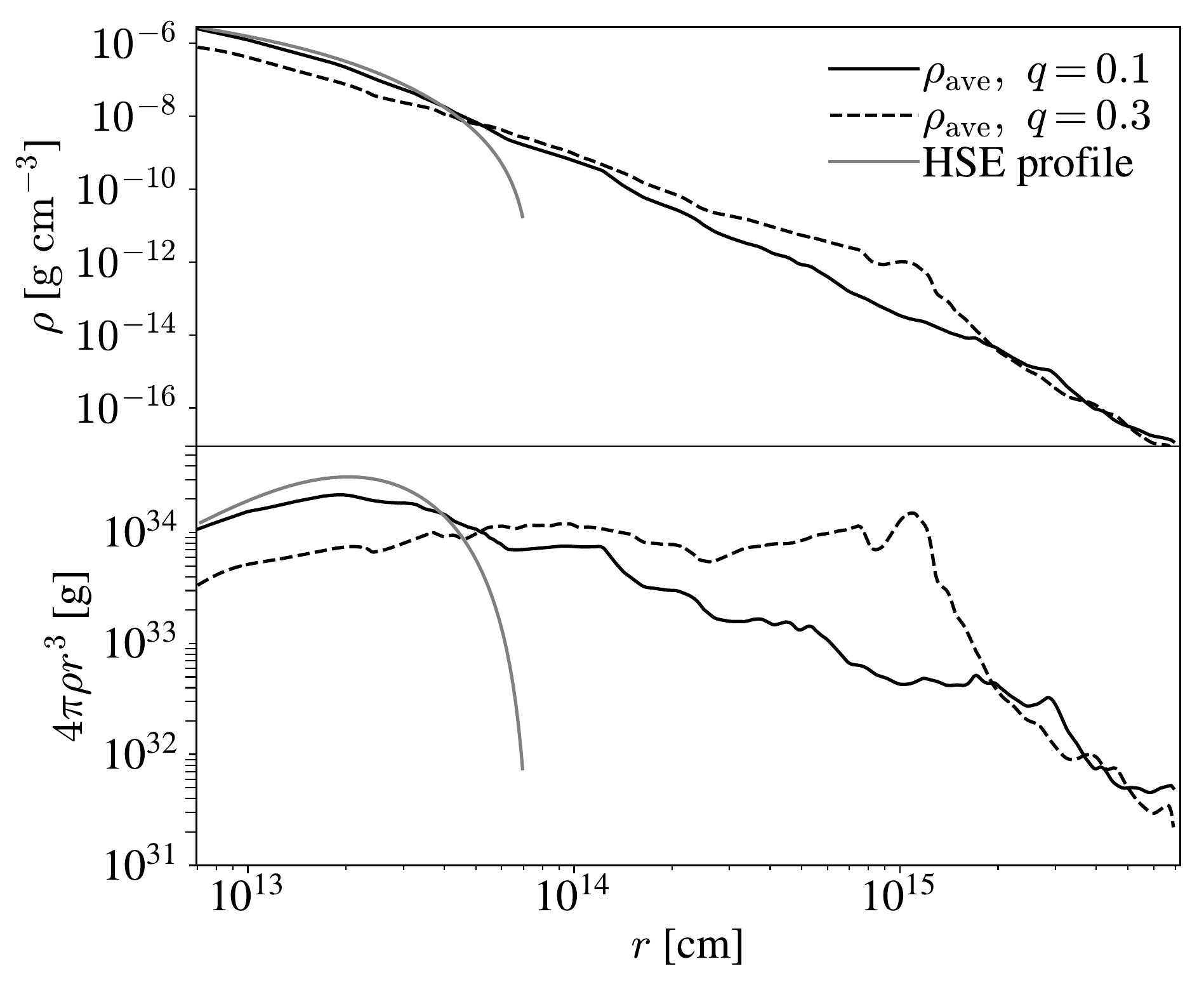}
\caption{Spherically-averaged density distributions, comparing the initial, hydrostatic polytrope (labeled HSE), with merger-simulation snapshots for $q=0.1$ and $q=0.3$. As in the previous Figures, we scale our models to a donor of $30 M_\odot$ and $1000R_\odot$.  The existence of significant quantities of mass near the type II supernova photosphere radius of $10^{15}$~cm implies that interaction with this medium will play an important role in explosive transient light curves.  }
\label{fig:1D}
\end{center}
\end{figure}

These 1D density profiles are shown in Figure \ref{fig:1D}, in which we compare the unperturbed envelope profile to the cases disturbed by binaries of $q=0.1$ and $q=0.3$. Where the hydrostatic profile has a distinct limb at the donor's radius, the post-merger profiles show a roughly power-law slope in radius, with approximate scaling of $r^{-3}$, as shown in the lower panel. Comparing the $q=0.1$ and $q=0.3$ results shows that in the higher mass-ratio coalescence, more of the envelope material has been expelled beyond the donor's original radius, yielding a shallower density fall-off with radius and a profile with more mass at large radii. In both cases, we see that the distribution of ejecta extends to roughly $10^{15}$~cm, with of order a solar mass  ($q=0.3$) or a tenth of a solar mass ($q=0.1$) on these scales.

\section{Merger-Driven Light Curves}\label{sec:lightcurves}

In this section we use analytic and numerical models to understand the properties of  merger-driven light curves.  We find that the CSM distribution plays a crucial role in shaping these light curves. In Section \ref{sec:analytic} we provide some analytic context for the potential role of CSM. In Section \ref{sec:SNEC}, we describe our numerical method for 1D radiative hydrodynamics calculations and light curve generation. Finally, in Sections \ref{sec:csm} and \ref{sec:param} we describe the key features and variations across parameter space of these numerical model light curves. 

\subsection{Analytic Context for Contribution to Radiative Efficiency from CSM Interaction}\label{sec:analytic}

In order to provide context for the interpretation of our numerical light curve models, in this section we analyze simplified analytic models of emission powered by CSM interaction.  In general, CSM interaction enhances the intrinsic luminosity of ``cooling emission" from heated, expanding material like supernova ejecta. As ejecta shock-heat by collisions with CSM,  their kinetic energy is dissipated and converted into radiation.  Depending on the location of the interaction (within or outside the photosphere), this radiation may either escape immediately or adiabatically decay with expansion in the outflow prior to being free to stream out. 

In particular, we focus on different radial density profiles of CSM material and the role this plays in shaping transient light curves. In doing so, we summarize and build on a considerable literature that describes how CSM interaction can form a significant contribution or even dominate the radiative power of a transient under certain conditions \citep[e.g.][]{1994MNRAS.268..173C,2007ApJ...671L..17S,2011ApJ...729L...6C,2012ApJ...752L...2C,2012ApJ...747L..17C,2012ApJ...757..178G,2012ApJ...756L..22M,2013MNRAS.428.1020M,2013MNRAS.435.1520M,2013MNRAS.433..838P,2014ApJ...788..154O,2014ApJ...780...18G,2015ApJ...814...63M,2017ApJ...838...28M,2018ApJ...858...15M,2018ApJ...867....4M,2018MNRAS.475.3152K, 2018SSRv..214...27C}. 

\subsubsection{Thin Shell of CSM}

In the simplest version of a CSM interaction, an additional internal energy $\Delta E$ is added to ejecta by sweeping up a thin shell of CSM,
\beq
\Delta E  \approx \frac{dM_{\rm CSM}}{M_{\rm tot}} E,
\eeq
where $M_{\rm tot}$ is the sum of the explosive ejecta and the swept-up CSM mass internal to the shell, $dM_{\rm CSM}$ is the CSM shell mass, and $E$ is the kinetic energy of the explosive ejecta. If this deposition of internal energy occurs in optically thin regions, all of this energy is radiated, and $\Delta E_{\rm rad} \approx \Delta E$. If the CSM shell lies interior to the photosphere radius, the heated ejecta must continue to expand, with gas internal energy decaying along an adiabat, before they are free to radiate. If we assume that gas specific internal energy decays adiabatically as $r^{-1}$ prior to reaching the photosphere  (which is the case when radiation pressure dominates and $P\propto \rho^{4/3}$ along an adiabat), then $\Delta E_{\rm rad} \approx (r/R_{\rm ph}) \Delta E$.

\subsubsection{Continuous Distributions of CSM}
The differential radiated energy due to sweeping up a CSM mass $dM_{\rm CSM}$ at radius $r$ interior to the photosphere radius is 
\beq
\frac{dE_{\rm rad}}{dM_{\rm CSM}} \approx \frac{E}{M_{\rm tot}}\frac{r}{R_{\rm ph}},
\eeq
where $E_{\rm rad}$ is the contribution to the radiated energy due to CSM interaction alone. As before, we are also assuming that gas specific internal energy decays adiabatically as $r^{-1}$.  Given a continuous distribution of CSM material distributed between $R_\ast$ and $R_{\rm ph}$, we can integrate this expression over radius. In what follows, we will assume that $E$ is constant, which is justified only if $M_{\rm CSM} \ll M_{\rm tot}$ and $E_{\rm rad} \ll E$. Otherwise, the losses in kinetic energy to thermal energy or radiation must be taken into account.   We replace $dM_{\rm CSM}=4\pi r^2 \rho dr$, where $\rho$ is the CSM density, to write
\begin{align}
\frac{E_{\rm rad}}{E} &\approx \int_{R_\ast}^{R_{\rm ph}} \frac{r}{R_{\rm ph}} \frac{4\pi r^2 \rho }{M_{\rm tot}} dr, \nonumber \\
&\approx \frac{4\pi}{R_{\rm ph} M_{\rm tot} } \int_{R_\ast}^{R_{\rm ph}} \rho r^3  dr. 
\label{eq:Erad}
\end{align}
 This integral shows that the dependence of $\rho(r)$ will be critical in determining the CSM contribution to the radiated  luminosity.

Let us write a general, power law density form for the CSM that applies from the stellar radius, $R_\ast$, to the eventual photosphere radius, $R_{\rm ph}$, 
\beq
\rho(r) = \rho_{\rm ph} \left( \frac{r}{R_{\rm ph}} \right)^{-n},
\eeq
 where $\rho_{\rm ph}$ is the density at the photosphere radius, and we have chosen $R_{\rm ph}$ as a characteristic radius to normalize the power law. We will further adopt the approximation that $R_{\rm ph}\gg R_\ast$, under which the total CSM mass can be written,
\begin{equation}\label{Mcsm}
M_{\rm CSM} \approx 4\pi\rho_{\rm ph} R_{\rm ph}^3 \times 
    \begin{cases}
    1  & n=2, \\
    \ln ( R_{\rm ph} / R_\ast ) & n=3, \\
    R_{\rm ph}/R_\ast & n=4,
    \end{cases}
\end{equation}
for several representative values of $n$. 

The most frequently considered form of $\rho$ is that of a steady, spherical wind, $n=2$. Then, equation \eqref{eq:Erad} becomes 
\begin{align}
\frac{E_{\rm rad}}{E} &\approx \frac{4\pi \rho_{\rm ph} R_{\rm ph} }{ M_{\rm tot} } \int_{R_\ast}^{R_{\rm ph}} r  dr, \nonumber \\ 
&\approx \frac{4\pi \rho_{\rm ph} R_{\rm ph}^3 }{2 M_{\rm tot} }, \nonumber \\ 
& \approx \frac{1}{2} \frac{M_{\rm CSM}}{M_{\rm tot}} \ \ \ \  ({\rm for} \ n=2),
\label{eq:Erad2}
\end{align}
thus retrieving the often quoted result of the increase in radiated energy scaling with the CSM mass as a fraction of the total mass \citep{2011ApJ...729L...6C, 2012ApJ...757..178G, 2013MNRAS.433..838P}.

In our $q=0.3$ model, the one-dimensional profile approximates $n=3$, which yeilds constant mass per logarithmic increase in radius. For $n=3$, equation \eqref{eq:Erad} evaluates to
\begin{align}
\frac{E_{\rm rad}}{E} &\approx \frac{4\pi \rho_{\rm ph} R_{\rm ph}^2 }{ M_{\rm tot} } \int_{R_\ast}^{R_{\rm ph}}   dr, \nonumber \\ 
&\approx \frac{4\pi \rho_{\rm ph} R_{\rm ph}^3 }{ M_{\rm tot} }, \nonumber \\ 
& \approx   \frac{1}{ \ln \left( \frac{R_{\rm ph}}{R_\ast}  \right)} \frac{M_{\rm CSM}}{M_{\rm tot}} \ \ \ \  ({\rm for} \ n=3),
\label{eq:Erad3}
\end{align}
where, in the last line, we have used $M_{\rm CSM}$ from equation \eqref{Mcsm}. First, we emphasize that the ratio of $R_{\rm ph}$ to $R_\ast$ now affects the radiated luminosity arising from CSM interaction, which is not the case for $n=2$.  Therefore, when $R_{\rm ph} \gg R_\ast$, the radiated energy from this CSM profile is considerably less than that of the $n=2$ profile. 
A final way to interpret this result is in terms of the CSM mass at radii similar to the photosphere radius, $r\sim R_{\rm ph}$. From equation \eqref{Mcsm}, this is approximately $M_{\rm CSM} / \ln ( R_{\rm ph} / R_\ast)$. Equation \eqref{eq:Erad3} shows that this is the fraction of the CSM mass that contributes significantly to the radiated energy. 

In our $q=0.1$ merger model, the scaling of the CSM is steeper, approximately  $\rho \propto r^{-4}$. We reevaluate equation \eqref{eq:Erad} for $n=4$ to find, 
\begin{align}
\frac{E_{\rm rad}}{E} &\approx \frac{4\pi \rho_{\rm ph} R_{\rm ph}^3 }{ M_{\rm tot} } \int_{R_\ast}^{R_{\rm ph}}  \frac{1}{r} dr, \nonumber \\ 
&\approx \frac{4\pi \rho_{\rm ph} R_{\rm ph}^3 }{ M_{\rm tot} } \ln \left( \frac{R_{\rm ph}}{R_\ast}  \right), \nonumber \\ 
& \approx   \left(\frac{R_\ast}{R_{\rm ph}}\right) \ln \left( \frac{R_{\rm ph}}{R_\ast}  \right) \frac{M_{\rm CSM}}{M_{\rm tot}} \ \ \ \  ({\rm for} \ n=4).
\label{eq:Erad4}
\end{align}
Thus, for $R_{\rm ph} \gg R_\ast$, the CSM contribution to radiated luminosity is less for $n=4$ than either $n=3$ or $n=2$ given a CSM mass. This result can be interperted in light of equation \ref{Mcsm}, which shows that the fraction of CSM mass with $r\sim R_{\rm ph}$ is $R_\ast/R_{\rm ph}$ for $n=4$.

\subsubsection{Interpretation}\label{sec:csm_interp}
 
In the preceding subsection, we have shown that for CSM density profiles that are sufficiently steep, $n\geq3$,  the CSM contribution to the radiated luminosity $E_{\rm rad}/E$ depends on the ratio of the stellar radius over the photosphere radius -- the radial extent of the CSM. This important ratio varies  in explosions with different size stars of similar mass, or over the time evolution of a given transient as the photosphere radius increases.
If $E_{\rm rad}/E$ becomes too small, then CSM interaction does not contribute significantly to the light curve of the transient at a given phase and the bulk of the radiated luminosity comes instead from the adiabatically-expanding blast wave. In this case, the transient assumes more typical supernovae type IIP properties.

 These scalings indicate that we expect the CSM to be an important contribution to the $q=0.3$ merger case light curve, because the mass in the CSM is a significant fraction of the total envelope mass, and with $n=3$ there is only logarithmic dependence on $R_{\rm ph}/R_\ast$, equation \eqref{eq:Erad3}.  In the $q=0.1$ merger case, in which $n=4$, we expect preferential contribution from CSM interaction at early times in the transient light curves or for particularly extended donors, because either situation maximizes the ratio $R_\ast/R_{\rm ph}$, see equation \eqref{eq:Erad4}. 
 
  Finally, we have so far discussed the case in which the CSM distribution extends out to, and perhaps beyond the photosphere radius. This is not necessarily realized. In the case where the CSM terminates at a radius $R_0$, for which $R_0<R_{\rm ph}$, the radiated luminosity from CSM interaction is computed much as before, but only integrating the mass distribution out to $R_0$. This reduces the radiated luminosity due to CSM interaction by a factor similar to $R_0/R_{\rm ph} < 1$.  
 In the sections that follow, we use this framework to interpret 1D radiative transfer models of explosions interacting with our model CSM distributions.

\subsection{1D Radiation Hydrodynamic Models}\label{sec:SNEC}

While the analytic approach highlighted above is useful, it is necessarily simplified. To extend these calculations of the CSM imprint on transient light curves to slightly more realistic scenarios, we need to perform the associated integrations numerically.   We utilize the publicly available spherically symmetric (1D) Lagrangian hydrodynamics code SuperNova Explosion Code (SNEC) to calculate  bolometric and filtered light curves \citep{2015ascl.soft05033M,2015ApJ...814...63M,2017ApJ...838...28M}. The code uses equilibrium-diffusion radiation transport to follow the time dependent radiation hydrodynamics of the expanding blast wave. We choose to set the equation of state using the built-in version of the \citet{1983ApJ...267..315P} equation of state, which includes contributions to the total pressure from radiation, ions and electrons based on the composition.

We map our one-dimensional, spherically-averaged profiles, shown in Figure \ref{fig:1D}, into SNEC. Mass grid cells are customized by the user's choice of binning in mass, and we have found that the density profiles' steep decline is best simulated with increasingly fine mass resolution at larger radii.   This means that the shock break out is not well resolved \citep{1992ApJ...393..742E,2015ascl.soft05033M}, but light curve calculations after the first day are robust as shown by \citet{2015ApJ...814...63M}. We run the simulations for this paper with 456 grid cells.

Even so, the code does not function well for the lowest densities. We therefore are required to restrict the CSM  density profile to $\rho > 10^{-12}$~g~cm$^{-3}$ \citep{2018ApJ...867....4M}. The outer radius is therefore $2.6 \times 10^{14}$~cm for the fiducial simulation with $q=0.1$ and $1.0 \times 10^{15}$~cm for the simulation with $q=0.3$.  We note that this restriction is not ideal because it limits the potential interaction-driven luminosity of our models, see equations \eqref{eq:Erad3} and \eqref{eq:Erad4}.  In practice, this implies that the CSM maximum radius is often less than the photosphere radius, $R_0<R_{\rm ph}$, and the CSM contribution to the eventual radiated luminosity is reduced accordingly, see the discussion of Section \ref{sec:csm_interp}.   Nonetheless, the truncated profiles do retain more than 95\% of the CSM mass in all parameter variations. 
We assume a roughly solar isotopic composition that matches the hydrogen envelope of a presupernova stellar model of an initially $15 M_\odot$ star evolved with the MESA code that is included in the SNEC distribution.

To drive the explosion of our models, we adopt a thermal bomb at the inner edge of the envelope domain.  This broadly mimics the energy deposition of the quenched jet into the hydrogen envelope, as described in Section \ref{sec:explosion}. Based on \citet{2015ApJ...814...63M}, we deposit the energy over the innermost $\Delta \mathrm{M_{bomb}} = 0.1 \mathrm{M_\odot}$ over a duration $\Delta \mathrm{t_{bomb}} = 0.1$~s \citep[][has shown that the model light curves are not particularly sensitive to these parameters, see their Figure 5]{2015ApJ...814...63M}. The code also offers the option of adding Nickel to the composition. Though Nickel and lanthanide production is possible in merger-driven explosions \citep[e.g.][]{2018arXiv181000098S,2018arXiv181003889G}, the quantities are uncertain. Because decay of radioactive $^{56}$Ni mainly powers the late-time emission, here we choose to focus on Ni-free models of the early light curve dominated by the CSM and the hydrogen envelope \citep{2015ApJ...814...63M}. Lacking radioactive material in the ejecta, our models decline rapidly after the ejecta become fully transparent.  Emission from the photosphere in SNEC is assumed to follow a thermal blackbody, and thus neglects some line-blanketing effects that may be important for iron-rich ejecta in the U and B bands. 

Finally, in Appendix \ref{sec:appendix}, we test the sensitivity of our model results to these choices by varying the inner mass (or equivalently, radius) at which energy is deposited, as well as the mass-resolution of the SNEC calculation.

%Finally, we note that the initial  shock breakout is not well resolved, because it happens in the outermost cell \citep{1992ApJ...393..742E,2015ascl.soft05033M}, however, as the photosphere proceeds inward in mass, the model light curves become robust.

\subsection{Imprint of CSM}\label{sec:csm}

\begin{figure*}[tbp]
\begin{center}
\includegraphics[width=0.46\textwidth]{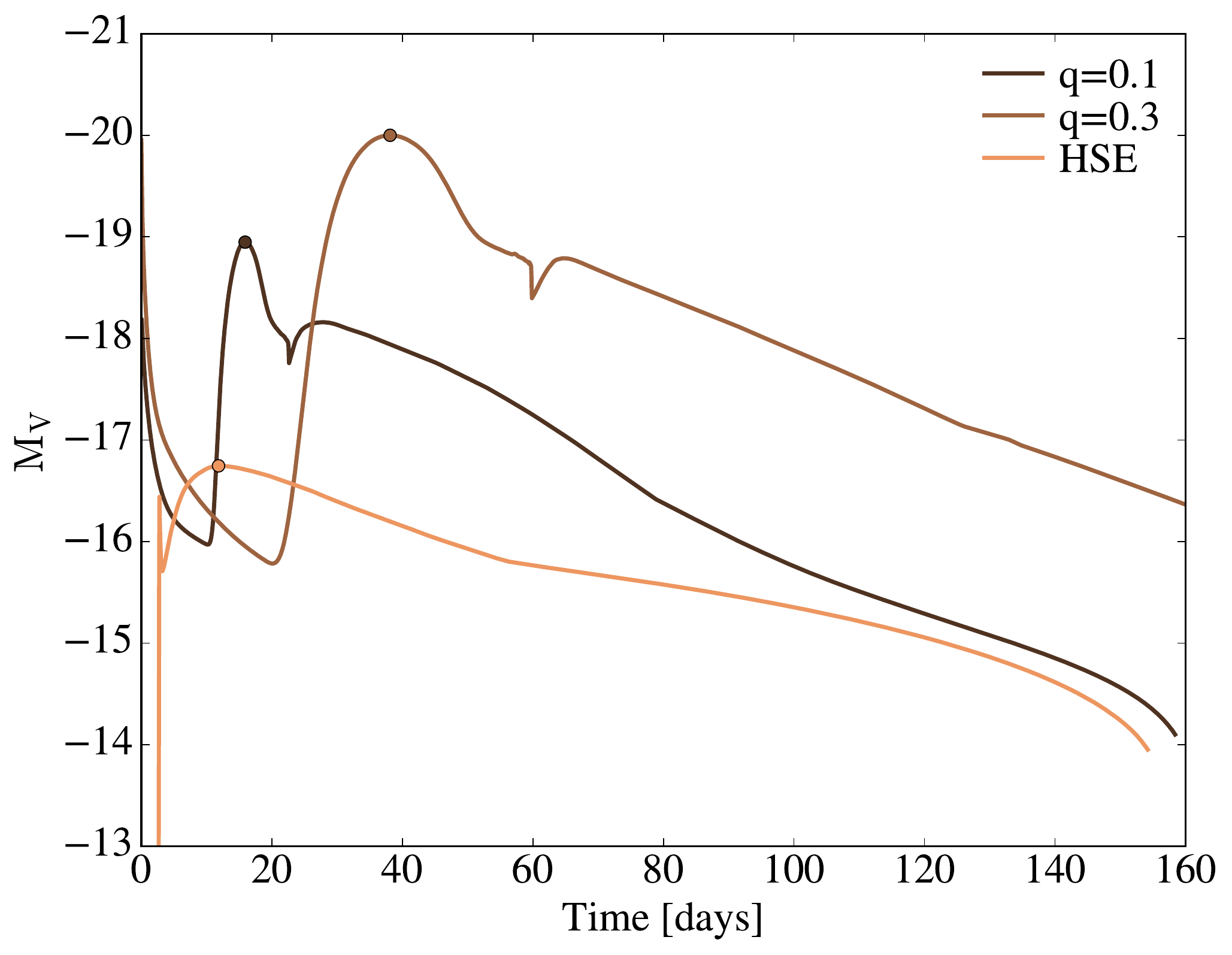}
\includegraphics[width=0.5\textwidth]{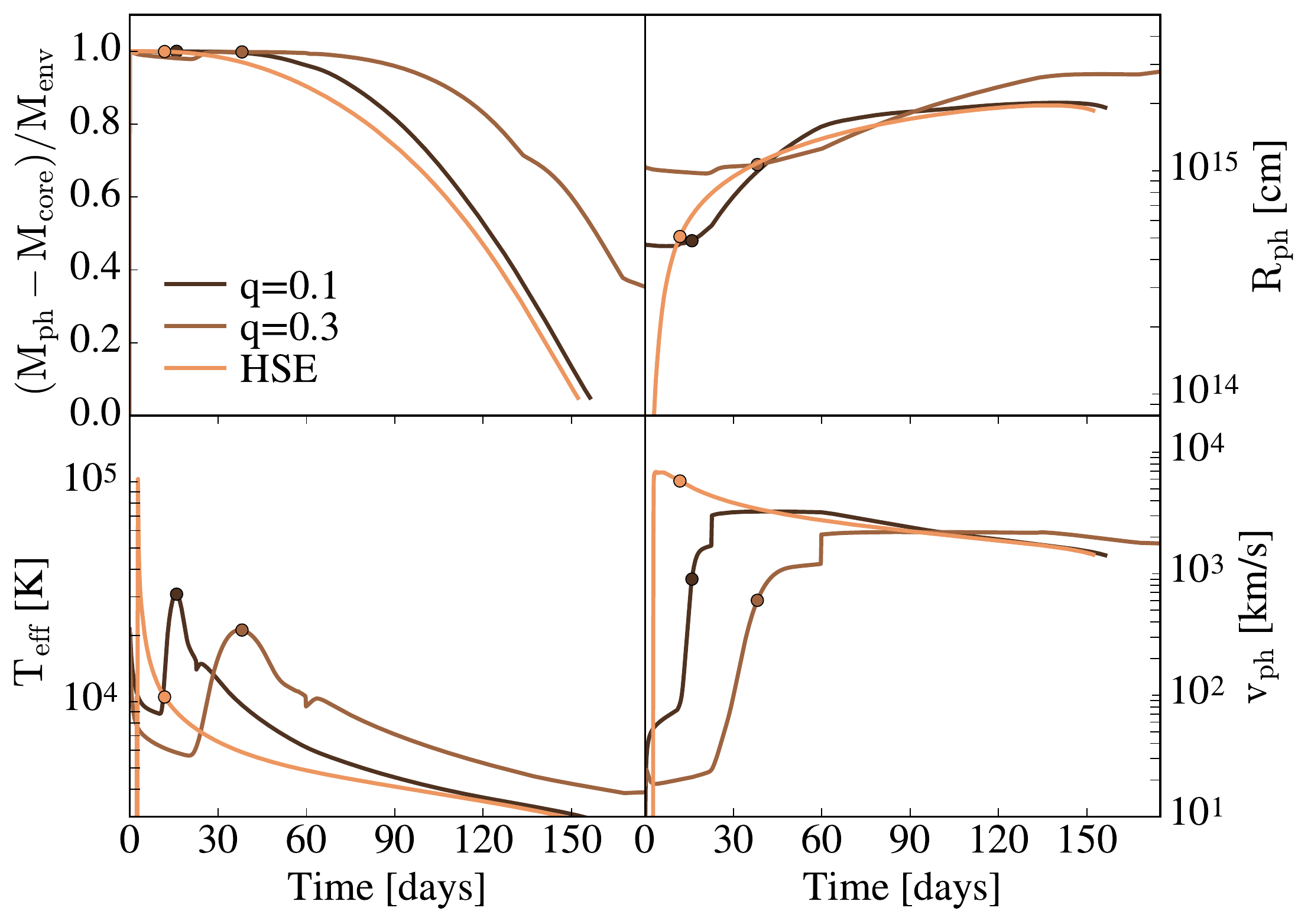}
\caption{Fiducial, $30M_\odot$, $1000R_\odot$ model star undergoing explosion of $10^{51}$~erg in three scenarios: in its initial hydrostatic state (HSE), after merging with a $3M_\odot$ black hole ($q=0.1$), and after merging with a $9M_\odot$ black hole ($q=0.3$).  The large panel shows bolometric luminosity, while the smaller panels track the photosphere's properties -- its mass element within the ejecta, radius, effective temperature, and bulk velocity at its location. CSM interaction brightens the merger models significantly as compared to the HSE case. Points mark the time of peak V-band brightness. }
\label{fig:q_HSE}
\end{center}
\end{figure*}

We begin to explore the imprint of the CSM mass distribution on the explosion light curve in Figure \ref{fig:q_HSE}, in which we plot luminosities, along with photosphere radii, effective temperature, and gas bulk velocity at the photosphere radius for our fiducial case of a $30 M_\odot$ and $1000R_\odot$ donor star in its initial, hydrostatic equilibrium state (labeled HSE), and following merger with a $3M_\odot$ ($q=0.1$) or $9M_\odot$ ($q=0.3$) black hole. In each case, the explosion energy is taken to be $10^{51}$~erg, and is injected at 0.1 times the radius of the original donor star, the innermost radius resolved in our hydrodynamical models.

As predicted by the analytic scalings in Section \ref{sec:analytic}, the models with CSM are significantly more luminous at peak (by a factor of roughly 100) than the hydrostatic model.  The CSM interaction models also show  delayed time of peak brightness,  modified colors, and light curve shapes, as we discuss in what follows. \citet{2017ApJ...838...28M} have recently discussed how rapid mass loss immediately pre-supernova can transform light curves from a IIP shape (at low pre-supernova mass loss rates) to the more luminous type IIL (at higher mass loss rates).  This occurs when additional internal energy is added to the ejecta by shock-heating due to the CSM mass as it is swept up. Because the CSM lies outside the stellar radius, this new internal energy does not adiabatically decay as much prior to being radiated from the transient's photosphere. As a result, the radiative efficiency of the models, $E_{\rm rad}/E$, ranges from 1.4\% for the HSE model, to 8.5\% for the $q=0.1$ model, to 23\% for the $q=0.3$ model.  We note that these radiative efficiencies are with a factor of two of those prediced by the scaling models of Section \ref{sec:analytic}, for $n=4$ and $n=3$, respectively, equations \eqref{eq:Erad3} and \eqref{eq:Erad4}. 

Many features of our model light curves with merger-ejecta are similar to \citet{2017ApJ...838...28M}'s model suites including dense CSM distributions of varying mass and power-law slope. In particular, elevated early ``plateau'' luminosities that decay down to the unperturbed plateau are representative of significant CSM at radii less than the transient's eventual maximum photosphere radius of approximately $10^{15}$~cm.    Comparing to Figure \ref{fig:1D}, we note that large masses of relatively close-in CSM are the distinguishing features of our models. The resultant light curves, therefore, have typical duration of hundreds of days like normal IIP, not the thousands of days observed in some IIn supernovae with extended CSM distributions like that observed for SN 1988Z and SN 2005ip \citep{2017hsn..book..403S}.

Comparing the two mergers, the $q=0.3$ scenario with larger $M_{\rm CSM}$ has a later and more luminous peak, along with a higher luminosity during the plateau.  Because the CSM mass is related to the merger, we find $M_{\rm CSM}/M_{\rm tot} \sim 1.5 q$, see Section \ref{sec:merger}.
The models of Figure \ref{fig:q_HSE} for a hydrostatic explosion, $q=0.1$, and $q=0.3$, thus provide a context for interpreting the apparent variations in CSM contribution.  While the CSM mass plays a primary role in determining the light curve brightness, the distribution of CSM is crucial in shaping the light curves.  In the case of the hydrostatic explosion (labeled HSE in Figure \ref{fig:q_HSE}), there is no contribution from CSM interaction. The $q=0.3$ model shows a light curve that is always elevated  by approximately three magnitudes above the HSE model due to CSM interaction (at $t>30$~d). By contrast, the $q=0.1$ model is significantly elevated above the HSE model only earlier in the lightcurve, and converges to the HSE plateau luminosity around $100$~d. The distinction between these cases lies in the slope of the CSM density  and in the outer CSM radius.

For $q=0.3$, the CSM has $\rho \propto r^{-3}$ ($n=3$). The total radiative efficiency due to the CSM, $E_{\rm rad}/E$, equation \eqref{eq:Erad3}, only decreases with the logarithm of the ratio of the expanding ejetca photosphere radius, seen in the right hand panels of Figure \ref{fig:q_HSE}, to initial stellar radius (which is the base of the CSM distribution). Further, the outermost radius of the CSM at the moment of energy injection in our numerical model is $R_0 = 10^{15}$~cm. This is larger than the ejecta photosphere radius early in the light curve, and similar to the ejecta photosphere radius later in the light curve, implying that there is not a significant adiabatic degradation of the CSM contribution before light can escape from the expanding ejecta. 

We can compare these trends to the  $q=0.1$ model, in which case the overall CSM mass is lower and $\rho \propto r^{-4}$ ($n=4$). As the photosphere grows with time across the transient duration, the contribution of CSM interaction decreases approximately as $R_\ast/R_{\rm ph}$, see equation \eqref{eq:Erad4}. Of similar importance, the outermost CSM radius at the moment of energy deposition, $R_0=2.6\times10^{14}$~cm is a factor of a few less than the ejecta photosphere radius late in the plateau phase. This effect also decreases the contribution of the CSM to the late-stage light curve compared to an $n=4$ CSM of infinite extension, e.g. equation \eqref{eq:Erad4}. As a result, the light curve converges to a similar magnitude as the plateau of the hydrostatic explosion late in the light curve.

In addition to brightening the explosion, CSM interaction modifies the object's colors at timescales of days to weeks on which transients are typically discovered. CSM interaction yields bluer colors at time of peak (effective temperatures of several $10^4$~K on timescales of tens of days). The photosphere cools to more typical IIP temperatures of thousands of Kelvin only after 50 to 100 days (for the $q=0.1$ and $q=0.3$ models, respectively). These higher temperatures are directly representative of the extra internal energy injection due to shock heating of the ejecta by the CSM density distribution.

\subsection{Implications of Varying Energetics and Donor Star Properties}\label{sec:param}
We expect mergers between compact objects and giant stars to occur at a wide range of donor star and compact object properties because the binaries from which they form have broad distributions of mass, semi-major axis, and mass ratio.  Further, the energy of the central engine is unknown, and may, in fact, vary from merger to merger. Here we explore the implications of the parameter space of merger properties on the resultant light curves. 

\begin{figure}[tbp]
\begin{center}
\includegraphics[width=\linewidth]{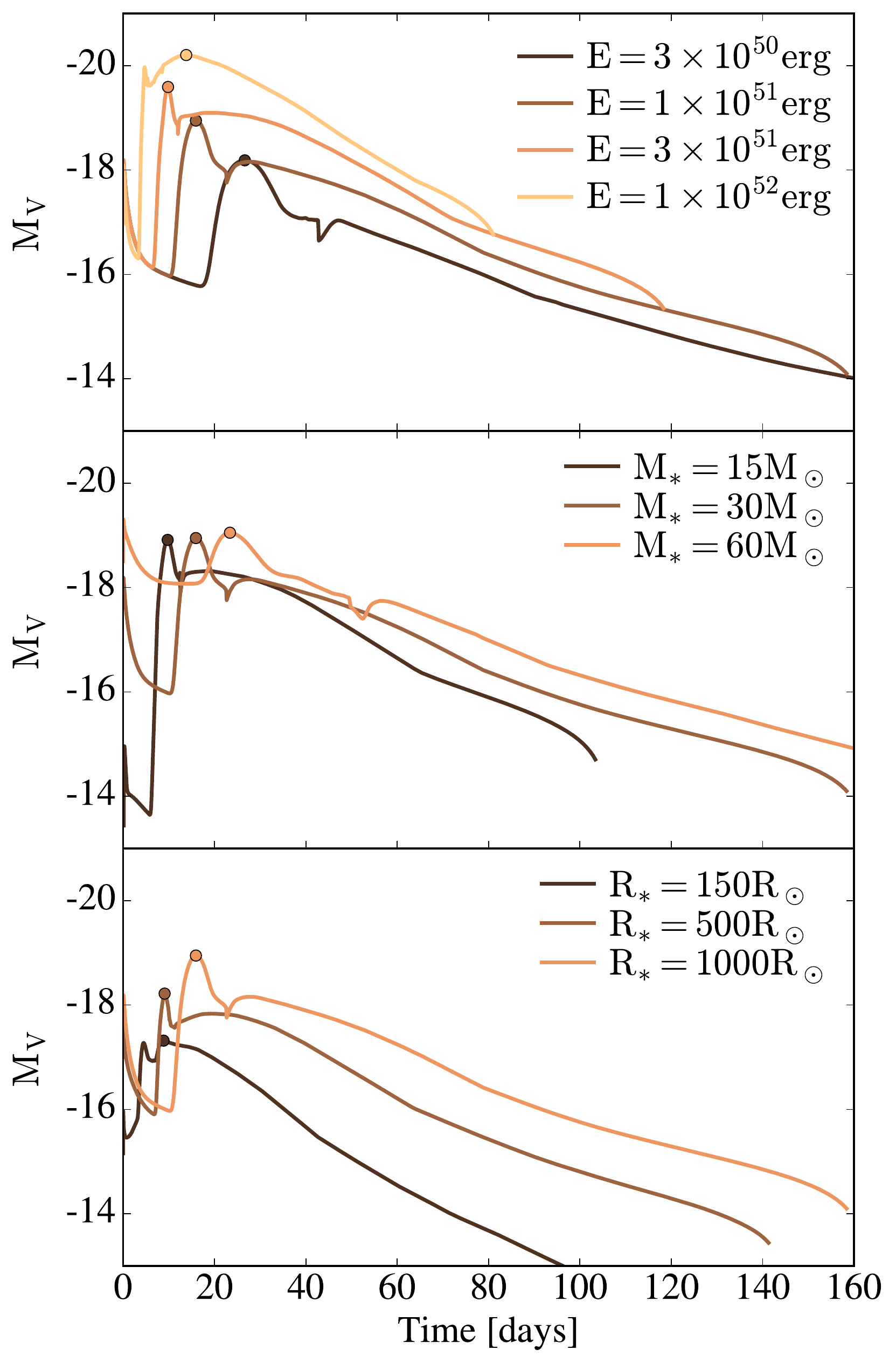}
\caption{Bolometric light curves for $q=0.1$ transients with varying energy (top panel) mass (center panel) and radius (lower panel). Unless specifically modified, we adopt our fiducial values of a $30 M_\odot$ and $1000R_\odot$ donor star and $10^{51}$~erg explosion. The location of the $V$-band peak is marked with a dot. Varying energetics and donor star properties create light curves of different duration, peak brightness, and degree of CSM contribution. }
\label{fig:Lvar}
\end{center}
\end{figure}

Figure \ref{fig:Lvar} shows V-band light curves for a range of models, all $q=0.1$, in which we vary energy (top panel), mass (center panel) and radius (bottom panel) around our fiducial, $30M_\odot$, $1000R_\odot$ case with $10^{51}$~erg explosion energy.   In all of these cases, because $q=0.1$, the CSM density profile is roughly $\rho \propto r^{-4}$, and equation \eqref{eq:Erad4} predicts the approximate contribution of CSM interaction to the radiated energy.

Varying explosion energy with other properties kept fixed yields the qualitatively expected variation in light curve luminosity and duration -- higher energy explosions give rise to faster ejecta, with more luminous but shorter duration transients. We note that the relative contribution of the CSM interaction, shown by the early bump in the lightcurve, decreases in the more energetic supernovae.  When the explosion energy changes, one consequence is that the photosphere radius during the plateau phase changes, roughly as $R_{\rm ph} \propto E^{5/12}$ \citep{1993ApJ...414..712P,2009ApJ...703.2205K}.  For higher energies, the larger photosphere radii imply smaller contributions from CSM interaction,  because the photosphere is further outside the outermost CSM radius, $R_0$.  In Figure \ref{fig:Lvar}, we observe that the light curve shape transforms as the energy increases. This is reflective of the decreasing contribution of CSM interaction to the light curve as $E$ increases. Consequently, the radiative efficiency decreases from 10\% for the $3\times 10^{50}$~erg explosion to 4.7\% for the  the $10^{52}$~erg explosion. With this smaller CSM contribution, the $10^{52}$~erg explosion light curve shows a relatively typical IIP shape, with a small, early bump due to the CSM. 

Varying donor star properties, in the form of mass and radius, similarly changes light curve duration, peak brightness, and shape. More massive donor stars yield higher ejecta masses,  but constant $M_{\rm CSM}/M_{\rm tot}\propto q$. At fixed energy, the ejecta velocities are lower and light curve durations are correspondingly longer. As mass varies in Figure \ref{fig:Lvar}, the plateau photosphere radius varies only mildly because the higher ejecta masses are balanced by lower ejecta velocities. Therefore, though these models show different characteristic timescales, they all have very similar peak magnitudes and degrees of CSM contribution to their overall radiated luminosity (total radiative efficiencies range from 11.7\% to 9.0\%).

Varying donor star radius changes not only the extent of the donor itself but the extent of the CSM, which extends to tens of stellar radii. This, in turn, varies the crucial ratio of maximum CSM radius to transient photosphere radius (because varying donor radius has little effect on $R_{\rm ph}$). When the donor is more compact, for example $150R_\odot$, the CSM extends to approximately $10^{14}$~cm, and largely affects only the early lightcurve. Progressively larger donors of $500R_\odot$ and $1000R_\odot$ scale the radial size of the CSM distribution. This scaling yields more CSM material at radii closer to the transient's photosphere radius at later times (for example near peak), in turn implying higher radiative efficiencies, and brighter transients.  For example, the $150R_\odot$ model has a radiative efficiency of only 1.9\%, while the $500R_\odot$ model radiates 5.3\% of the explosion energy and the $1000R_\odot$ model radiates 8.5\% of the explosion energy.

\begin{figure}[tbp]
\begin{center}
\includegraphics[width=0.5\textwidth]{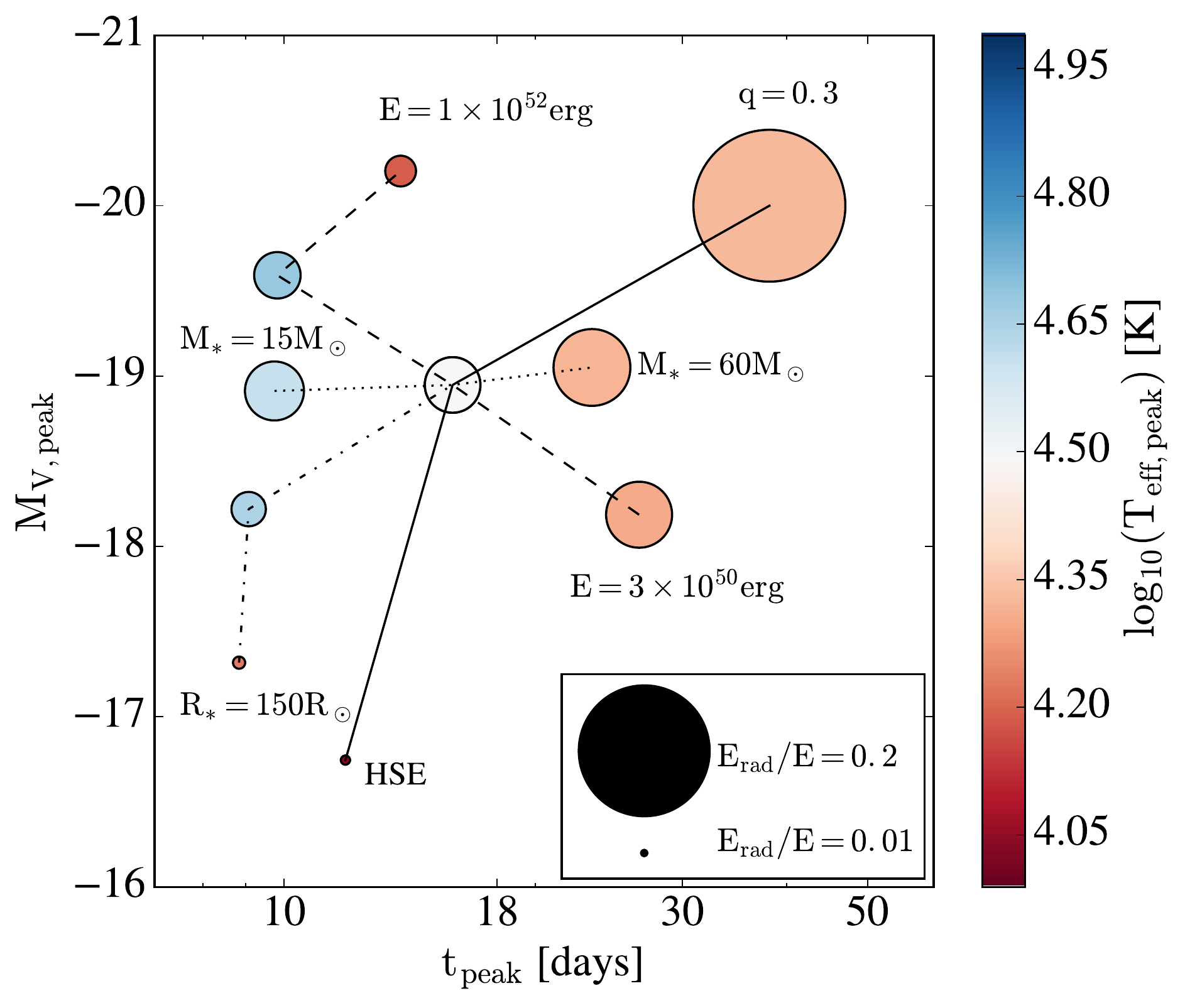}
\caption{Luminosity and timescale of the optical light curves of  merger-driven explosions. Here we additionally summarize radiative efficiency with size of the marker and effective temperature at peak with color. The central point is our fiducial model; lines connect the isolated variations of energy (dashed), mass (dotted), radius (dot-dash), and mass ratio (solid).  }
\label{fig:summary}
\end{center}
\end{figure}

Figure \ref{fig:summary} summarizes the parameter space of merger-driven explosions in luminosity, timescale, effective temperature, and radiative efficiency. The majority of merger-driven explosions have radiative efficiency on the order of 10\%, much higher than the hydrostatic model with no CSM. The $q=0.3$ model has even higher radiative efficiency of 25\%. However, the more compact $150R_\odot$ donor model and the highest explosion energy model, $10^{52}$~erg, both show relatively minimal CSM-interaction features in Figure \ref{fig:Lvar} and have somewhat lower radiative efficiency  (because $R_\ast/R_{\rm ph}$  and $R_0/R_{\rm ph}$ are reduced, see section \ref{sec:analytic}).  

Together, Figure \ref{fig:summary} shows that merger-driven explosions occupy a somewhat restricted phase space of luminosity and timescale. Typical models are more luminous than standard type IIP, but less luminous than super-luminous supernovae. In all of the models bearing significant CSM-interaction features, effective temperature varies systematically with time of peak brightness with longer-duration transients appearing redder and shorter-duration transients appearing bluer.

\section{Population Synthesis of Merger-Driven Explosions}\label{sec:pop}

We use population synthesis models of stellar binary evolution to explore the statistical properties of binary systems at the time of a common envelope phase leading to merger between a compact object and a giant star's core. We then use these models to estimate the population of observable merger-driven explosions. 

\subsection{Population Model}

We analyze rapid population synthesis models from the Compact Object Mergers: Population, Astrophysics and Statistics (COMPAS) suite \citep{2017NatCo...814906S,2018MNRAS.477.4685B,2018MNRAS.481.4009V}. These models employ approximate stellar evolution tracks and parameterized physics in order to facilitate exploration of the statistical properties of binary stellar evolution -- including rare outcomes like the formation of double compact object binaries \citep[for a full description of the approach, see ][]{2017NatCo...814906S,2018MNRAS.477.4685B,2018MNRAS.481.4009V}.  In particular, we adopt the model parameters of \citet{2018MNRAS.481.4009V}'s ``Fiducial" case and we capitalize on a recent development by Vigna-Gomez et. al. (in preparation) to record the characteristics and outcomes of all common envelope phases experienced by modeled binaries.

Initial distributions of binary properties are sampled at the zero-age main sequence (ZAMS) in COMPAS. In the models we study, the mass of the primary star is drawn from an initial mass function in the form $dN/dm \propto m^{-2.3}$ \citep{salpeter1955luminosity} with masses between $5 \leq m/\rm{M_{\odot}} \leq 100$. The mass of the secondary star is then chosen from a flat distribution in mass ratio with $0.1 < q_{\rm{ZAMS}} \leq 1$ \citep{2012Sci...337..444S}. The initial separation is drawn from a flat-in-the-log distribution, $dN/da \propto a^{-1}$, with separations between $0.01 < a_{\rm{ZAMS}}/\rm{AU} < 1000$ \citep{opik1924statistical,2012Sci...337..444S}. All stars in our model population adopt solar metalicity ($Z=0.0142$). A total of $10^6$ binary systems are simulated. 

Common envelope phases are identified by conditions for dynamically unstable mass transfer in COMPAS. When a common envelope episode occurs, an energy criterion is used to evaluate the outcome. In particular, the final change in orbital energy is related to the energy needed to unbind the giant star's hydrogen envelope from its core, $\Delta E_{\rm orb} = - \alpha E_{\rm bind}$, where $\alpha=1$ is an  efficiency parameter \citep{1984ApJ...277..355W}.  If the maximal change in orbital energy (defined on the basis of the minimal separation at which the core fills its Roche lobe) is insufficient to unbind the envelope, $|\Delta E_{\rm orb}| <  \alpha |E_{\rm bind}|$, then a merger between the companion and the core is assumed to result. This scaling implies that more compact stars have higher binding energies and, for a given companion mass, are more likely to result in merger. More extend stars (nearer to the tip of their giant-branch evolution) have lower binding energies and their common envelope phases are more likely to result in envelope ejection \citep{de1990common,2016A&A...596A..58K}.

\subsection{Compact Object-Core Mergers}

\begin{figure*}[tbp]
\begin{center}
\includegraphics[width=0.45\linewidth]{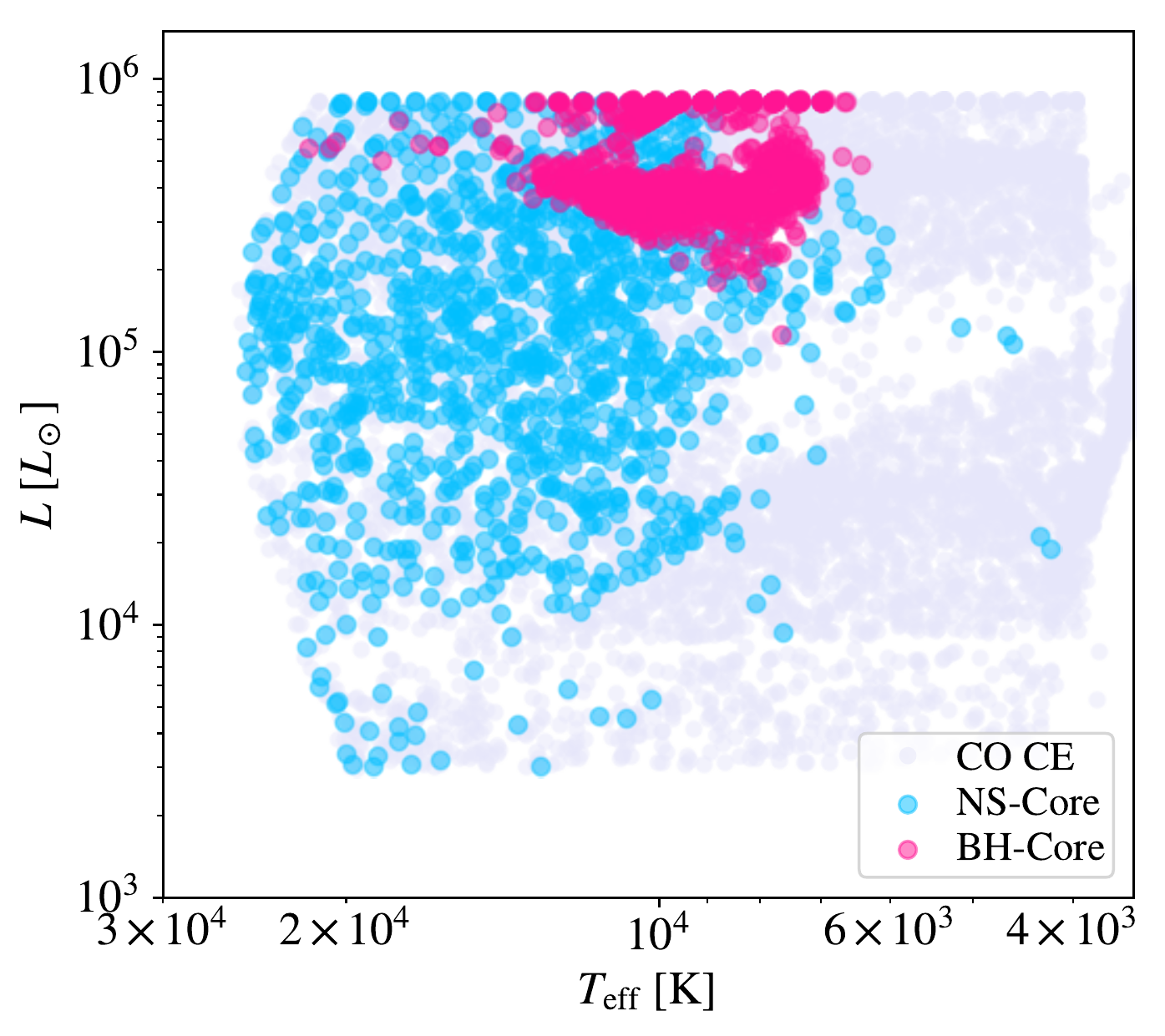}
\hspace{0.5cm}
\includegraphics[width=0.45\linewidth]{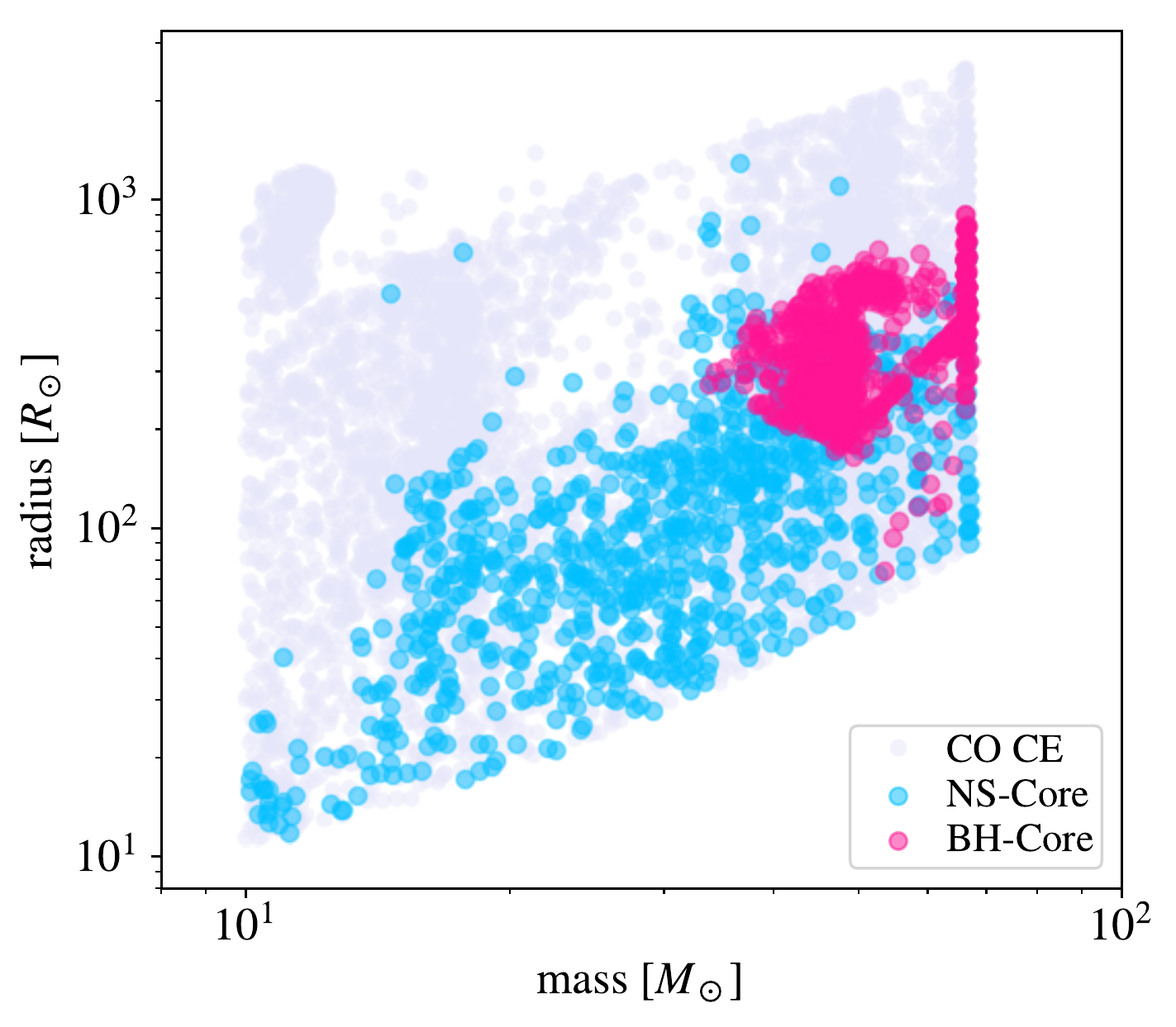}
\includegraphics[width=0.95\linewidth]{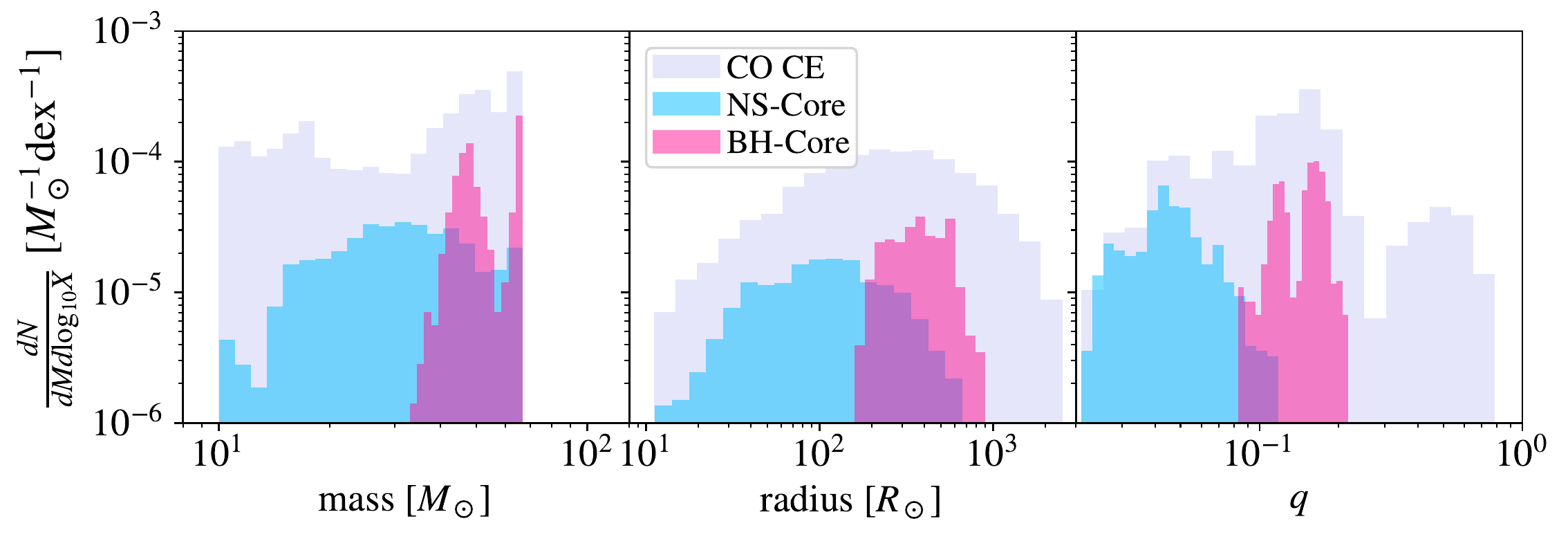}
\caption{Distributions of binary properties at the onset of common envelope phases involving black holes or neutron stars interacting with evolved, massive star donors with mass greater than $10M_\odot$. Here we distinguish between all common envelope phases involving compact objects (labeled CO CE), and cases in which a neutron star merges with the donor's helium core (NS-Core) or a black hole merges with the donor's helium core (BH-Core). Mergers occur in roughly 22\% of the compact object common envelope phases, and are split relatively equally between black hole and neutron star events. The companion mass distribution, especially for black hole mergers, favors massive companions.  Histograms are plotted in units of events per solar mass of stars formed per logarithmic bin in x-value (mass, radius, or mass ratio).}  
\label{fig:pop}
\end{center}
\end{figure*}

The most common evolutionary channel leading to a compact - giant star merger and a merger-driven explosion is as follows. A binary pair in an initially relatively wide orbit evolves, likely going through a dynamically stable mass transfer from the initially more-massive star onto its companion. That initially more massive star undergoes core collapse, leaving behind either a neutron star or black hole remnant. Because neutron star kicks tend to be large in magnitude, a relatively small fraction of systems containing newly-formed neutron stars -- less than 4\% \citep{2018MNRAS.481.4009V} -- remain bound following the supernova. Of those that remain binaries, a large fraction will undergo a common envelope phase during a reverse episode of mass transfer onto the compact object, initiated by the expansion of the intially less massive companion after it completes core hydrogen fusion. This may result in either a merger or a common envelope ejection. Those that eject their envelopes may go on to form a double compact object binary, as discussed by \citet{2018MNRAS.481.4009V}.

The population of common envelope phases involving compact objects in these models is depicted in Figure \ref{fig:pop}. In what follows, we report on and show only events that involve post-main-sequence donor stars more massive than $10M_\odot$. We show the distribution of these sources in the Hertzsprung-Russell Diagram (HRD) as well as in mass, radius, and mass ratio. We highlight the distinction between all common envelope phases involving compact objects (labeled CO CE), and events resulting in mergers between a neutron star and the donor core (labeled NS-Core) and a black hole and the donor core (labeled BH-Core). 

A number of interesting trends emerge from these distributions. While common envelope phases occur throughout the donor star's post-main sequence evolution, and therefore also the HRD, particular criteria are most likely to result in a merger. Merging sources tend to have the more compact radii compared to the overall distribution of common envelope phases ($T_{\rm eff} \gtrsim 10^4$~K). Neutron stars interact with a broad range of stellar companion masses, while black holes common envelope phases tend to involve massive $M\gtrsim30M_\odot$ and thus luminous donors. Typical mass ratios of compact-object common envelope phases range from $0.02\lesssim q\lesssim 0.6$; those resulting in mergers tend to have $q\lesssim 0.2$. The upper limits of these ranges reflect the conditions of dynamical mass transfer stability and envelope ejection, respectively. Of the mergers, the black holes form the higher mass-ratio population, $0.1 \lesssim q \lesssim 0.2$, while neutron stars typically have $q\lesssim 0.1$.

\subsection{Event Rate}

We can estimate the event rate of compact object - giant star mergers using the results of these population synthesis models. We simulate $10^6$ binary systems, or approximately  $1.96\times 10^7 M_\odot$ of binary mass. Each solar mass of modeled stars represents $3.8 M_\odot$ of stars formed \citep{2018MNRAS.481.4009V}. From our models, common envelope phases involving compact objects and donors more massive than $10M_\odot$ occur with a frequency of $1.5\times 10^{-4} M_\odot^{-1}$, where the unit denotes mergers per solar mass of stars formed. Of these, approximately 22\% result in mergers. Among the mergers, 49\% involve black holes, while 51\% involve neutron stars. The rate of black hole-core mergers and neutron-star mergers are thus each approximately $1.6\times10^{-5}M_\odot^{-1}$. These events occur in a spread of ages between 3 and 40 Myr, and are thus strongly correlated with recent star formation. By comparison, core collapse supernovae occur with a frequency of $5.8\times10^{-3}M_\odot^{-1}$ in the model systems. Merger-driven explosions therefore represent on the order of 0.6\% of all core collapse events.

\subsection{Outburst Population}
Having assessed the population of donor stars and compact object companions that undergo mergers, we now extend the results of our light curve models to estimate the properties of the population of observable transients. Guided by the results of Sections \ref{sec:param} and \ref{sec:analytic}, we note that CSM interaction is most important when the binary mass ratio is larger (yielding more merger ejecta and higher CSM mass) and when the radius is extended (yielding less adiabatic degradation of CSM-interaction energy, proportional to $R_\ast/R_{\rm ph}$). Comparison to the population properties in Figure \ref{fig:pop} shows that the systems that tend to have high mass ratios and large radii are predominantly the BH-Core merger group, in which a black hole merges with its giant star companion. By contrast, the typical radii, $R_\ast \sim 100R_\odot$, and mass ratios, $q<0.1$, for the neutron star-core mergers are such that we expect less dramatic signatures of CSM interaction (see Figure \ref{fig:Lvar}). 

To map our parameter variations onto the modeled population, we estimate the following scalings of $M_V$ with changing model parameters from the results of Figure \ref{fig:summary},
\begin{align}
\label{eq:Mvpeak}
M_{V,{\rm peak}} \approx  -18.9 & - 2.42 \log_{\rm 10}(R_\ast/1000 R_\odot) \nonumber \\
&- 0.229 \log_{\rm 10}(M_\ast/30M_\odot) \nonumber \\
&- 1.41 \log_{\rm 10}(E/10^{51}~{\rm erg}) \nonumber \\
&- 2.20 \log_{\rm 10}(q/0.1)
\end{align}
An important caveat is that, given our limited model parameter coverage, these numerical scalings represent the individual dependencies on binary properties about our fiducial model rather than the full parameter covariance. We will compare these luminosities to those of standard type IIP supernovae \citep{1993ApJ...414..712P}, 
\begin{align}
\label{eq:MvPopov}
M_V \approx  -11.42 &-1.67 \log_{\rm 10}(R_\ast/R_\odot) \nonumber \\
&+ 1.25 \log_{\rm 10}(M_\ast/M_\odot) \nonumber \\
&- 2.08 \log_{\rm 10}(E/10^{50}~{\rm erg}).
\end{align}
We note that the \citet{1993ApJ...414..712P} model accurately predicts the peak V-band luminosity of our hydrostatic model (Figures \ref{fig:q_HSE} and \ref{fig:summary}).

\begin{figure}[tbp]
\begin{center}
\includegraphics[width=\linewidth]{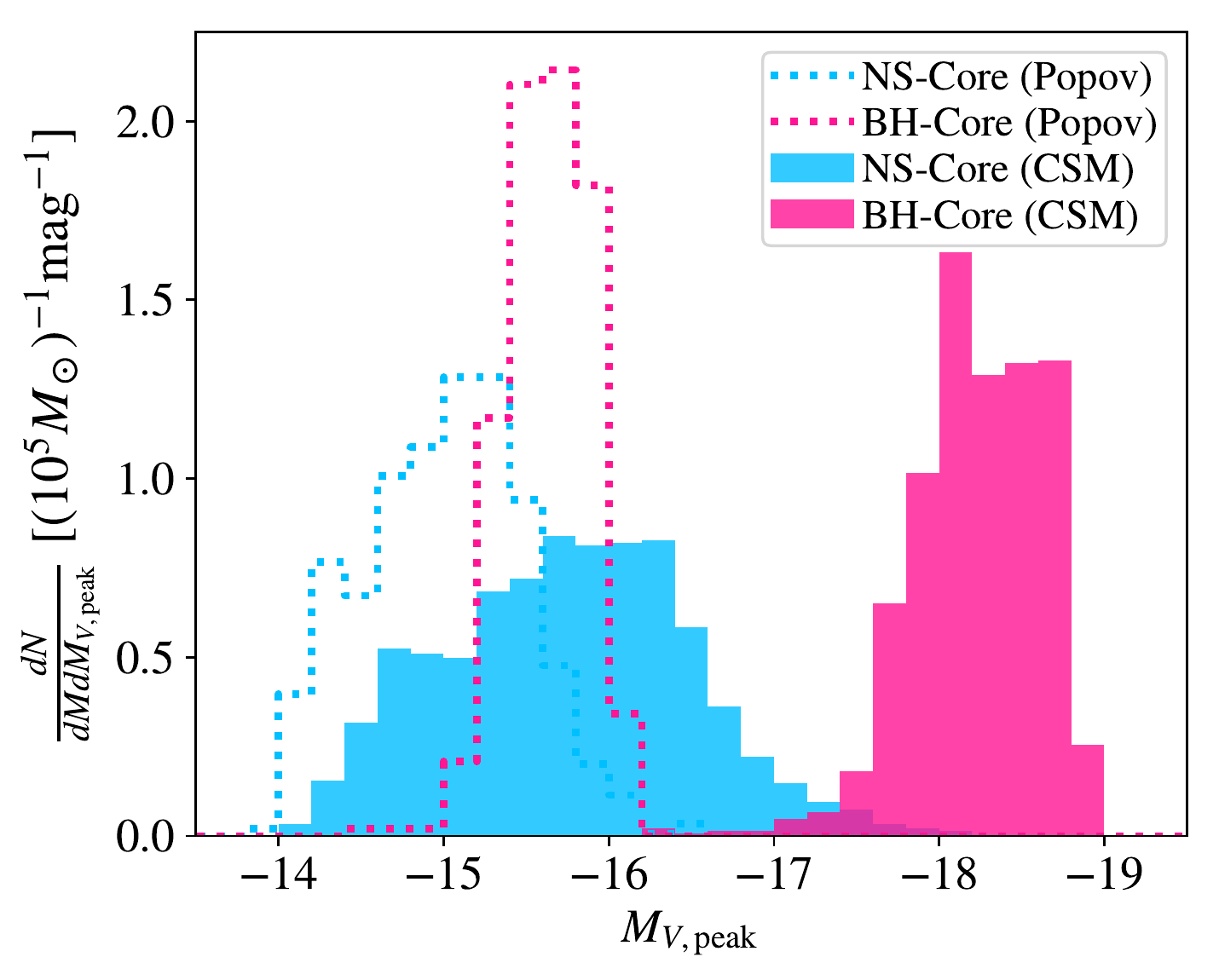}
\caption{ Transformation of merger-driven explosions by CSM interaction. We show peak V-band magnitudes of the population of merger-driven explosions using the \citet{1993ApJ...414..712P} model (equation \eqref{eq:MvPopov}; dashed lines), then apply our results (equation \eqref{eq:Mvpeak}; filled histograms) to derive the peak magnitudes including CSM interaction with merger-expelled ejecta. Black hole merger-driven explosions, in particular, form a distinct and luminous group that comprises 49\% of the merger-driven explosion transients.  The $y$-axis is shown in units of events per magnitude per $10^5M_\odot$ of stars formed.}  
\label{fig:popMv}
\end{center}
\end{figure}

In Figure \ref{fig:popMv}, we apply these scalings to the population of compact-object core mergers. Again we divide the population on the basis of whether a neutron star or a black hole is merging with the core. We assume that all events have $10^{51}$~erg explosion energy for the sake of this illustration. We find that CSM interaction (as predicted by equation \eqref{eq:Mvpeak}) brightens all merger-driven explosions relative to their hydrostatic equivalents (as estimated from equation \eqref{eq:MvPopov}). The neutron star-core mergers are brighter by approximately 1 magnitude than their Popov-model equivalents. However, the black hole mergers are brightened significantly more, by approximately 3 magnitudes. In this diagram, we observe that the black hole merger-driven explosions form a distinct population more luminous than the non-CSM-interacting IIP population.

\section{Discussion}\label{sec:discussion}

\subsection{Production of Supernovae-like Transients With Massive, Close CSM}

It has recently become apparent that a large fraction of type II supernovae show signs of interaction with CSM of densities much larger than that implied by nominal stellar-wind mass loss. Type IIn supernovae have long been acknowledged to have CSM due to the persistent narrow lines in their spectra. This otherwise diverse class of supernovae occupies approximately 10\% of the overall core collapse rate \citep[e.g.][]{2012ApJ...744...10K}.  More recently, evidence has been emerging that a majority (up to 70\%) of type II supernovae show evidence of having at least $0.1M_\odot$ of CSM  imprinted on their light curves \citep{2018ApJ...858...15M,2018ApJ...867....4M}. For example, \citet{2018NatAs...2..808F} has argued for systematic evidence that most type II shock breakouts are delayed by interaction with dense CSM.  A shared feature of the CSM in many  type IIP and IIL supernovae is that it is very close to the donor star, indicating its loss in the years immediately prior to the explosion \citep[e.g.][]{ 2013Natur.494...65O,2014ApJ...785...82S}. 

One proposed explanation for the presence of pre-supernova CSM ejection lies in the phenomenological comparison to luminous blue variable (LBV) outbursts, which are non-terminal outbursts of massive O-type stars. Though the precise cause of these outbursts remains uncertain \citep[e.g.][]{2014ApJ...796..121J}, as does their potential correlation with the evolutionary trend of the core toward collapse, in at least one dramatic example, SN2009ip, both LBV outbursts and a terminal supernova were observed in the same object over the course of a decade \citep{2010AJ....139.1451S,2013ApJ...763L..27P,2013ApJ...768...47O,2014MNRAS.438.1191S,2014ApJ...780...21M,2014MNRAS.442.1166M}.   

Another possible explanation links the CSM to the vigorous convection due to accelerating nuclear burning in the pre-supernova core. In this case, convection launches gravity waves at the interface between the convective core and an overlying radiative layer. These waves propagate through the radiative zone and dissipate near the base of the convective hydrogen envelope \citep{2012MNRAS.423L..92Q,2014ApJ...780...96S,2017MNRAS.470.1642F}. The luminosity of these dissipating waves can be highly super Eddington in the year prior to core-collapse, driving extensive mass loss \citep{2016MNRAS.458.1214Q} or outbursts \citep{2017MNRAS.470.1642F}. 

An explosion driven by the merger itself also naturally links merger ejecta and CSM with the explosive fate of the star, as we have described in the preceding sections.  However, a merger-driven model cannot explain the full diversity of type II supernovae or their CSM properties. In practice, some combination of these processes must be at play in order to explain the abundance and diversity of CSM observed in type II supernovae. 

\subsection{Comparison to Observed Supernovae}

We compare our model light curves to two representative, well-studied supernovae.  Photometric similarity is insufficient to demonstrate the origin of a given transient,  as we discuss further in Section \ref{sec:identify}. In this section, we contextualize our model merger-driven explosions by showing that they bear similarities to transients already in the supernovae archives.

\subsubsection{1979c}
SN1979c, classified as a type IIL, was discovered in April 1979, several weeks after explosion \citep{1979IAUC.3348....1M}.
It has been observed extensively at radio wavelengths \citep{1986ApJ...301..790W,2000ApJ...532.1124M,2008ApJ...682.1065B,2009A&A...503..869M}, and early modeling suggested a very large progenitor radius of $R \sim 6000 R_\odot$ and CSM extending out to $R \sim 10^5 R_\odot$ \citep{1992SvAL...18...43B}.

Observations from the following 20 years gave rise to more theoretical discussion. \citet{2003ApJ...591..301B} suggested that the remnant is expanding into low density CSM with $\rho \sim r^{-n}$, with $n = 1.94^{+0.10}_{-0.05}$ decreasing to $n< 1.5$ at larger radii.  Later \citet{2010ApJ...717..245K} suggested that the light curve was brightened by the spindown of a magnetar at the center of the SN remnant. \citet{2011NewA...16..187P} note that,  rather than a magnetar,  a $5-10 M_\odot$ BH accreting from fallback material can also explain X-ray data seen from 1995 and 2007.

\begin{figure}[tbp]
\begin{center}
 \includegraphics[width=0.98\linewidth]{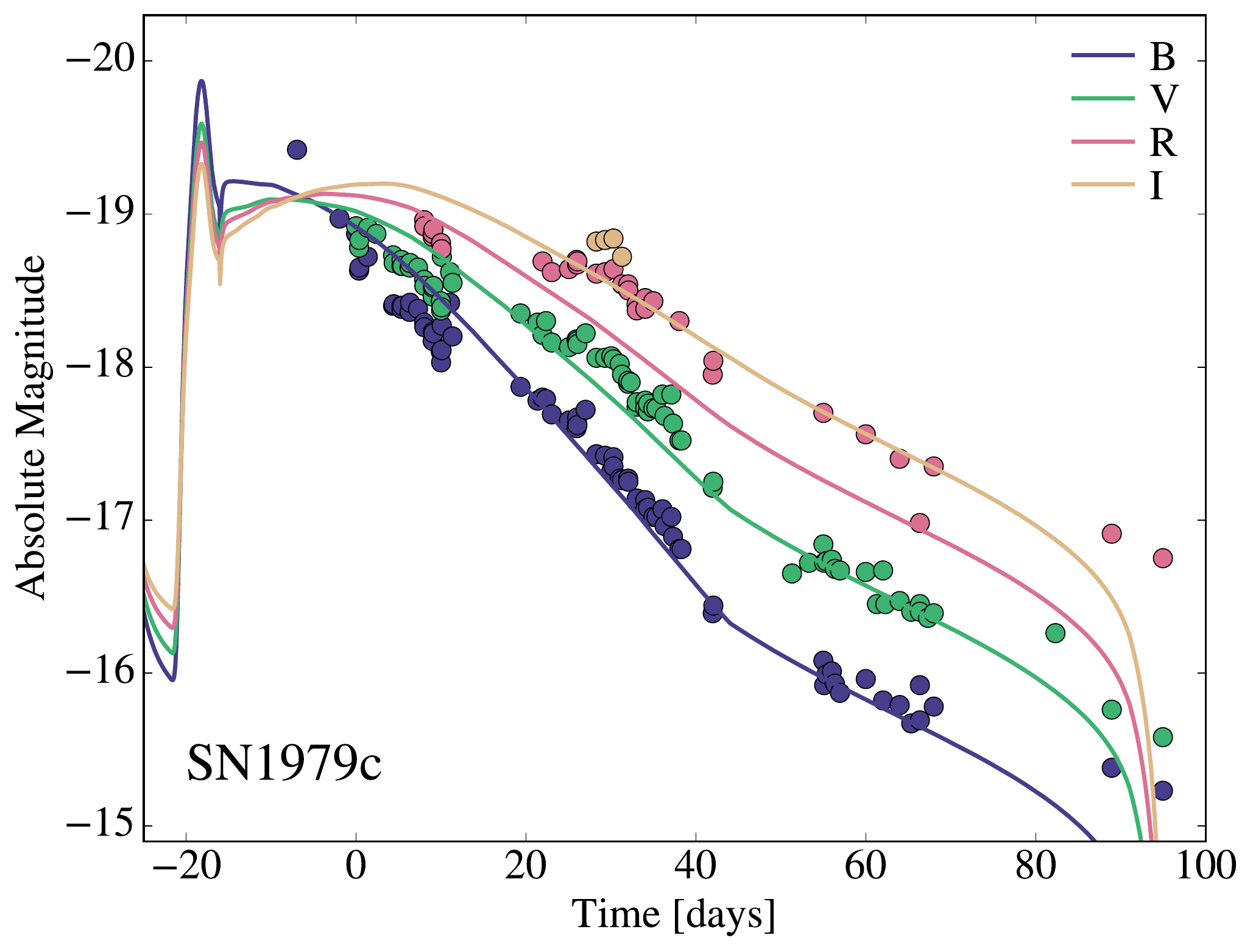}
  \includegraphics[width=0.98\linewidth]{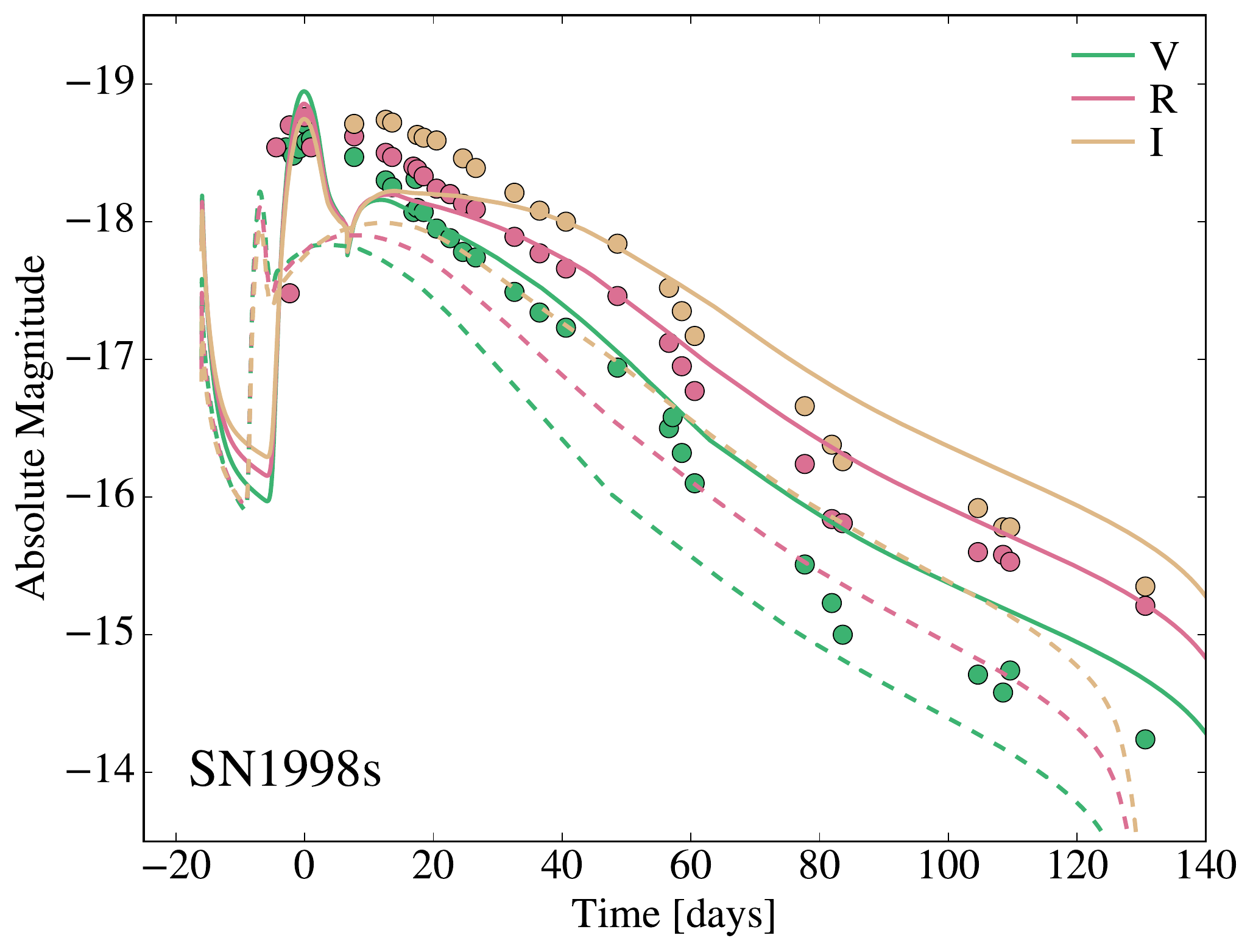}
\caption{{\it Upper panel:} SN 1979c plotted on top of absolute magnitude from simulations with $M_\ast  = 30 M_\odot$ and $R_\ast = 1000 R_\odot$ and $E_{SN} = 3 \times 10^{51} erg$. {\it Lower panel:} SN 1998s plotted on top of absolute magnitude from simulations with $M_\ast  = 30 M_\odot$ and radius scaled to $R_\ast = 1000 R_\odot$ (solid lines) and  $R_\ast = 500 R_\odot$ (dashed lines).}
\label{fig:SN_lightcurve}
\end{center}
\end{figure}

In the top panel of figure \ref{fig:SN_lightcurve} we have plotted optical data from the first 100 days of observation  \citep[from the open supernova catalog;][]{2017ApJ...835...64G}. On top we plot absolute magnitudes from our simulations with $M_\ast  = 30 M_\odot$ and $R_\ast = 1000 R_\odot$ and $E = 3 \times 10^{51} $erg. The plot is not a fit, but shows that the outcome of our simulations can closely replicate observed transients. This, in addition to the potential for a remnant black hole \citep{2011NewA...16..187P} make SN1979c an interesting candidate for further investigation under the merger-driven hypothesis. 

 \subsubsection{1998s}
 
SN1998s is one of the most studied type IIn supernovae \citep{2015ApJ...806..213S}. %A non-detection three days prior to discovery indicated that  the event was discovered within a few days of explosion \citep{2000ApJ...536..239L}.
From spectral lines, two shells of CSM were identified. \citet{2001MNRAS.325..907F} found that the inner CSM was within 90AU from the center and the outer CSM extended from 185 AU to over 1800 AU. The classification of SN1998s as a IIn is a direct result of the very early spectral observations; the narrow line features disappeared and morphed into the broad lines of a type IIL or IIb within weeks \citet{2017hsn..book..403S}.

SN1998S has later been interpreted as having a red supergiant progenitor with possibly asymmetric CSM consisting of the two separate shells, caused by separate mass-loss events. \citep{2016MNRAS.456..323K} claims that this type of SN is very common and that many  IIL and IIP share spectral features with IIn in early spectra.

%The early spectra from the SN1998s suggested that the progenitor star was a Wolf-Rayet star stripped of its hydrogen envelope in at least two strong mass-loss episodes, the last still occurring at the time of explosion \citep{2000ApJ...536..239L}. Emission lines had intrinsic continuum polarization of $p \sim 3\%$, suggesting a global asphericity of $\sim 45\%$ from the oblate, electron-scattering dominated models of \citet{1991A&A...246..481H}. They speculated that the asymmetrically distributed material perhaps had a disk-like or ring-like morphology.

In the bottom panel of figure \ref{fig:SN_lightcurve} we plot the $V, R$ and $I$ band data from SN1998s \citep[data from the open supernova catalog;][]{2017ApJ...835...64G} plotted on top of absolute magnitude from simulations with $M_\ast  = 30 M_\odot$ and $E= 1\times 10^{51}$~erg, solid lines for radius scaled to $R_\ast = 1000 R_\odot$,  and dashed  lines for $R_\ast = 500 R_\odot$. The light curve we see from from our simulations with $R_\ast = 1000 R_\odot$ is similar to SN1998s, though the rate of decline perhaps fits better with our $R_\ast = 500 R_\odot$ simulation. Just as with SN1979c, the overall light curve shape, duration, and brightness are well-approximated by our models. The presence of nearby CSM is also consistent with a merger-driven explosion. However, the explanation for two distinct shells of CSM is not immediately apparent given our model predictions, and may be in tension with the merger-driven hypothesis for this transient \citep[though see the discussion of][]{2017MNRAS.470.1788C}.

\subsection{Identification in Optical Surveys}\label{sec:identify}

Having shown that merger-driven explosion models can reproduce the basic, photometric properties of several observed supernovae, we now focus on the prospects for their more secure identification. 

The prevalence of merger-driven explosions (of order 0.5\% of the core-collapse rate) begs questions about their prior detection in existing datasets and their imprints on future surveys. 
Current surveys, such as the Zwicky Transient Factory \citep{2017NatAs...1E..71B} and All-Sky Automated Survey for Supernovae \citep[e.g.][]{2019MNRAS.484.1899H} are presently discovering hundreds of new core-collapse supernovae per year. This discovery rate suggests that one or more merger-driven explosions is currently being discovered per year. Efforts at early discovery and spectroscopy of these transients aim to reveal CSM properties through ``flash spectroscopy" in which the CSM is ionized prior to being swept up by the blast wave.  The Large Synoptic Survey Telescope (LSST) will discover on the order of $10^5$ core collapse events per year \citep[chapter 8]{2009arXiv0912.0201L}, implying hundreds to thousands of merger-driven explosions detected in a given observing year. 

Among this flood of optical transients, the challenge will be unambiguous identification of merger-driven explosions rather than detection. A full consideration is beyond the scope of our initial study, but we speculate on several potential signatures here.  As discussed in Section \ref{sec:merger}, the ejecta from the pre-merger common envelope phase are densest in the equatorial plane of the binary. When the supernova explodes into this aspherical density distribution, the blast wave will be shaped by these asymmetrical surroundings \citep{1996ApJ...472..257B}. Emission from the photosphere will, as a result, be polarized by one to several percent, as has been described in the case of SN2009ip \citep{2014MNRAS.442.1166M}.  

The interacting binary progenitor of the explosion may also offer clues in the identification of merger-driven supernovae, as in ongoing progenitor-monitoring efforts described by \citet{2008ApJ...684.1336K,2017MNRAS.469.1445A}. Drawing parallels to low-mass, Galactic stellar merger events like V1309 Sco \citep{2010A&A...516A.108M,2011A&A...528A.114T}, increasing rates of non-conservative mass transfer (seen in the panels of  Figure  \ref{fig:ts}) may enshroud the merging binary in dust and cause an optical fading of the progenitor star prior to merger. In V1309 Sco, such a phase of optical dimming was observed in the phase of 100 to 1000 orbital periods prior to coalescence. In the last orbits leading into the merger (the portion captured by Figure \ref{fig:ts}), V1309 Sco brightened in optical bands as more-and-more emission arose from the outflow from the binary \citep{2014ApJ...788...22P,2016MNRAS.455.4351P,2016MNRAS.461.2527P,2017ApJ...850...59P}. Future work is needed to extend these scenarios to detailed predictions for pre-explosive behavior in massive star coalescence. 

Multiwavelength, particularly X-ray, signatures, while less frequently available than optical photometry, provide a powerful tool for probing early CSM interaction \citep[e.g.][]{2012ApJ...747L..17C,2017ApJ...835..140M,2018ApJ...867....4M}. These data can probe the CSM distribuiton in great detail, including the density distribution through the hard to soft emission ratio \citep{2018ApJ...867....4M}. If the CSM is as steep as predicted  in the merger-driven models (steeper than $\rho\propto r^{-3}$), it will accelerate the leading edge of the ejecta  to high velocities and produce hard x-ray emission  \citep{2018ApJ...867....4M}.

Finally, merger-driven explosions will leave a black hole as the remnant of the rapid accretion phase following merger of the compact object with the stellar core. Though black hole formation is common in core-collapse events, it is typically believed to accompany implosion rather than explosions and luminous supernovae \citep[e.g.][]{2018ApJ...860...93S}. If detected, the coexistence of a supernova-like transient and a remnant black hole would thus be consistent with the merger-driven explosion scenario.  In theory we might distinguish neutron star and black hole central x-ray sources on the basis of their x-ray specta. In practice, this identification can be ambiguous when the surrounding, absorbing medium is substantial. One such example of a transient harboring an embedded x-ray source is AT 2018cow \citep{2019ApJ...872...18M}.

\section{Summary and Conclusion}\label{sec:conclusion}

In this paper, we have presented models for merger-driven explosions that arise from the plunge of a compact object within the helium core of its giant star companion following a common envelope phase \citep{2012ApJ...752L...2C}. When a compact object merges with the helium core of a massive, post main-sequence star, the conditions for rapid, neutrino-cooled accretion are met \citep[e.g.][]{2001ApJ...550..357Z}. The accompanying release of energy may deposit approximately $10^{51}$~erg into the surrounding hydrogen envelope, leading to a merger-driven explosion \citep{2012ApJ...752L...2C}.  Some key findings of our investigation are:

\begin{enumerate}
\item The binary coalescence leading to the merger of the compact object with the core expels slow-moving material into the surrounding environment, forming a dense, toroidal CSM (Figures \ref{fig:ts} and \ref{fig:largescale}). The spherically-averaged density profile has a steep radial slope of $\rho\propto r^{-3}$ or $\rho\propto r^{-4}$ (Figure \ref{fig:1D}). 
\item Using 1D radiation hydrodynamic models of the explosions, we find that the CSM distribution is crucial in shaping the transient light curves. Merger-driven explosions are brightened by up to three magnitudes relative to their counterparts in hydrostatic stars (Figure \ref{fig:q_HSE}), with timescale and light curve shape that vary with donor-star mass and radius and explosion energy (Figures \ref{fig:Lvar} and \ref{fig:summary}). 
\item  From population models, we find that black hole and neutron star mergers with giant star companions occur with similar frequency, each with a rate per mass of stars formed of $1.6\times10^{-5}M_\odot^{-1}$. The combined rate is 0.6\% of the core-collapse rate in our models. Merger-driven explosions occur across a roughly flat distribution of donor-star masses from $10M_\odot$ to $100M_\odot$ (Figure \ref{fig:pop}). CSM interaction brightens neutron star mergers by approximately one magnitude, but brightens the population of black hole mergers by approximately three magnitudes relative to type  IIP models with the same energy injection and pre-supernova stellar mass and radius (Figure \ref{fig:popMv}).  
\item The most luminous transients, those involving black hole mergers, are at least as common as their less luminous neutron star counterparts. Black hole mergers have $M_{V,{\rm peak}}\sim-18$ to $-19$ with $t_{\rm peak}\sim 20$ to $30$~d. The implication for optical surveys is that the brightest, easiest-to-detect events comprise a significant fraction of the entire population. 
\end{enumerate}

The calculations presented in this paper have demonstrated that  merger-driven explosions provide a natural mechanism for the production of supernovae-like transients with close-in, slow-moving CSM. Future work could improve on the treatment of the stellar model (a polytropic envelope in our approximation) and the details of energy injection into this envelope. At present, we inject energy spherically into the envelope at one tenth the star's overall radius. In practice, the unknown location and asymmetry of energy injection might play a key role in shaping transient light curves, colors, and peak luminosities with respect to the estimates of our current models.

We compare our models to two representative supernovae, SN1979c and SN1998s in Figure \ref{fig:SN_lightcurve}. However, we note that more work is needed to provide unambiguous confirmations of merger-driven explosions.   In Section \ref{sec:identify}, we discuss additional strategies for the identification of merger-driven explosions including their asymmetry and polarization due to the toroidal CSM, the properties of their progenitor binaries, and their early spectra and X-ray emission. In future work, these signatures can be investigated through multi-dimensional calculations of the explosive evolution and emergent light curve, as well as more detailed modeling of the progenitor system's plunge toward merger. 

\acknowledgments
We thank R. Margutti, E. Ramirez-Ruiz, and M. Rees for advice and helpful discussions in the development of this work. 
We gratefully acknowledge the support of E. Ostriker and J. Stone in the development and analysis of the hydrodynamic models, and V. Morozova for support with SNEC. 
S. S. acknowledges support by the Danish National Research Foundation (DNRF132). Part of the simulations used in this paper were performed on the University of Copenhagen high-performance computing cluster funded by a grant from VILLUM FONDEN (project number 16599). 
M.M. is grateful for support for this work provided by NASA through Einstein Postdoctoral Fellowship grant number PF6-170169 awarded by the Chandra X-ray Center, which is operated by the Smithsonian Astrophysical Observatory for NASA under contract NAS8-03060. 
This work was supported in part by the Black Hole Initiative at Harvard University, which is funded by a JTF grant. 
Resources supporting this work were provided by the NASA High-End Computing (HEC) Program through the NASA Advanced Supercomputing (NAS) Division at Ames Research Center.

\software{  Athena++, Stone et al. (in preparation) \url{http://princetonuniversity.github.io/athena}, Astropy \citep{2013A&A...558A..33A}, SNEC \citep{2015ascl.soft05033M}, COMPAS \citep{2017NatCo...814906S,2018MNRAS.477.4685B,2018MNRAS.481.4009V} }

\appendix
\section{Validation of Light Curve Calculations}\label{sec:appendix}

In this appendix, we discuss the validation of several numerical choices in the 1D radiation hydrodynamics calculations with SNEC that we use to produce model light curves. 

\begin{figure}[tbp]
\begin{center}
\includegraphics[width=0.46\linewidth]{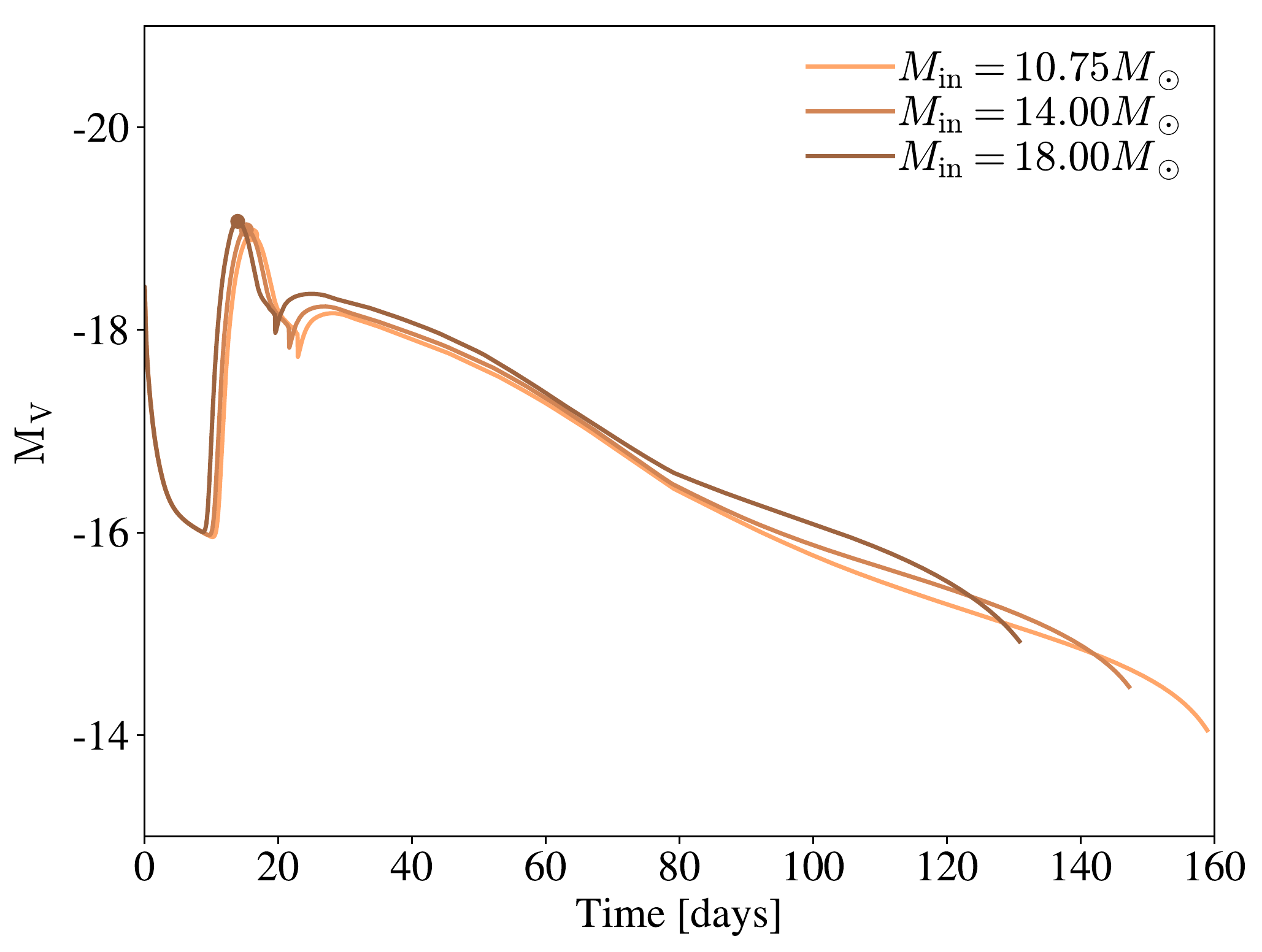}
\includegraphics[width=0.5\linewidth]{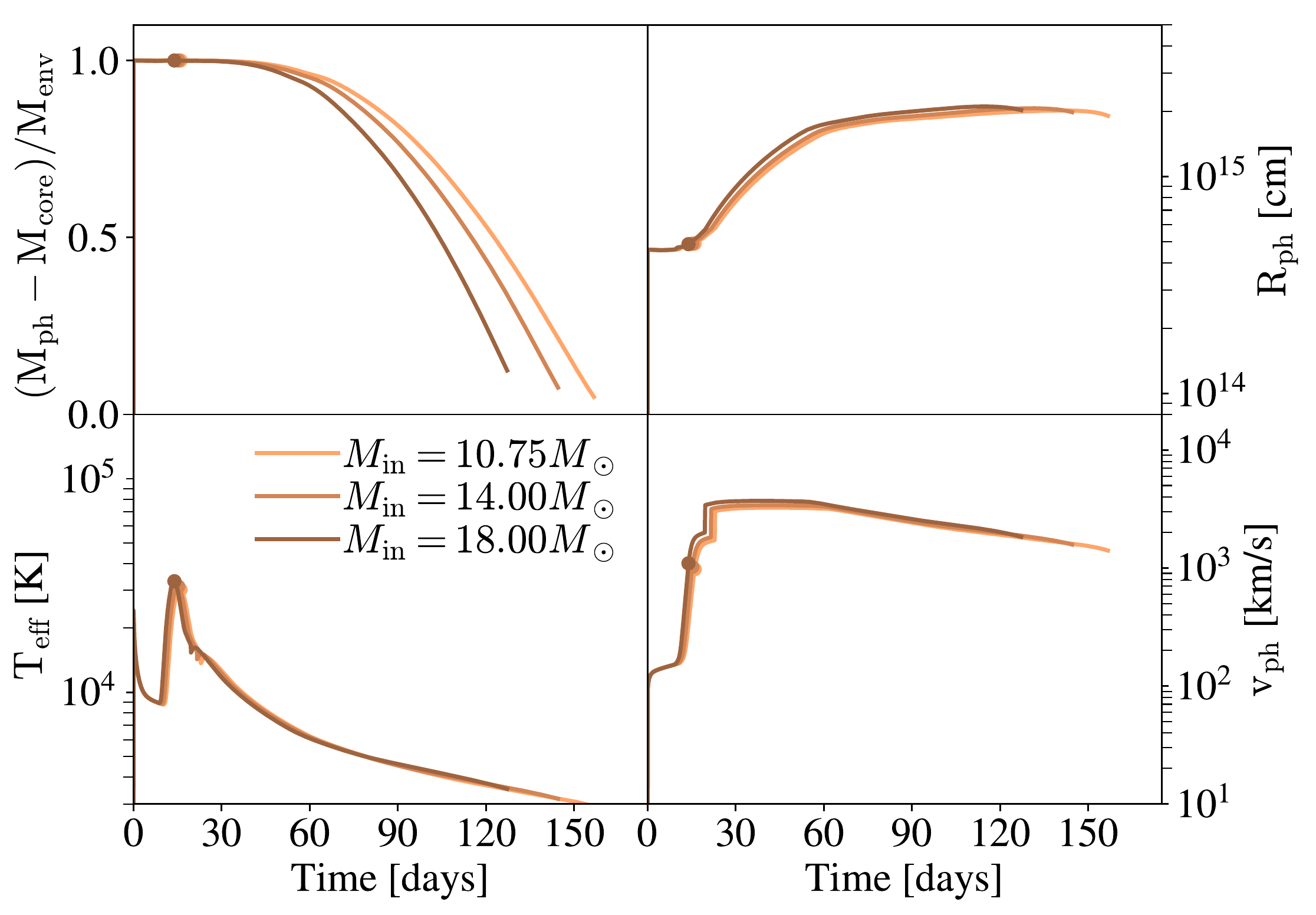}
\caption{V-band absolute magnitude (left) and photosphere properties (right) for a $30M_\odot$ donor star involved in a $q=0.1$ merger, in which we vary the inner mass coordinate of the $10^{51}$~erg of energy deposition and subsequent ejection within the Hydrogen envelope. Material inside $M_{\rm in}$ does not explode and acts as a gravitational point mass, while material outside $M_{\rm in}$ is expelled. We find that within a factor of two in $M_{\rm in}$, model light curves are very similar, varying primarily in total plateau duration (which results from differing ejecta masses).  }
\label{fig:Mcut_Mv}
\end{center}
\end{figure}

In mapping the 3D hydrodynamics calculation of the merger (Section \ref{sec:merger}) to the 1D explosive calculation, we need to make an assumption about the location (described by radius or mass coordinate) where the explosion energy is deposited. We have argued in Section  \ref{sec:explosion} that this deposition location is somewhere within the hydrogen envelope. Here we explore the sensitivity to that choice as follows. Beginning with our fiducial case of a $30M_\odot$ donor star and $3M_\odot$ black hole in a  $q=0.1$ merger, we deposit $10^{51}$~erg of thermal energy spread over $0.1M_\odot$ at different mass coordinate locations. Our default assumption is $M_{\rm in}=10.75M_\odot$, which corresponds to the enclosed mass of at $0.1R_\ast$ of $7.75M_\odot$ plus a $3M_\odot$ black hole. This model in Figure \ref{fig:Mcut_Mv} corresponds to the fiducial simulation presented in Figure \ref{fig:q_HSE}, which is labeled $q=0.1$. We then vary the mass coordinate at which thermal energy is deposited, moving outward in the star's Hydrogen envelope. Material inside $M_{\rm in}$ acts as a gravitational point mass for the remainder of the calculation. We find that for $M_{\rm in}=10.75M_\odot$, $M_{\rm in}=14M_\odot$, and $M_{\rm in}=18M_\odot$ (corresponding to radius coordinates of 0.1,  0.16, and 0.24 times the donor star's original radius) the model light curves are very similar indicating only weak dependence on how the energy is spatially deposited within the hydrogen envelope. All of these radii are significantly outside the star's more compact Helium core. We note that for $M_{\rm in}>20M_\odot$, the case in which $>2/3$ of the donor star forms a black hole while only $<1/3$ is expelled, we do observe departures in the model light curves, with the bulk of the thermal energy radiated early, and not coupling efficiently to driving envelope expansion. 

\begin{figure}[tbp]
\begin{center}
\includegraphics[width=0.5\linewidth]{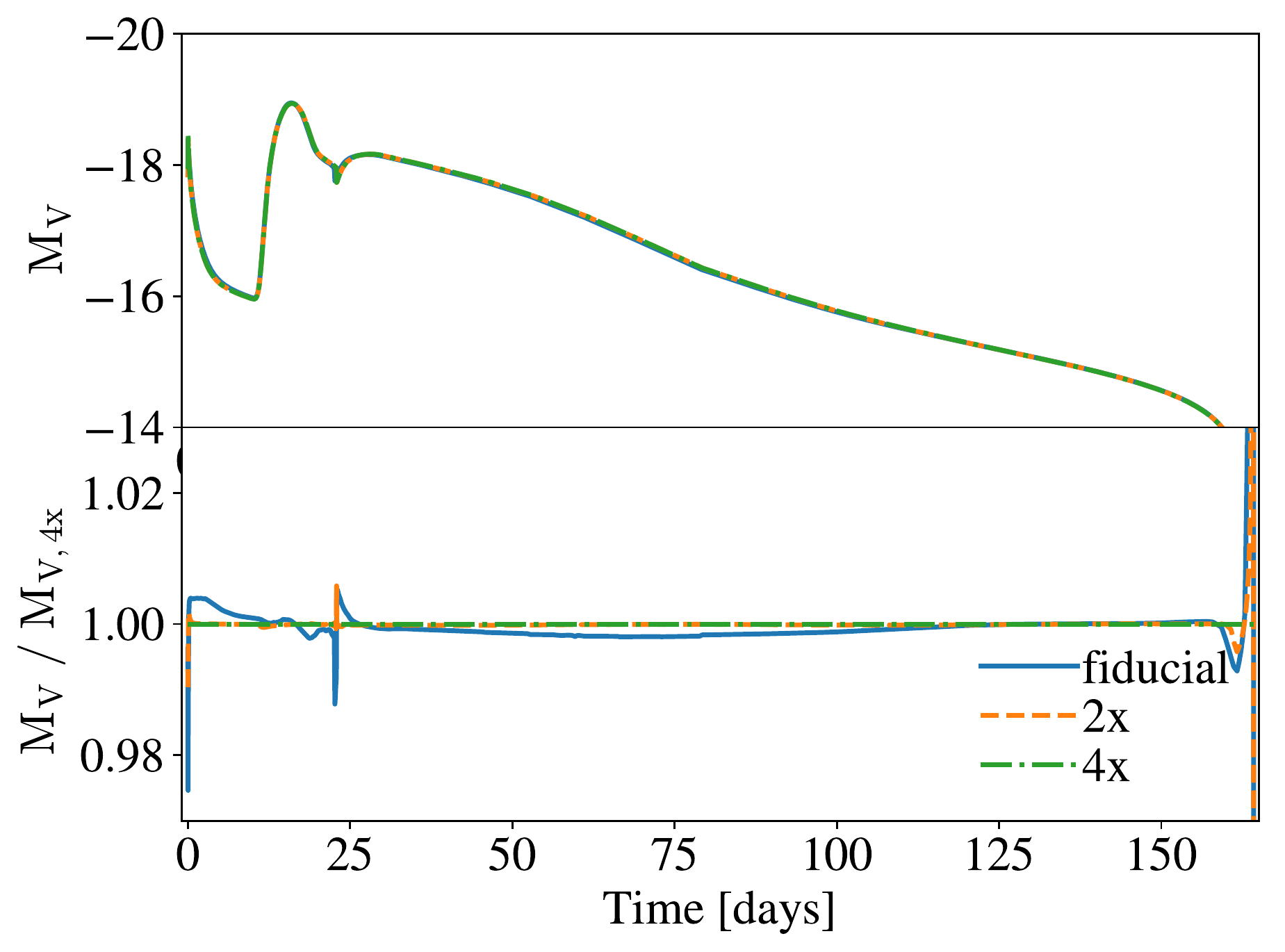}
\caption{ Absolute V-band magnitude  for a $30M_\odot$ donor star involved in a $q=0.1$ merger. We show models with a fiducial resolution of 456 mass zones, and for mass resolutions of twice and four times as many zones. We find that the light curves are converged to within 1\% for the bulk of the model light curves, with the largest variations at shock breakout and near the end of the light curve.  }
\label{fig:Res_Mv}
\end{center}
\end{figure}

We also test the dependence of our model results on spatial resolution within the 1D SNEC calculations. Our fiducial case divides the mass into 456 elements. Figure \ref{fig:Res_Mv} compares this case to models with twice and four times as many zones (912, 1824, respectively). These tests confirm that our results are converged to within 1\% across the light curve duration with any of these resolution choices. 

\bibliographystyle{aasjournal}
\bibliography{ce,sne,catalogueDNS,mSN2}

@ARTICLE{2018PASA...35...49E,
       author = {{Eldridge}, J.~J. and {Xiao}, L. and {Stanway}, E.~R. and
         {Rodrigues}, N. and {Guo}, N. -Y.},
        title = "{Supernova lightCURVE POPulation Synthesis I: Including interacting binaries is key to understanding the diversity of type II supernova lightcurves}",
      journal = {Publications of the Astronomical Society of Australia},
     keywords = {binaries: general, stars: massive, supernovae: general, Astrophysics - Solar and Stellar Astrophysics, Astrophysics - High Energy Astrophysical Phenomena},
         year = "2018",
        month = "Dec",
       volume = {35},
        pages = {49},
          doi = {10.1017/pasa.2018.47},
archivePrefix = {arXiv},
       eprint = {1811.00282},
 primaryClass = {astro-ph.SR},
       adsurl = {https://ui.adsabs.harvard.edu/\#abs/2018PASA...35...49E},
      adsnote = {Provided by the SAO/NASA Astrophysics Data System}
}

@ARTICLE{2018ApJ...867....4M,
       author = {{Morozova}, Viktoriya and {Stone}, James M.},
        title = "{Theoretical X-Ray Light Curves of Young SNe. II. The Example of SN 2013ej}",
      journal = {\apj},
     keywords = {hydrodynamics, supernovae: general, Astrophysics - High Energy Astrophysical Phenomena},
         year = "2018",
        month = "Nov",
       volume = {867},
          eid = {4},
        pages = {4},
          doi = {10.3847/1538-4357/aae2b3},
archivePrefix = {arXiv},
       eprint = {1804.07312},
 primaryClass = {astro-ph.HE},
       adsurl = {https://ui.adsabs.harvard.edu/\#abs/2018ApJ...867....4M},
      adsnote = {Provided by the SAO/NASA Astrophysics Data System}
}

@ARTICLE{2018MNRAS.477...74A,
       author = {{Andrews}, Jennifer E. and {Smith}, Nathan},
        title = "{Strong late-time circumstellar interaction in the peculiar supernova iPTF14hls}",
      journal = {\mnras},
     keywords = {circumstellar matter, supernovae: general, supernovae: individual: iPTF14hls, stars: winds, outflows, Astrophysics - High Energy Astrophysical Phenomena},
         year = "2018",
        month = "Jun",
       volume = {477},
        pages = {74-79},
          doi = {10.1093/mnras/sty584},
archivePrefix = {arXiv},
       eprint = {1712.00514},
 primaryClass = {astro-ph.HE},
       adsurl = {https://ui.adsabs.harvard.edu/\#abs/2018MNRAS.477...74A},
      adsnote = {Provided by the SAO/NASA Astrophysics Data System}
}

@ARTICLE{2018ApJ...858...15M,
       author = {{Morozova}, Viktoriya and {Piro}, Anthony L. and {Valenti}, Stefano},
        title = "{Measuring the Progenitor Masses and Dense Circumstellar Material of Type II Supernovae}",
      journal = {\apj},
     keywords = {hydrodynamics, radiative transfer, supernovae: general, Astrophysics - High Energy Astrophysical Phenomena},
         year = "2018",
        month = "May",
       volume = {858},
          eid = {15},
        pages = {15},
          doi = {10.3847/1538-4357/aab9a6},
archivePrefix = {arXiv},
       eprint = {1709.04928},
 primaryClass = {astro-ph.HE},
       adsurl = {https://ui.adsabs.harvard.edu/\#abs/2018ApJ...858...15M},
      adsnote = {Provided by the SAO/NASA Astrophysics Data System}
}

@ARTICLE{2018MNRAS.475.3152K,
       author = {{Kleiser}, Io K.~W. and {Kasen}, Daniel and {Duffell}, Paul C.},
        title = "{Models of bright nickel-free supernovae from stripped massive stars with circumstellar shells}",
      journal = {\mnras},
     keywords = {binaries: general, circumstellar matter, stars: general, supernovae: general, supernovae: individual: SN 2010X, supernovae: individual: SN 2015U, supernovae: individual: SN 2002bj, Astrophysics - High Energy Astrophysical Phenomena},
         year = "2018",
        month = "Apr",
       volume = {475},
        pages = {3152-3164},
          doi = {10.1093/mnras/stx3321},
archivePrefix = {arXiv},
       eprint = {1801.01943},
 primaryClass = {astro-ph.HE},
       adsurl = {https://ui.adsabs.harvard.edu/\#abs/2018MNRAS.475.3152K},
      adsnote = {Provided by the SAO/NASA Astrophysics Data System}
}

@ARTICLE{2018ApJ...856...29M,
       author = {{McDowell}, Austin T. and {Duffell}, Paul C. and {Kasen}, Daniel},
        title = "{Interaction of a Supernova with a Circumstellar Disk}",
      journal = {\apj},
     keywords = {hydrodynamics, ISM: jets and outflows, shock waves, supernovae: general, Astrophysics - High Energy Astrophysical Phenomena},
         year = "2018",
        month = "Mar",
       volume = {856},
          eid = {29},
        pages = {29},
          doi = {10.3847/1538-4357/aaa96e},
archivePrefix = {arXiv},
       eprint = {1802.05152},
 primaryClass = {astro-ph.HE},
       adsurl = {https://ui.adsabs.harvard.edu/\#abs/2018ApJ...856...29M},
      adsnote = {Provided by the SAO/NASA Astrophysics Data System}
}

@ARTICLE{2017ApJ...851..138D,
       author = {{Das}, Sanskriti and {Ray}, Alak},
        title = "{Modeling Type II-P/II-L Supernovae Interacting with Recent Episodic Mass Ejections from Their Presupernova Stars with MESA and SNEC}",
      journal = {\apj},
     keywords = {circumstellar matter, hydrodynamics, radiative transfer, stars: mass-loss, supernovae: general, supernovae: individual: SN 2013ej, Astrophysics - High Energy Astrophysical Phenomena, Astrophysics - Solar and Stellar Astrophysics},
         year = "2017",
        month = "Dec",
       volume = {851},
          eid = {138},
        pages = {138},
          doi = {10.3847/1538-4357/aa97e1},
archivePrefix = {arXiv},
       eprint = {1711.02102},
 primaryClass = {astro-ph.HE},
       adsurl = {https://ui.adsabs.harvard.edu/\#abs/2017ApJ...851..138D},
      adsnote = {Provided by the SAO/NASA Astrophysics Data System}
}

@ARTICLE{2017ApJ...849..109P,
       author = {{Patnaude}, Daniel J. and {Lee}, Shiu-Hang and {Slane}, Patrick O. and
         {Badenes}, Carles and {Nagataki}, Shigehiro and {Ellison}, Donald C. and
         {Milisavljevic}, Dan},
        title = "{The Impact of Progenitor Mass Loss on the Dynamical and Spectral Evolution of Supernova Remnants}",
      journal = {\apj},
     keywords = {circumstellar matter, ISM: supernova remnants, nuclear reactions, nucleosynthesis, abundances, stars: mass-loss, supernovae: general, X-rays: general, Astrophysics - High Energy Astrophysical Phenomena},
         year = "2017",
        month = "Nov",
       volume = {849},
          eid = {109},
        pages = {109},
          doi = {10.3847/1538-4357/aa9189},
archivePrefix = {arXiv},
       eprint = {1708.04984},
 primaryClass = {astro-ph.HE},
       adsurl = {https://ui.adsabs.harvard.edu/\#abs/2017ApJ...849..109P},
      adsnote = {Provided by the SAO/NASA Astrophysics Data System}
}

@ARTICLE{2017MNRAS.470.1642F,
       author = {{Fuller}, Jim},
        title = "{Pre-supernova outbursts via wave heating in massive stars - I. Red supergiants}",
      journal = {\mnras},
     keywords = {waves, stars: evolution, stars: massive, stars: mass-loss, supergiants, supernovae: general, Astrophysics - Solar and Stellar Astrophysics, Astrophysics - High Energy Astrophysical Phenomena},
         year = "2017",
        month = "Sep",
       volume = {470},
        pages = {1642-1656},
          doi = {10.1093/mnras/stx1314},
archivePrefix = {arXiv},
       eprint = {1704.08696},
 primaryClass = {astro-ph.SR},
       adsurl = {https://ui.adsabs.harvard.edu/\#abs/2017MNRAS.470.1642F},
      adsnote = {Provided by the SAO/NASA Astrophysics Data System}
}

@ARTICLE{2017ApJ...838...28M,
       author = {{Morozova}, Viktoriya and {Piro}, Anthony L. and {Valenti}, Stefano},
        title = "{Unifying Type II Supernova Light Curves with Dense Circumstellar Material}",
      journal = {\apj},
     keywords = {hydrodynamics, radiative transfer, supernovae: general, supernovae: individual: SN 2013by, SN 2013ej, SN 2013fs, Astrophysics - High Energy Astrophysical Phenomena, Astrophysics - Solar and Stellar Astrophysics},
         year = "2017",
        month = "Mar",
       volume = {838},
          eid = {28},
        pages = {28},
          doi = {10.3847/1538-4357/aa6251},
archivePrefix = {arXiv},
       eprint = {1610.08054},
 primaryClass = {astro-ph.HE},
       adsurl = {https://ui.adsabs.harvard.edu/\#abs/2017ApJ...838...28M},
      adsnote = {Provided by the SAO/NASA Astrophysics Data System}
}

@INBOOK{2017hsn..book..403S,
       author = {{Smith}, Nathan},
        title = "{Interacting Supernovae: Types IIn and Ibn}",
     keywords = {Physics, Astrophysics - High Energy Astrophysical Phenomena, Astrophysics - Solar and Stellar Astrophysics},
    booktitle = {Handbook of Supernovae, ISBN 978-3-319-21845-8. Springer International Publishing AG, 2017, p. 403},
         year = "2017",
        pages = {403},
          doi = {10.1007/978-3-319-21846-5_38},
       adsurl = {https://ui.adsabs.harvard.edu/\#abs/2017hsn..book..403S},
      adsnote = {Provided by the SAO/NASA Astrophysics Data System}
}

@ARTICLE{2017ApJ...835...64G,
       author = {{Guillochon}, James and {Parrent}, Jerod and {Kelley}, Luke Zoltan and
         {Margutti}, Raffaella},
        title = "{An Open Catalog for Supernova Data}",
      journal = {\apj},
     keywords = {catalogs, ISM: supernova remnants, supernovae: general, Astrophysics - Solar and Stellar Astrophysics, Astrophysics - High Energy Astrophysical Phenomena, Astrophysics - Instrumentation and Methods for Astrophysics},
         year = "2017",
        month = "Jan",
       volume = {835},
          eid = {64},
        pages = {64},
          doi = {10.3847/1538-4357/835/1/64},
archivePrefix = {arXiv},
       eprint = {1605.01054},
 primaryClass = {astro-ph.SR},
       adsurl = {https://ui.adsabs.harvard.edu/\#abs/2017ApJ...835...64G},
      adsnote = {Provided by the SAO/NASA Astrophysics Data System}
}

@ARTICLE{2016arXiv161005323S,
       author = {{Shussman}, Tomer and {Waldman}, Roni and {Nakar}, Ehud},
        title = "{Type II supernovae Early Light Curves}",
      journal = {arXiv e-prints},
     keywords = {Astrophysics - High Energy Astrophysical Phenomena},
         year = "2016",
        month = "Oct",
          eid = {arXiv:1610.05323},
        pages = {arXiv:1610.05323},
archivePrefix = {arXiv},
       eprint = {1610.05323},
 primaryClass = {astro-ph.HE},
       adsurl = {https://ui.adsabs.harvard.edu/\#abs/2016arXiv161005323S},
      adsnote = {Provided by the SAO/NASA Astrophysics Data System}
}

@ARTICLE{2016ApJ...829..109M,
       author = {{Morozova}, Viktoriya and {Piro}, Anthony L. and {Renzo}, Mathieu and
         {Ott}, Christian D.},
        title = "{Numerical Modeling of the Early Light Curves of Type IIP Supernovae}",
      journal = {\apj},
     keywords = {hydrodynamics, radiative transfer, supernovae: general, Astrophysics - High Energy Astrophysical Phenomena},
         year = "2016",
        month = "Oct",
       volume = {829},
          eid = {109},
        pages = {109},
          doi = {10.3847/0004-637X/829/2/109},
archivePrefix = {arXiv},
       eprint = {1603.08530},
 primaryClass = {astro-ph.HE},
       adsurl = {https://ui.adsabs.harvard.edu/\#abs/2016ApJ...829..109M},
      adsnote = {Provided by the SAO/NASA Astrophysics Data System}
}

@ARTICLE{2016MNRAS.456..323K,
       author = {{Kangas}, T. and {Mattila}, S. and {Kankare}, E. and {Lundqvist}, P. and
         {V{\"a}is{\"a}nen}, P. and {Childress}, M. and {Pignata}, G. and
         {McCully}, C. and {Valenti}, S. and {Vink{\'o}}, J. and
         {Pastorello}, A. and {Elias-Rosa}, N. and {Fraser}, M. and
         {Gal-Yam}, A. and {Kotak}, R. and {Kotilainen}, J.~K. and
         {Smartt}, S.~J. and {Galbany}, L. and {Harmanen}, J. and
         {Howell}, D.~A. and {Inserra}, C. and {Marion}, G.~H. and
         {Quimby}, R.~M. and {Silverman}, J.~M. and {Szalai}, T. and
         {Wheeler}, J.~C. and {Ashall}, C. and {Benetti}, S. and
         {Romero-Ca{\~n}izales}, C. and {Smith}, K.~W. and {Sullivan}, M. and
         {Tak{\'a}ts}, K. and {Young}, D.~R.},
        title = "{Supernova 2013fc in a circumnuclear ring of a luminous infrared galaxy: the big brother of SN 1998S}",
      journal = {\mnras},
     keywords = {supernovae: general, supernovae: individual: 2013fc, galaxies: starburst, Astrophysics - High Energy Astrophysical Phenomena, Astrophysics - Solar and Stellar Astrophysics},
         year = "2016",
        month = "Feb",
       volume = {456},
        pages = {323-346},
          doi = {10.1093/mnras/stv2567},
archivePrefix = {arXiv},
       eprint = {1509.05389},
 primaryClass = {astro-ph.HE},
       adsurl = {https://ui.adsabs.harvard.edu/\#abs/2016MNRAS.456..323K},
      adsnote = {Provided by the SAO/NASA Astrophysics Data System}
}

@ARTICLE{2015ApJ...814...63M,
       author = {{Morozova}, Viktoriya and {Piro}, Anthony L. and {Renzo}, Mathieu and
         {Ott}, Christian D. and {Clausen}, Drew and {Couch}, Sean M. and
         {Ellis}, Justin and {Roberts}, Luke F.},
        title = "{Light Curves of Core-collapse Supernovae with Substantial Mass Loss Using the New Open-source SuperNova Explosion Code (SNEC)}",
      journal = {\apj},
     keywords = {hydrodynamics, radiative transfer, supernovae: general, Astrophysics - High Energy Astrophysical Phenomena, Astrophysics - Solar and Stellar Astrophysics},
         year = "2015",
        month = "Nov",
       volume = {814},
          eid = {63},
        pages = {63},
          doi = {10.1088/0004-637X/814/1/63},
archivePrefix = {arXiv},
       eprint = {1505.06746},
 primaryClass = {astro-ph.HE},
       adsurl = {https://ui.adsabs.harvard.edu/\#abs/2015ApJ...814...63M},
      adsnote = {Provided by the SAO/NASA Astrophysics Data System}
}

@ARTICLE{2015MNRAS.449.4304D,
       author = {{Dessart}, Luc and {Audit}, Edouard and {Hillier}, D. John},
        title = "{Numerical simulations of superluminous supernovae of type IIn}",
      journal = {\mnras},
     keywords = {polarization, radiative transfer, supernovae: general, supernovae: individual: 2010jl, Astrophysics - Solar and Stellar Astrophysics, Astrophysics - High Energy Astrophysical Phenomena},
         year = "2015",
        month = "Jun",
       volume = {449},
        pages = {4304-4325},
          doi = {10.1093/mnras/stv609},
archivePrefix = {arXiv},
       eprint = {1503.05463},
 primaryClass = {astro-ph.SR},
       adsurl = {https://ui.adsabs.harvard.edu/\#abs/2015MNRAS.449.4304D},
      adsnote = {Provided by the SAO/NASA Astrophysics Data System}
}

@ARTICLE{2015ApJ...806..213S,
       author = {{Shivvers}, Isaac and {Groh}, Jose H. and {Mauerhan}, Jon C. and
         {Fox}, Ori D. and {Leonard}, Douglas C. and {Filippenko}, Alexei V.},
        title = "{Early Emission from the Type IIn Supernova 1998S at High Resolution}",
      journal = {\apj},
     keywords = {stars: winds, outflows, supernovae: general, supernovae: individual: SN 1998S, techniques: spectroscopic, Astrophysics - High Energy Astrophysical Phenomena, Astrophysics - Solar and Stellar Astrophysics},
         year = "2015",
        month = "Jun",
       volume = {806},
          eid = {213},
        pages = {213},
          doi = {10.1088/0004-637X/806/2/213},
archivePrefix = {arXiv},
       eprint = {1408.1404},
 primaryClass = {astro-ph.HE},
       adsurl = {https://ui.adsabs.harvard.edu/\#abs/2015ApJ...806..213S},
      adsnote = {Provided by the SAO/NASA Astrophysics Data System}
}

@MISC{2015ascl.soft05033M,
       author = {{Morozova}, V. and {Ott}, C.~D. and {Piro}, A.~L.},
        title = "{SNEC: SuperNova Explosion Code}",
     keywords = {Software},
         year = "2015",
        month = "May",
          eid = {ascl:1505.033},
        pages = {ascl:1505.033},
archivePrefix = {ascl},
       eprint = {1505.033},
       adsurl = {https://ui.adsabs.harvard.edu/\#abs/2015ascl.soft05033M},
      adsnote = {Provided by the SAO/NASA Astrophysics Data System}
}

@ARTICLE{2014ApJ...780...96S,
       author = {{Shiode}, Joshua H. and {Quataert}, Eliot},
        title = "{Setting the Stage for Circumstellar Interaction in Core-Collapse Supernovae. II. Wave-driven Mass Loss in Supernova Progenitors}",
      journal = {\apj},
     keywords = {stars: mass-loss, stars: winds, outflows, supernovae: general, Astrophysics - Solar and Stellar Astrophysics},
         year = "2014",
        month = "Jan",
       volume = {780},
          eid = {96},
        pages = {96},
          doi = {10.1088/0004-637X/780/1/96},
archivePrefix = {arXiv},
       eprint = {1308.5978},
 primaryClass = {astro-ph.SR},
       adsurl = {https://ui.adsabs.harvard.edu/\#abs/2014ApJ...780...96S},
      adsnote = {Provided by the SAO/NASA Astrophysics Data System}
}

@ARTICLE{2013MNRAS.435.1520M,
       author = {{Moriya}, Takashi J. and {Maeda}, Keiichi and {Taddia}, Francesco and
         {Sollerman}, Jesper and {Blinnikov}, Sergei I. and {Sorokina}, Elena I.},
        title = "{An analytic bolometric light curve model of interaction-powered supernovae and its application to Type IIn supernovae}",
      journal = {\mnras},
     keywords = {circumstellar matter, stars: mass-loss, supernovae: general, supernovae: individual: SN 2005ip, supernovae: individual: SN 2006jd, supernovae: individual: SN 2010jl, Astrophysics - High Energy Astrophysical Phenomena, Astrophysics - Solar and Stellar Astrophysics},
         year = "2013",
        month = "Oct",
       volume = {435},
        pages = {1520-1535},
          doi = {10.1093/mnras/stt1392},
archivePrefix = {arXiv},
       eprint = {1307.2644},
 primaryClass = {astro-ph.HE},
       adsurl = {https://ui.adsabs.harvard.edu/\#abs/2013MNRAS.435.1520M},
      adsnote = {Provided by the SAO/NASA Astrophysics Data System}
}

@ARTICLE{2013A&A...557L...2M,
       author = {{Moriya}, Takashi J. and {Groh}, Jose H. and {Meynet}, Georges},
        title = "{Episodic modulations in supernova radio light curves from luminous blue variable supernova progenitor models}",
      journal = {\aap},
     keywords = {circumstellar matter, stars: mass-loss, supernovae: general, supernovae: individual: SN 2001ig, supernovae: individual: SN 2003bg, Astrophysics - Solar and Stellar Astrophysics, Astrophysics - High Energy Astrophysical Phenomena},
         year = "2013",
        month = "Sep",
       volume = {557},
          eid = {L2},
        pages = {L2},
          doi = {10.1051/0004-6361/201322012},
archivePrefix = {arXiv},
       eprint = {1306.0605},
 primaryClass = {astro-ph.SR},
       adsurl = {https://ui.adsabs.harvard.edu/\#abs/2013A&A...557L...2M},
      adsnote = {Provided by the SAO/NASA Astrophysics Data System}
}

@ARTICLE{2013MNRAS.433..838P,
       author = {{Pan}, Tony and {Patnaude}, Daniel and {Loeb}, Abraham},
        title = "{Superluminous X-ray emission from the interaction of supernova ejecta with dense circumstellar shells}",
      journal = {\mnras},
     keywords = {shock waves, circumstellar matter, supernovae: general, stars: winds, outflows, X-rays: general, Astrophysics - High Energy Astrophysical Phenomena},
         year = "2013",
        month = "Jul",
       volume = {433},
        pages = {838-848},
          doi = {10.1093/mnras/stt780},
archivePrefix = {arXiv},
       eprint = {1303.6958},
 primaryClass = {astro-ph.HE},
       adsurl = {https://ui.adsabs.harvard.edu/\#abs/2013MNRAS.433..838P},
      adsnote = {Provided by the SAO/NASA Astrophysics Data System}
}

@ARTICLE{2013A&A...555A..10T,
       author = {{Taddia}, F. and {Stritzinger}, M.~D. and {Sollerman}, J. and
         {Phillips}, M.~M. and {Anderson}, J.~P. and {Boldt}, L. and
         {Campillay}, A. and {Castell{\'o}n}, S. and {Contreras}, C. and
         {Folatelli}, G. and {Hamuy}, M. and {Heinrich-Josties}, E. and
         {Krzeminski}, W. and {Morrell}, N. and {Burns}, C.~R. and
         {Freedman}, W.~L. and {Madore}, B.~F. and {Persson}, S.~E. and
         {Suntzeff}, N.~B.},
        title = "{Carnegie Supernova Project: Observations of Type IIn supernovae}",
      journal = {\aap},
     keywords = {supernovae: general, stars: winds, outflows, stars: massive, Astrophysics - Cosmology and Nongalactic Astrophysics, Astrophysics - Solar and Stellar Astrophysics},
         year = "2013",
        month = "Jul",
       volume = {555},
          eid = {A10},
        pages = {A10},
          doi = {10.1051/0004-6361/201321180},
archivePrefix = {arXiv},
       eprint = {1304.3038},
 primaryClass = {astro-ph.CO},
       adsurl = {https://ui.adsabs.harvard.edu/\#abs/2013A&A...555A..10T},
      adsnote = {Provided by the SAO/NASA Astrophysics Data System}
}

@INPROCEEDINGS{2013msao.confE.171M,
       author = {{Moriya}, Takashi},
        title = "{Type IIn Supernovae and Mass Loss of their Progenitors}",
    booktitle = {Massive Stars: From alpha to Omega},
         year = "2013",
        month = "Jun",
          eid = {171},
        pages = {171},
       adsurl = {https://ui.adsabs.harvard.edu/\#abs/2013msao.confE.171M},
      adsnote = {Provided by the SAO/NASA Astrophysics Data System}
}

@ARTICLE{2013MNRAS.430.1402M,
       author = {{Moriya}, Takashi J. and {Blinnikov}, Sergei I. and {Baklanov}, Petr V. and
         {Sorokina}, Elena I. and {Dolgov}, Alexander D.},
        title = "{Synthetic light curves of shocked dense circumstellar shells}",
      journal = {\mnras},
     keywords = {supernovae: general, supernovae: individual: SN 1988Z, supernovae: individual: SN 2003ma, supernovae: individual: SN 2006gy, Astrophysics - High Energy Astrophysical Phenomena},
         year = "2013",
        month = "Apr",
       volume = {430},
        pages = {1402-1407},
          doi = {10.1093/mnras/stt011},
archivePrefix = {arXiv},
       eprint = {1301.0355},
 primaryClass = {astro-ph.HE},
       adsurl = {https://ui.adsabs.harvard.edu/\#abs/2013MNRAS.430.1402M},
      adsnote = {Provided by the SAO/NASA Astrophysics Data System}
}

@ARTICLE{2013MNRAS.428.1020M,
       author = {{Moriya}, Takashi J. and {Blinnikov}, Sergei I. and {Tominaga}, Nozomu and
         {Yoshida}, Naoki and {Tanaka}, Masaomi and {Maeda}, Keiichi and
         {Nomoto}, Ken'ichi},
        title = "{Light-curve modelling of superluminous supernova 2006gy: collision between supernova ejecta and a dense circumstellar medium}",
      journal = {\mnras},
     keywords = {circumstellar matter, stars: mass-loss, supernovae: individual: SN 2006gy, early Universe, Astrophysics - High Energy Astrophysical Phenomena},
         year = "2013",
        month = "Jan",
       volume = {428},
        pages = {1020-1035},
          doi = {10.1093/mnras/sts075},
archivePrefix = {arXiv},
       eprint = {1204.6109},
 primaryClass = {astro-ph.HE},
       adsurl = {https://ui.adsabs.harvard.edu/\#abs/2013MNRAS.428.1020M},
      adsnote = {Provided by the SAO/NASA Astrophysics Data System}
}

@ARTICLE{2013ApJ...763...42O,
       author = {{Ofek}, E.~O. and {Fox}, D. and {Cenko}, S.~B. and {Sullivan}, M. and
         {Gnat}, O. and {Frail}, D.~A. and {Horesh}, A. and {Corsi}, A. and
         {Quimby}, R.~M. and {Gehrels}, N. and {Kulkarni}, S.~R. and
         {Gal-Yam}, A. and {Nugent}, P.~E. and {Yaron}, O. and
         {Filippenko}, A.~V. and {Kasliwal}, M.~M. and {Bildsten}, L. and
         {Bloom}, J.~S. and {Poznanski}, D. and {Arcavi}, I. and {Laher}, R.~R. and
         {Levitan}, D. and {Sesar}, B. and {Surace}, J.},
        title = "{X-Ray Emission from Supernovae in Dense Circumstellar Matter Environments: A Search for Collisionless Shocks}",
      journal = {\apj},
     keywords = {stars: mass-loss, supernovae: general, supernovae: individual: SN 2006jc SN 2010jl, Astrophysics - High Energy Astrophysical Phenomena},
         year = "2013",
        month = "Jan",
       volume = {763},
          eid = {42},
        pages = {42},
          doi = {10.1088/0004-637X/763/1/42},
archivePrefix = {arXiv},
       eprint = {1206.0748},
 primaryClass = {astro-ph.HE},
       adsurl = {https://ui.adsabs.harvard.edu/\#abs/2013ApJ...763...42O},
      adsnote = {Provided by the SAO/NASA Astrophysics Data System}
}

@ARTICLE{2012ApJ...757..178G,
       author = {{Ginzburg}, Sivan and {Balberg}, Shmuel},
        title = "{Superluminous Light Curves from Supernovae Exploding in a Dense Wind}",
      journal = {\apj},
     keywords = {circumstellar matter, shock waves, supernovae: general, supernovae: individual: SN 2005ap SN 2006gy SN 2010gx, Astrophysics - Solar and Stellar Astrophysics},
         year = "2012",
        month = "Oct",
       volume = {757},
          eid = {178},
        pages = {178},
          doi = {10.1088/0004-637X/757/2/178},
archivePrefix = {arXiv},
       eprint = {1205.3455},
 primaryClass = {astro-ph.SR},
       adsurl = {https://ui.adsabs.harvard.edu/\#abs/2012ApJ...757..178G},
      adsnote = {Provided by the SAO/NASA Astrophysics Data System}
}

@ARTICLE{2012MNRAS.423L..92Q,
       author = {{Quataert}, E. and {Shiode}, J.},
        title = "{Wave-driven mass loss in the last year of stellar evolution: setting the stage for the most luminous core-collapse supernovae}",
      journal = {\mnras},
     keywords = {stars: mass loss, supernovae: general, Astrophysics - Solar and Stellar Astrophysics},
         year = "2012",
        month = "Jun",
       volume = {423},
        pages = {L92-L96},
          doi = {10.1111/j.1745-3933.2012.01264.x},
archivePrefix = {arXiv},
       eprint = {1202.5036},
 primaryClass = {astro-ph.SR},
       adsurl = {https://ui.adsabs.harvard.edu/\#abs/2012MNRAS.423L..92Q},
      adsnote = {Provided by the SAO/NASA Astrophysics Data System}
}

@ARTICLE{2012ApJ...750...68L,
       author = {{Lazzati}, Davide and {Morsony}, Brian J. and
         {Blackwell}, Christopher H. and {Begelman}, Mitchell C.},
        title = "{Unifying the Zoo of Jet-driven Stellar Explosions}",
      journal = {\apj},
     keywords = {gamma-ray burst: general, hydrodynamics, supernovae: general, supernovae: individual: SN2009bb, Astrophysics - High Energy Astrophysical Phenomena, Astrophysics - Cosmology and Nongalactic Astrophysics, Astrophysics - Solar and Stellar Astrophysics},
         year = "2012",
        month = "May",
       volume = {750},
          eid = {68},
        pages = {68},
          doi = {10.1088/0004-637X/750/1/68},
archivePrefix = {arXiv},
       eprint = {1111.0970},
 primaryClass = {astro-ph.HE},
       adsurl = {https://ui.adsabs.harvard.edu/\#abs/2012ApJ...750...68L},
      adsnote = {Provided by the SAO/NASA Astrophysics Data System}
}

@ARTICLE{2012ApJ...744...10K,
       author = {{Kiewe}, Michael and {Gal-Yam}, Avishay and {Arcavi}, Iair and
         {Leonard}, Douglas C. and {Emilio Enriquez}, J. and
         {Cenko}, S. Bradley and {Fox}, Derek B. and {Moon}, Dae-Sik and {Sand
        }, David J. and {Soderberg}, Alicia M. and {CCCP}, The},
        title = "{Caltech Core-Collapse Project (CCCP) Observations of Type IIn Supernovae: Typical Properties and Implications for Their Progenitor Stars}",
      journal = {\apj},
     keywords = {stars: mass-loss, supernovae: general, supernovae: individual: SN 2005bx SN 2005cl SN 2005cp SN 2005db, Astrophysics - Cosmology and Nongalactic Astrophysics, Astrophysics - Solar and Stellar Astrophysics},
         year = "2012",
        month = "Jan",
       volume = {744},
          eid = {10},
        pages = {10},
          doi = {10.1088/0004-637X/744/1/10},
archivePrefix = {arXiv},
       eprint = {1010.2689},
 primaryClass = {astro-ph.CO},
       adsurl = {https://ui.adsabs.harvard.edu/\#abs/2012ApJ...744...10K},
      adsnote = {Provided by the SAO/NASA Astrophysics Data System}
}

@ARTICLE{2011NewA...16..187P,
       author = {{Patnaude}, D.~J. and {Loeb}, A. and {Jones}, C.},
        title = "{Evidence for a possible black hole remnant in the Type IIL Supernova 1979C}",
      journal = {\na},
     keywords = {Astrophysics - High Energy Astrophysical Phenomena},
         year = "2011",
        month = "Apr",
       volume = {16},
        pages = {187-190},
          doi = {10.1016/j.newast.2010.09.004},
archivePrefix = {arXiv},
       eprint = {0912.1571},
 primaryClass = {astro-ph.HE},
       adsurl = {https://ui.adsabs.harvard.edu/\#abs/2011NewA...16..187P},
      adsnote = {Provided by the SAO/NASA Astrophysics Data System}
}

@ARTICLE{2011ApJ...729L...6C,
       author = {{Chevalier}, Roger A. and {Irwin}, Christopher M.},
        title = "{Shock Breakout in Dense Mass Loss: Luminous Supernovae}",
      journal = {\apj},
     keywords = {circumstellar matter, shock waves, supernovae: general, supernovae: individual: SN 2006gy, Astrophysics - High Energy Astrophysical Phenomena},
         year = "2011",
        month = "Mar",
       volume = {729},
          eid = {L6},
        pages = {L6},
          doi = {10.1088/2041-8205/729/1/L6},
archivePrefix = {arXiv},
       eprint = {1101.1111},
 primaryClass = {astro-ph.HE},
       adsurl = {https://ui.adsabs.harvard.edu/\#abs/2011ApJ...729L...6C},
      adsnote = {Provided by the SAO/NASA Astrophysics Data System}
}

@ARTICLE{2010ApJ...717..245K,
       author = {{Kasen}, Daniel and {Bildsten}, Lars},
        title = "{Supernova Light Curves Powered by Young Magnetars}",
      journal = {\apj},
     keywords = {radiative transfer, stars: neutron, supernovae: general, supernovae: individual: SN 2005ap SN 2008es SN 2007bi, Astrophysics - High Energy Astrophysical Phenomena},
         year = "2010",
        month = "Jul",
       volume = {717},
        pages = {245-249},
          doi = {10.1088/0004-637X/717/1/245},
archivePrefix = {arXiv},
       eprint = {0911.0680},
 primaryClass = {astro-ph.HE},
       adsurl = {https://ui.adsabs.harvard.edu/\#abs/2010ApJ...717..245K},
      adsnote = {Provided by the SAO/NASA Astrophysics Data System}
}

@ARTICLE{2009A&A...503..869M,
       author = {{Marcaide}, J.~M. and {Mart{\'\i}-Vidal}, I. and {Perez-Torres}, M.~A. and
         {Alberdi}, A. and {Guirado}, J.~C. and {Ros}, E. and {Weiler}, K.~W.},
        title = "{1.6 GHz VLBI observations of SN 1979C: almost-free expansion}",
      journal = {\aap},
     keywords = {radio continuum: stars, supernovae: individual: SN 1979C, Astrophysics - High Energy Astrophysical Phenomena, Astrophysics - Cosmology and Nongalactic Astrophysics},
         year = "2009",
        month = "Sep",
       volume = {503},
        pages = {869-872},
          doi = {10.1051/0004-6361/200912485},
archivePrefix = {arXiv},
       eprint = {0907.0780},
 primaryClass = {astro-ph.HE},
       adsurl = {https://ui.adsabs.harvard.edu/\#abs/2009A&A...503..869M},
      adsnote = {Provided by the SAO/NASA Astrophysics Data System}
}

@ARTICLE{2008ApJ...682.1065B,
       author = {{Bartel}, N. and {Bietenholz}, M.~F.},
        title = "{Shell Revealed in SN 1979C}",
      journal = {\apj},
     keywords = {astrometry, galaxies: individual: M100, radio continuum: stars, supernovae: individual: SN 1979C, techniques: interferometric, Astrophysics},
         year = "2008",
        month = "Aug",
       volume = {682},
        pages = {1065-1069},
          doi = {10.1086/589503},
archivePrefix = {arXiv},
       eprint = {0806.3482},
 primaryClass = {astro-ph},
       adsurl = {https://ui.adsabs.harvard.edu/\#abs/2008ApJ...682.1065B},
      adsnote = {Provided by the SAO/NASA Astrophysics Data System}
}

@ARTICLE{2007ApJ...671L..17S,
       author = {{Smith}, Nathan and {McCray}, Richard},
        title = "{Shell-shocked Diffusion Model for the Light Curve of SN 2006gy}",
      journal = {\apj},
     keywords = {Stars: Circumstellar Matter, Stars: Evolution, supernovae: individual (SN 2006gy), Astrophysics},
         year = "2007",
        month = "Dec",
       volume = {671},
        pages = {L17-L20},
          doi = {10.1086/524681},
archivePrefix = {arXiv},
       eprint = {0710.3428},
 primaryClass = {astro-ph},
       adsurl = {https://ui.adsabs.harvard.edu/\#abs/2007ApJ...671L..17S},
      adsnote = {Provided by the SAO/NASA Astrophysics Data System}
}

@ARTICLE{2004MNRAS.352.1213C,
       author = {{Chugai}, Nikolai N. and {Blinnikov}, Sergei I. and
         {Cumming}, Robert J. and {Lundqvist}, Peter and {Bragaglia}, Angela and
         {Filippenko}, Alexei V. and {Leonard}, Douglas C. and
         {Matheson}, Thomas and {Sollerman}, Jesper},
        title = "{The Type IIn supernova 1994W: evidence for the explosive ejection of a circumstellar envelope}",
      journal = {\mnras},
     keywords = {circumstellar matter, supernovae: general, supernovae: individual: SN 1994W, Astrophysics},
         year = "2004",
        month = "Aug",
       volume = {352},
        pages = {1213-1231},
          doi = {10.1111/j.1365-2966.2004.08011.x},
archivePrefix = {arXiv},
       eprint = {astro-ph/0405369},
 primaryClass = {astro-ph},
       adsurl = {https://ui.adsabs.harvard.edu/\#abs/2004MNRAS.352.1213C},
      adsnote = {Provided by the SAO/NASA Astrophysics Data System}
}

@ARTICLE{2003ApJ...591..301B,
       author = {{Bartel}, Norbert and {Bietenholz}, Michael F.},
        title = "{SN 1979C VLBI: 22 Years of Almost Free Expansion}",
      journal = {\apj},
     keywords = {Cosmology: Distance Scale, Galaxies: Distances and Redshifts, Galaxies: Individual: Messier Number: M100, Radio Continuum: Stars, Stars: Supernovae: Individual: Alphanumeric: SN 1979C},
         year = "2003",
        month = "Jul",
       volume = {591},
        pages = {301-315},
          doi = {10.1086/375267},
       adsurl = {https://ui.adsabs.harvard.edu/\#abs/2003ApJ...591..301B},
      adsnote = {Provided by the SAO/NASA Astrophysics Data System}
}

@ARTICLE{2002AJ....123..745R,
       author = {{Richardson}, Dean and {Branch}, David and {Casebeer}, Darrin and
         {Millard}, Jennifer and {Thomas}, R.~C. and {Baron}, E.},
        title = "{A Comparative Study of the Absolute Magnitude Distributions of Supernovae}",
      journal = {\aj},
     keywords = {Catalogs, Stars: Supernovae: General, Astrophysics},
         year = "2002",
        month = "Feb",
       volume = {123},
        pages = {745-752},
          doi = {10.1086/338318},
archivePrefix = {arXiv},
       eprint = {astro-ph/0112051},
 primaryClass = {astro-ph},
       adsurl = {https://ui.adsabs.harvard.edu/\#abs/2002AJ....123..745R},
      adsnote = {Provided by the SAO/NASA Astrophysics Data System}
}

@BOOK{2002apa..book.....F,
       author = {{Frank}, Juhan and {King}, Andrew and {Raine}, Derek J.},
        title = "{Accretion Power in Astrophysics: Third Edition}",
    booktitle = {Accretion Power in Astrophysics, by Juhan Frank and Andrew King and Derek Raine, pp. 398. ISBN 0521620538. Cambridge, UK: Cambridge University Press, February 2002.},
         year = "2002",
       adsurl = {https://ui.adsabs.harvard.edu/\#abs/2002apa..book.....F},
      adsnote = {Provided by the SAO/NASA Astrophysics Data System}
}

@ARTICLE{2001MNRAS.326.1448C,
       author = {{Chugai}, N.~N.},
        title = "{Broad emission lines from the opaque electron-scattering environment of SN 1998S}",
      journal = {\mnras},
     keywords = {line profiles, scattering, circumstellar matter, supernovae: general, supernovae: individual: SN 1998S, Astrophysics},
         year = "2001",
        month = "Oct",
       volume = {326},
        pages = {1448-1454},
          doi = {10.1111/j.1365-2966.2001.04717.x},
archivePrefix = {arXiv},
       eprint = {astro-ph/0106234},
 primaryClass = {astro-ph},
       adsurl = {https://ui.adsabs.harvard.edu/\#abs/2001MNRAS.326.1448C},
      adsnote = {Provided by the SAO/NASA Astrophysics Data System}
}

@ARTICLE{2001MNRAS.325..907F,
       author = {{Fassia}, A. and {Meikle}, W.~P.~S. and {Chugai}, N. and
         {Geballe}, T.~R. and {Lundqvist}, P. and {Walton}, N.~A. and
         {Pollacco}, D. and {Veilleux}, S. and {Wright}, G.~S. and
         {Pettini}, M. and {Kerr}, T. and {Puchnarewicz}, E. and {Puxley}, P. and
         {Irwin}, M. and {Packham}, C. and {Smartt}, S.~J. and {Harmer}, D.},
        title = "{Optical and infrared spectroscopy of the type IIn SN 1998S: days 3-127}",
      journal = {\mnras},
     keywords = {CIRCUMSTELLAR MATTER, SUPERNOVAE: INDIVIDUAL: SN 1998S, Astrophysics},
         year = "2001",
        month = "Aug",
       volume = {325},
        pages = {907-930},
          doi = {10.1046/j.1365-8711.2001.04282.x},
archivePrefix = {arXiv},
       eprint = {astro-ph/0011340},
 primaryClass = {astro-ph},
       adsurl = {https://ui.adsabs.harvard.edu/\#abs/2001MNRAS.325..907F},
      adsnote = {Provided by the SAO/NASA Astrophysics Data System}
}

@ARTICLE{2000MNRAS.318.1093F,
       author = {{Fassia}, A. and {Meikle}, W.~P.~S. and {Vacca}, W.~D. and
         {Kemp}, S.~N. and {Walton}, N.~A. and {Pollacco}, D.~L. and
         {Smartt}, S. and {Oscoz}, A. and {Arag{\'o}n-Salamanca}, A. and
         {Bennett}, S. and {Hawarden}, T.~G. and {Alonso}, A. and {Alcalde}, D. and
         {Pedrosa}, A. and {Telting}, J. and {Arevalo}, M.~J. and {Deeg}, H.~J. and
         {Garz{\'o}n}, F. and {G{\'o}mez-Rold{\'a}n}, A. and {G{\'o}mez}, G. and
         {Guti{\'e}rrez}, C. and {L{\'o}pez}, S. and {Rozas}, M. and
         {Serra-Ricart}, M. and {Zapatero-Osorio}, M.~R.},
        title = "{Optical and infrared photometry of the Type IIn SN 1998S: days 11-146}",
      journal = {\mnras},
     keywords = {CIRCUMSTELLAR MATTER, SUPERNOVAE: INDIVIDUAL: SN 19985, INFRARED: STARS, Astrophysics},
         year = "2000",
        month = "Nov",
       volume = {318},
        pages = {1093-1104},
          doi = {10.1046/j.1365-8711.2000.03797.x},
archivePrefix = {arXiv},
       eprint = {astro-ph/0006080},
 primaryClass = {astro-ph},
       adsurl = {https://ui.adsabs.harvard.edu/\#abs/2000MNRAS.318.1093F},
      adsnote = {Provided by the SAO/NASA Astrophysics Data System}
}

@ARTICLE{2000ApJ...536..239L,
       author = {{Leonard}, Douglas C. and {Filippenko}, Alexei V. and {Barth}, Aaron J. and
         {Matheson}, Thomas},
        title = "{Evidence for Asphericity in the Type IIN Supernova SN 1998S}",
      journal = {\apj},
     keywords = {Stars: Circumstellar Matter, Polarization, Stars: Mass Loss, supernovae: individual (SN 1998S), Techniques: Polarimetric, Astrophysics},
         year = "2000",
        month = "Jun",
       volume = {536},
        pages = {239-254},
          doi = {10.1086/308910},
archivePrefix = {arXiv},
       eprint = {astro-ph/9908040},
 primaryClass = {astro-ph},
       adsurl = {https://ui.adsabs.harvard.edu/\#abs/2000ApJ...536..239L},
      adsnote = {Provided by the SAO/NASA Astrophysics Data System}
}

@ARTICLE{2000ApJ...532.1124M,
       author = {{Montes}, Marcos J. and {Weiler}, Kurt W. and {Van Dyk}, Schuyler D. and
         {Panagia}, Nino and {Lacey}, Christina K. and {Sramek}, Richard A. and
         {Park}, Richard},
        title = "{Radio Observations of SN 1979C: Evidence for Rapid Presupernova Evolution}",
      journal = {\apj},
     keywords = {RADIO CONTINUUM: STARS, SHOCK WAVES, STARS: EVOLUTION, SUPERNOVAE: INDIVIDUAL (SN 1979C), Astrophysics},
         year = "2000",
        month = "Apr",
       volume = {532},
        pages = {1124-1131},
          doi = {10.1086/308602},
archivePrefix = {arXiv},
       eprint = {astro-ph/9911399},
 primaryClass = {astro-ph},
       adsurl = {https://ui.adsabs.harvard.edu/\#abs/2000ApJ...532.1124M},
      adsnote = {Provided by the SAO/NASA Astrophysics Data System}
}

@ARTICLE{1997ARep...41..672C,
       author = {{Chugai}, N.~N.},
        title = "{The origin of supernovae with dense winds}",
      journal = {Astronomy Reports},
         year = "1997",
        month = "Sep",
       volume = {41},
        pages = {672-681},
       adsurl = {https://ui.adsabs.harvard.edu/\#abs/1997ARep...41..672C},
      adsnote = {Provided by the SAO/NASA Astrophysics Data System}
}

@ARTICLE{1994MNRAS.268..173C,
       author = {{Chugai}, N.~N. and {Danziger}, I.~J.},
        title = "{SN 1988Z: low-mass ejecta colliding with the clumpy wind?}",
      journal = {\mnras},
         year = "1994",
        month = "May",
       volume = {268},
        pages = {173-180},
          doi = {10.1093/mnras/268.1.173},
       adsurl = {https://ui.adsabs.harvard.edu/\#abs/1994MNRAS.268..173C},
      adsnote = {Provided by the SAO/NASA Astrophysics Data System}
}

@ARTICLE{1993A&A...273..106B,
       author = {{Blinnikov}, S.~I. and {Bartunov}, O.~S.},
        title = "{Non-equilibrium radiative transfer in supernova theory : models of linear type II supernovae.}",
      journal = {\aap},
     keywords = {stars: supernovae: supernova light curves, supernovae: SN 1979C},
         year = "1993",
        month = "Jun",
       volume = {273},
        pages = {106-122},
       adsurl = {https://ui.adsabs.harvard.edu/\#abs/1993A&A...273..106B},
      adsnote = {Provided by the SAO/NASA Astrophysics Data System}
}

@ARTICLE{1992ApJ...393..742E,
       author = {{Ensman}, Lisa and {Burrows}, Adam},
        title = "{Shock Breakout in SN 1987A}",
      journal = {\apj},
     keywords = {Shock Wave Propagation, Stellar Temperature, Supernova Remnants, Supernova 1987a, Ultraviolet Radiation, Black Body Radiation, Light Curve, Stellar Color, Stellar Envelopes, Astrophysics},
         year = "1992",
        month = "Jul",
       volume = {393},
        pages = {742},
          doi = {10.1086/171542},
       adsurl = {https://ui.adsabs.harvard.edu/\#abs/1992ApJ...393..742E},
      adsnote = {Provided by the SAO/NASA Astrophysics Data System}
}

@ARTICLE{1992SvAL...18...43B,
       author = {{Bartunov}, O.~S. and {Blinnikov}, S.~I.},
        title = "{Model of supernova 1979C with radiative transfer in the envelope.}",
      journal = {Soviet Astronomy Letters},
     keywords = {Light Curve, Radiative Transfer, Stellar Envelopes, Stellar Models, Supernovae, Attenuation Coefficients, Equations Of State, Red Giant Stars, Stellar Magnitude, Astrophysics},
         year = "1992",
        month = "Feb",
       volume = {18},
        pages = {43},
       adsurl = {https://ui.adsabs.harvard.edu/\#abs/1992SvAL...18...43B},
      adsnote = {Provided by the SAO/NASA Astrophysics Data System}
}

@ARTICLE{1991A&A...246..481H,
       author = {{Hoflich}, P.},
        title = "{Asphericity effects in scatterring dominated photospheres.}",
      journal = {\aap},
         year = "1991",
        month = "Jun",
       volume = {246},
        pages = {481},
       adsurl = {https://ui.adsabs.harvard.edu/\#abs/1991A&A...246..481H},
      adsnote = {Provided by the SAO/NASA Astrophysics Data System}
}

@ARTICLE{1986ApJ...301..790W,
       author = {{Weiler}, K.~W. and {Sramek}, R.~A. and {Panagia}, N. and
         {van der Hulst}, J.~M. and {Salvati}, M.},
        title = "{Radio Supernovae}",
      journal = {\apj},
     keywords = {Calibrating, Radio Sources (Astronomy), Stellar Spectrophotometry, Supernovae, Variable Stars, Light Curve, Radiant Flux Density, Spectral Energy Distribution, Spectrum Analysis, Stellar Models, Supernova Remnants, Tables (Data), Astrophysics, NEBULAE: SUPERNOVA REMNANTS, RADIO SOURCES: VARIABLE, STARS: SUPERNOVAE},
         year = "1986",
        month = "Feb",
       volume = {301},
        pages = {790},
          doi = {10.1086/163944},
       adsurl = {https://ui.adsabs.harvard.edu/\#abs/1986ApJ...301..790W},
      adsnote = {Provided by the SAO/NASA Astrophysics Data System}
}

@ARTICLE{1983ApJ...267..315P,
       author = {{Paczynski}, B.},
        title = "{Models of X-ray bursters with radius expansion}",
      journal = {\apj},
     keywords = {Neutron Stars, Stellar Envelopes, Stellar Evolution, Stellar Models, Thermonuclear Reactions, X Ray Sources, Bursts, Electron Scattering, Gas Expansion, Helium Plasma, Stellar Luminosity, Stellar Mass Accretion, Stellar Temperature, Astrophysics},
         year = "1983",
        month = "Apr",
       volume = {267},
        pages = {315-321},
          doi = {10.1086/160870},
       adsurl = {https://ui.adsabs.harvard.edu/\#abs/1983ApJ...267..315P},
      adsnote = {Provided by the SAO/NASA Astrophysics Data System}
}

@ARTICLE{1979IAUC.3348....1M,
       author = {{Mattei}, J. and {Johnson}, G.~E. and {Rosino}, L. and {Rafanelli}, P. and
         {Kirshner}, R.},
        title = "{Supernova in NGC 4321}",
      journal = {International Astronomical Union Circular},
         year = "1979",
        month = "Apr",
       volume = {3348},
        pages = {1},
       adsurl = {https://ui.adsabs.harvard.edu/\#abs/1979IAUC.3348....1M},
      adsnote = {Provided by the SAO/NASA Astrophysics Data System}
}



\begin{thebibliography}{}
\expandafter\ifx\csname natexlab\endcsname\relax\def\natexlab#1{#1}\fi
\providecommand{\url}[1]{\href{#1}{#1}}

\bibitem[{{Abbott} {et~al.}(2016{\natexlab{a}}){Abbott}, {Abbott}, {Abbott},
  {Abernathy}, {Acernese}, {Ackley}, {Adams}, {Adams}, {Addesso}, {Adhikari},
  {Adya}, {Affeldt}, {Agathos}, {Agatsuma}, {Aggarwal}, {Aguiar}, {Aiello},
  {Ain}, {Ajith}, {Allen}, {Allocca}, {Altin}, {Anderson}, {Anderson}, {Arai},
  {Arain}, {Araya}, {Arceneaux}, {Areeda}, {Arnaud}, {Arun}, {Ascenzi},
  {Ashton}, {Ast}, {Aston}, {Astone}, {Aufmuth}, {Aulbert}, {Babak}, {Bacon},
  {Bader}, {Baker}, {Baldaccini}, {Ballardin}, {Ballmer}, {Barayoga},
  {Barclay}, {Barish}, {Barker}, {Barone}, {Barr}, {Barsotti}, {Barsuglia},
  {Barta}, {Bartlett}, {Barton}, {Bartos}, {Bassiri}, {Basti}, {Batch},
  {Baune}, {Bavigadda}, {Bazzan}, {Behnke}, {Bejger}, {Belczynski}, {Bell},
  {Bell}, {Berger}, {Bergman}, {Bergmann}, {Berry}, {Bersanetti}, {Bertolini},
  {Betzwieser}, {Bhagwat}, {Bhandare}, {Bilenko}, {Billingsley}, {Birch},
  {Birney}, {Birnholtz}, {Biscans}, {Bisht}, {Bitossi}, {Biwer}, {Bizouard},
  {Blackburn}, {Blair}, {Blair}, {Blair}, {Bloemen}, {Bock}, {Bodiya}, {Boer},
  {Bogaert}, {Bogan}, {Bohe}, {Bojtos}, {Bond}, {Bondu}, {Bonnand}, {Boom},
  {Bork}, {Boschi}, {Bose}, {Bouffanais}, {Bozzi}, {Bradaschia}, {Brady},
  {Braginsky}, {Branchesi}, {Brau}, {Briant}, {Brillet}, {Brinkmann},
  {Brisson}, {Brockill}, {Brooks}, {Brown}, {Brown}, {Brown}, {Buchanan},
  {Buikema}, {Bulik}, {Bulten}, {Buonanno}, {Buskulic}, {Buy}, {Byer},
  {Cabero}, {Cadonati}, {Cagnoli}, {Cahillane}, {Bustillo}, {Callister},
  {Calloni}, {Camp}, {Cannon}, {Cao}, {Capano}, {Capocasa}, {Carbognani},
  {Caride}, {Casanueva Diaz}, {Casentini}, {Caudill}, {Cavagli{\`a}},
  {Cavalier}, {Cavalieri}, {Cella}, {Cepeda}, {Baiardi}, {Cerretani},
  {Cesarini}, {Chakraborty}, {Chalermsongsak}, {Chamberlin}, {Chan}, {Chao},
  {Charlton}, {Chassand e-Mottin}, {Chen}, {Chen}, {Cheng}, {Chincarini},
  {Chiummo}, {Cho}, {Cho}, {Chow}, {Christensen}, {Chu}, {Chua}, {Chung},
  {Ciani}, {Clara}, {Clark}, {Cleva}, {Coccia}, {Cohadon}, {Colla}, {Collette},
  {Cominsky}, {Constancio}, {Conte}, {Conti}, {Cook}, {Corbitt}, {Cornish},
  {Corsi}, {Cortese}, {Costa}, {Coughlin}, {Coughlin}, {Coulon}, {Countryman},
  {Couvares}, {Cowan}, {Coward}, {Cowart}, {Coyne}, {Coyne}, {Craig},
  {Creighton}, {Creighton}, {Cripe}, {Crowder}, {Cruise}, {Cumming},
  {Cunningham}, {Cuoco}, {Dal Canton}, {Danilishin}, {D'Antonio}, {Danzmann},
  {Darman}, {Da Silva Costa}, {Dattilo}, {Dave}, {Daveloza}, {Davier},
  {Davies}, {Daw}, {Day}, {De}, {DeBra}, {Debreczeni}, {Degallaix}, {De
  Laurentis}, {Del{\'e}glise}, {Del Pozzo}, {Denker}, {Dent}, {Dereli},
  {Dergachev}, {DeRosa}, {De Rosa}, {DeSalvo}, {Dhurandhar}, {D{\'\i}az}, {Di
  Fiore}, {Di Giovanni}, {Di Lieto}, {Di Pace}, {Di Palma}, {Di Virgilio},
  {Dojcinoski}, {Dolique}, {Donovan}, {Dooley}, {Doravari}, {Douglas},
  {Downes}, {Drago}, {Drever}, {Driggers}, {Du}, {Ducrot}, {Dwyer}, {Edo},
  {Edwards}, {Effler}, {Eggenstein}, {Ehrens}, {Eichholz}, {Eikenberry},
  {Engels}, {Essick}, {Etzel}, {Evans}, {Evans}, {Everett}, {Factourovich},
  {Fafone}, {Fair}, {Fairhurst}, {Fan}, {Fang}, {Farinon}, {Farr}, {Farr},
  {Favata}, {Fays}, {Fehrmann}, {Fejer}, {Feldbaum}, {Ferrante}, {Ferreira},
  {Ferrini}, {Fidecaro}, {Finn}, {Fiori}, {Fiorucci}, {Fisher}, {Flaminio},
  {Fletcher}, {Fong}, {Fournier}, {Franco}, {Frasca}, {Frasconi}, {Frede},
  {Frei}, {Freise}, {Frey}, {Frey}, {Fricke}, {Fritschel}, {Frolov}, {Fulda},
  {Fyffe}, {Gabbard}, {Gair}, {Gammaitoni}, {Gaonkar}, {Garufi}, {Gatto},
  {Gaur}, {Gehrels}, {Gemme}, {Gendre}, {Genin}, {Gennai}, {George}, {Gergely},
  {Germain}, {Ghosh}, {Ghosh}, {Ghosh}, {Giaime}, {Giardina}, {Giazotto},
  {Gill}, {Glaefke}, {Gleason}, {Goetz}, {Goetz}, {Gondan}, {Gonz{\'a}lez},
  {Castro}, {Gopakumar}, {Gordon}, {Gorodetsky}, {Gossan}, {Gosselin},
  {Gouaty}, {Graef}, {Graff}, {Granata}, {Grant}, {Gras}, {Gray}, {Greco},
  {Green}, {Greenhalgh}, {Groot}, {Grote}, {Grunewald}, {Guidi}, {Guo},
  {Gupta}, {Gupta}, {Gushwa}, {Gustafson}, {Gustafson}, {Hacker}, {Hall},
  {Hall}, {Hammond}, {Haney}, {Hanke}, {Hanks}, {Hanna}, {Hannam}, {Hanson},
  {Hardwick}, {Harms}, {Harry}, {Harry}, {Hart}, {Hartman}, {Haster},
  {Haughian}, {Healy}, {Heefner}, {Heidmann}, {Heintze}, {Heinzel}, {Heitmann},
  {Hello}, {Hemming}, {Hendry}, {Heng}, {Hennig}, {Heptonstall}, {Heurs},
  {Hild}, {Hoak}, {Hodge}, {Hofman}, {Hollitt}, {Holt}, {Holz}, {Hopkins},
  {Hosken}, {Hough}, {Houston}, {Howell}, {Hu}, {Huang}, {Huerta}, {Huet},
  {Hughey}, {Husa}, {Huttner}, {Huynh-Dinh}, {Idrisy}, {Indik}, {Ingram},
  {Inta}, {Isa}, {Isac}, {Isi}, {Islas}, {Isogai}, {Iyer}, {Izumi}, {Jacobson},
  {Jacqmin}, {Jang}, {Jani}, {Jaranowski}, {Jawahar}, {Jim{\'e}nez-Forteza},
  {Johnson}, {Johnson-McDaniel}, {Jones}, {Jones}, {Jonker}, {Ju}, {Haris},
  {Kalaghatgi}, {Kalogera}, {Kandhasamy}, {Kang}, {Kanner}, {Karki},
  {Kasprzack}, {Katsavounidis}, {Katzman}, {Kaufer}, {Kaur}, {Kawabe},
  {Kawazoe}, {K{\'e}f{\'e}lian}, {Kehl}, {Keitel}, {Kelley}, {Kells},
  {Kennedy}, {Keppel}, {Key}, {Khalaidovski}, {Khalili}, {Khan}, {Khan},
  {Khan}, {Khazanov}, {Kijbunchoo}, {Kim}, {Kim}, {Kim}, {Kim}, {Kim}, {Kim},
  {King}, {King}, {Kinzel}, {Kissel}, {Kleybolte}, {Klimenko}, {Koehlenbeck},
  {Kokeyama}, {Koley}, {Kondrashov}, {Kontos}, {Koranda}, {Korobko}, {Korth},
  {Kowalska}, {Kozak}, {Kringel}, {Krishnan}, {Kr{\'o}lak}, {Krueger}, {Kuehn},
  {Kumar}, {Kumar}, {Kuo}, {Kutynia}, {Kwee}, {Lackey}, {Landry}, {Lange},
  {Lantz}, {Lasky}, {Lazzarini}, {Lazzaro}, {Leaci}, {Leavey}, {Lebigot},
  {Lee}, {Lee}, {Lee}, {Lee}, {Lenon}, {Leonardi}, {Leong}, {Leroy},
  {Letendre}, {Levin}, {Levine}, {Li}, {Libson}, {Littenberg}, {Lockerbie},
  {Logue}, {Lombardi}, {London}, {Lord}, {Lorenzini}, {Loriette}, {Lormand},
  {Losurdo}, {Lough}, {Lousto}, {Lovelace}, {L{\"u}ck}, {Lundgren}, {Luo},
  {Lynch}, {Ma}, {MacDonald}, {Machenschalk}, {MacInnis}, {Macleod},
  {Maga{\~n}a-Sandoval}, {Magee}, {Mageswaran}, {Majorana}, {Maksimovic},
  {Malvezzi}, {Man}, {Mandel}, {Mandic}, {Mangano}, {Mansell}, {Manske},
  {Mantovani}, {Marchesoni}, {Marion}, {M{\'a}rka}, {M{\'a}rka}, {Markosyan},
  {Maros}, {Martelli}, {Martellini}, {Martin}, {Martin}, {Martynov}, {Marx},
  {Mason}, {Masserot}, {Massinger}, {Masso-Reid}, {Matichard}, {Matone},
  {Mavalvala}, {Mazumder}, {Mazzolo}, {McCarthy}, {McClelland}, {McCormick},
  {McGuire}, {McIntyre}, {McIver}, {McManus}, {McWilliams}, {Meacher},
  {Meadors}, {Meidam}, {Melatos}, {Mendell}, {Mendoza-Gandara}, {Mercer},
  {Merilh}, {Merzougui}, {Meshkov}, {Messenger}, {Messick}, {Meyers},
  {Mezzani}, {Miao}, {Michel}, {Middleton}, {Mikhailov}, {Milano}, {Miller},
  {Millhouse}, {Minenkov}, {Ming}, {Mirshekari}, {Mishra}, {Mitra},
  {Mitrofanov}, {Mitselmakher}, {Mittleman}, {Moggi}, {Mohan}, {Mohapatra},
  {Montani}, {Moore}, {Moore}, {Moraru}, {Moreno}, {Morriss}, {Mossavi},
  {Mours}, {Mow-Lowry}, {Mueller}, {Mueller}, {Muir}, {Mukherjee}, {Mukherjee},
  {Mukherjee}, {Mukund}, {Mullavey}, {Munch}, {Murphy}, {Murray}, {Mytidis},
  {Nardecchia}, {Naticchioni}, {Nayak}, {Necula}, {Nedkova}, {Nelemans},
  {Neri}, {Neunzert}, {Newton}, {Nguyen}, {Nielsen}, {Nissanke}, {Nitz},
  {Nocera}, {Nolting}, {Normandin}, {Nuttall}, {Oberling}, {Ochsner}, {O'Dell},
  {Oelker}, {Ogin}, {Oh}, {Oh}, {Ohme}, {Oliver}, {Oppermann}, {Oram},
  {O'Reilly}, {O'Shaughnessy}, {Ott}, {Ottaway}, {Ottens}, {Overmier}, {Owen},
  {Pai}, {Pai}, {Palamos}, {Palashov}, {Palomba}, {Pal-Singh}, {Pan}, {Pan},
  {Pankow}, {Pannarale}, {Pant}, {Paoletti}, {Paoli}, {Papa}, {Paris},
  {Parker}, {Pascucci}, {Pasqualetti}, {Passaquieti}, {Passuello},
  {Patricelli}, {Patrick}, {Pearlstone}, {Pedraza}, {Pedurand }, {Pekowsky},
  {Pele}, {Penn}, {Perreca}, {Pfeiffer}, {Phelps}, {Piccinni}, {Pichot},
  {Pickenpack}, {Piergiovanni}, {Pierro}, {Pillant}, {Pinard}, {Pinto},
  {Pitkin}, {Poeld}, {Poggiani}, {Popolizio}, {Post}, {Powell}, {Prasad},
  {Predoi}, {Premachandra}, {Prestegard}, {Price}, {Prijatelj}, {Principe},
  {Privitera}, {Prix}, {Prodi}, {Prokhorov}, {Puncken}, {Punturo}, {Puppo},
  {P{\"u}rrer}, {Qi}, {Qin}, {Quetschke}, {Quintero}, {Quitzow-James}, {Raab},
  {Rabeling}, {Radkins}, {Raffai}, {Raja}, {Rakhmanov}, {Ramet}, {Rapagnani},
  {Raymond}, {Razzano}, {Re}, {Read}, {Reed}, {Regimbau}, {Rei}, {Reid},
  {Reitze}, {Rew}, {Reyes}, {Ricci}, {Riles}, {Robertson}, {Robie}, {Robinet},
  {Rocchi}, {Rolland}, {Rollins}, {Roma}, {Romano}, {Romano}, {Romanov},
  {Romie}, {Rosi{\'n}ska}, {Rowan}, {R{\"u}diger}, {Ruggi}, {Ryan}, {Sachdev},
  {Sadecki}, {Sadeghian}, {Salconi}, {Saleem}, {Salemi}, {Samajdar}, {Sammut},
  {Sampson}, {Sanchez}, {Sandberg}, {Sandeen}, {Sand ers}, {Sanders},
  {Sassolas}, {Sathyaprakash}, {Saulson}, {Sauter}, {Savage}, {Sawadsky},
  {Schale}, {Schilling}, {Schmidt}, {Schmidt}, {Schnabel}, {Schofield},
  {Sch{\"o}nbeck}, {Schreiber}, {Schuette}, {Schutz}, {Scott}, {Scott},
  {Sellers}, {Sengupta}, {Sentenac}, {Sequino}, {Sergeev}, {Serna},
  {Setyawati}, {Sevigny}, {Shaddock}, {Shaffer}, {Shah}, {Shahriar}, {Shaltev},
  {Shao}, {Shapiro}, {Shawhan}, {Sheperd}, {Shoemaker}, {Shoemaker}, {Siellez},
  {Siemens}, {Sigg}, {Silva}, {Simakov}, {Singer}, {Singer}, {Singh}, {Singh},
  {Singhal}, {Sintes}, {Slagmolen}, {Smith}, {Smith}, {Smith}, {Smith}, {Son},
  {Sorazu}, {Sorrentino}, {Souradeep}, {Srivastava}, {Staley}, {Steinke},
  {Steinlechner}, {Steinlechner}, {Steinmeyer}, {Stephens}, {Stevenson},
  {Stone}, {Strain}, {Straniero}, {Stratta}, {Strauss}, {Strigin}, {Sturani},
  {Stuver}, {Summerscales}, {Sun}, {Sutton}, {Swinkels}, {Szczepa{\'n}czyk},
  {Tacca}, {Talukder}, {Tanner}, {T{\'a}pai}, {Tarabrin}, {Taracchini},
  {Taylor}, {Theeg}, {Thirugnanasambandam}, {Thomas}, {Thomas}, {Thomas},
  {Thorne}, {Thorne}, {Thrane}, {Tiwari}, {Tiwari}, {Tokmakov}, {Tomlinson},
  {Tonelli}, {Torres}, {Torrie}, {T{\"o}yr{\"a}}, {Travasso}, {Traylor},
  {Trifir{\`o}}, {Tringali}, {Trozzo}, {Tse}, {Turconi}, {Tuyenbayev},
  {Ugolini}, {Unnikrishnan}, {Urban}, {Usman}, {Vahlbruch}, {Vajente},
  {Valdes}, {Vallisneri}, {van Bakel}, {van Beuzekom}, {van den Brand}, {Van
  Den Broeck}, {Vand er-Hyde}, {van der Schaaf}, {van Heijningen}, {van
  Veggel}, {Vardaro}, {Vass}, {Vas{\'u}th}, {Vaulin}, {Vecchio}, {Vedovato},
  {Veitch}, {Veitch}, {Venkateswara}, {Verkindt}, {Vetrano}, {Vicer{\'e}},
  {Vinciguerra}, {Vine}, {Vinet}, {Vitale}, {Vo}, {Vocca}, {Vorvick}, {Voss},
  {Vousden}, {Vyatchanin}, {Wade}, {Wade}, {Wade}, {Waldman}, {Walker},
  {Wallace}, {Walsh}, {Wang}, {Wang}, {Wang}, {Wang}, {Wang}, {Ward}, {Ward},
  {Warner}, {Was}, {Weaver}, {Wei}, {Weinert}, {Weinstein}, {Weiss}, {Welborn},
  {Wen}, {We{\ss}els}, {Westphal}, {Wette}, {Whelan}, {Whitcomb}, {White},
  {Whiting}, {Wiesner}, {Wilkinson}, {Willems}, {Williams}, {Williams},
  {Williamson}, {Willis}, {Willke}, {Wimmer}, {Winkelmann}, {Winkler}, {Wipf},
  {Wiseman}, {Wittel}, {Woan}, {Worden}, {Wright}, {Wu}, {Yablon}, {Yakushin},
  {Yam}, {Yamamoto}, {Yancey}, {Yap}, {Yu}, {Yvert}, {Zadro{\.Z}ny},
  {Zangrando}, {Zanolin}, {Zendri}, {Zevin}, {Zhang}, {Zhang}, {Zhang},
  {Zhang}, {Zhao}, {Zhou}, {Zhou}, {Zhu}, {Zucker}, {Zuraw}, {Zweizig}, {LIGO
  Scientific Collaboration}, \& {Virgo Collaboration}}]{2016PhRvL.116f1102A}
{Abbott}, B.~P., {Abbott}, R., {Abbott}, T.~D., {et~al.} 2016{\natexlab{a}},
  \prl, 116, 061102

\bibitem[{{Abbott} {et~al.}(2016{\natexlab{b}}){Abbott}, {Abbott}, {Abbott},
  {Abernathy}, {Acernese}, {Ackley}, {Adams}, {Adams}, {Addesso}, {Adhikari},
  {Adya}, {Affeldt}, {Agathos}, {Agatsuma}, {Aggarwal}, {Aguiar}, {Aiello},
  {Ain}, {Ajith}, {Allen}, {Allocca}, {Altin}, {Anderson}, {Anderson}, {Arai},
  {Araya}, {Arceneaux}, {Areeda}, {Arnaud}, {Arun}, {Ascenzi}, {Ashton}, {Ast},
  {Aston}, {Astone}, {Aufmuth}, {Aulbert}, {Babak}, {Bacon}, {Bader}, {Baker},
  {Baldaccini}, {Ballardin}, {Ballmer}, {Barayoga}, {Barclay}, {Barish},
  {Barker}, {Barone}, {Barr}, {Barsotti}, {Barsuglia}, {Barta}, {Bartlett},
  {Bartos}, {Bassiri}, {Basti}, {Batch}, {Baune}, {Bavigadda}, {Bazzan},
  {Bejger}, {Bell}, {Berger}, {Bergmann}, {Berry}, {Bersanetti}, {Bertolini},
  {Betzwieser}, {Bhagwat}, {Bhandare}, {Bilenko}, {Billingsley}, {Birch},
  {Birney}, {Birnholtz}, {Biscans}, {Bisht}, {Bitossi}, {Biwer}, {Bizouard},
  {Blackburn}, {Blair}, {Blair}, {Blair}, {Bloemen}, {Bock}, {Boer}, {Bogaert},
  {Bogan}, {Bohe}, {Bond}, {Bondu}, {Bonnand}, {Boom}, {Bork}, {Boschi},
  {Bose}, {Bouffanais}, {Bozzi}, {Bradaschia}, {Brady}, {Braginsky},
  {Branchesi}, {Brau}, {Briant}, {Brillet}, {Brinkmann}, {Brisson}, {Brockill},
  {Broida}, {Brooks}, {Brown}, {Brown}, {Brown}, {Brunett}, {Buchanan},
  {Buikema}, {Bulik}, {Bulten}, {Buonanno}, {Buskulic}, {Buy}, {Byer},
  {Cabero}, {Cadonati}, {Cagnoli}, {Cahillane}, {Calder{\'o}n Bustillo},
  {Callister}, {Calloni}, {Camp}, {Cannon}, {Cao}, {Capano}, {Capocasa},
  {Carbognani}, {Caride}, {Casanueva Diaz}, {Casentini}, {Caudill},
  {Cavagli{\`a}}, {Cavalier}, {Cavalieri}, {Cella}, {Cepeda}, {Cerboni
  Baiardi}, {Cerretani}, {Cesarini}, {Chamberlin}, {Chan}, {Chao}, {Charlton},
  {Chassande-Mottin}, {Cheeseboro}, {Chen}, {Chen}, {Cheng}, {Chincarini},
  {Chiummo}, {Cho}, {Cho}, {Chow}, {Christensen}, {Chu}, {Chua}, {Chung},
  {Ciani}, {Clara}, {Clark}, {Cleva}, {Coccia}, {Cohadon}, {Colla}, {Collette},
  {Cominsky}, {Constancio}, {Conte}, {Conti}, {Cook}, {Corbitt}, {Cornish},
  {Corsi}, {Cortese}, {Costa}, {Coughlin}, {Coughlin}, {Coulon}, {Countryman},
  {Couvares}, {Cowan}, {Coward}, {Cowart}, {Coyne}, {Coyne}, {Craig},
  {Creighton}, {Cripe}, {Crowder}, {Cumming}, {Cunningham}, {Cuoco}, {Dal
  Canton}, {Danilishin}, {D'Antonio}, {Danzmann}, {Darman}, {Dasgupta}, {Da
  Silva Costa}, {Dattilo}, {Dave}, {Davier}, {Davies}, {Daw}, {Day}, {De},
  {DeBra}, {Debreczeni}, {Degallaix}, {De Laurentis}, {Del{\'e}glise}, {Del
  Pozzo}, {Denker}, {Dent}, {Dergachev}, {De Rosa}, {DeRosa}, {DeSalvo},
  {Devine}, {Dhurand har}, {D{\'\i}az}, {Di Fiore}, {Di Giovanni}, {Di
  Girolamo}, {Di Lieto}, {Di Pace}, {Di Palma}, {Di Virgilio}, {Dolique},
  {Donovan}, {Dooley}, {Doravari}, {Douglas}, {Downes}, {Drago}, {Drever},
  {Driggers}, {Ducrot}, {Dwyer}, {Edo}, {Edwards}, {Effler}, {Eggenstein},
  {Ehrens}, {Eichholz}, {Eikenberry}, {Engels}, {Essick}, {Etzel}, {Evans},
  {Evans}, {Everett}, {Factourovich}, {Fafone}, {Fair}, {Fairhurst}, {Fan},
  {Fang}, {Farinon}, {Farr}, {Farr}, {Favata}, {Fays}, {Fehrmann}, {Fejer},
  {Fenyvesi}, {Ferrante}, {Ferreira}, {Ferrini}, {Fidecaro}, {Fiori},
  {Fiorucci}, {Fisher}, {Flaminio}, {Fletcher}, {Fong}, {Fournier}, {Frasca},
  {Frasconi}, {Frei}, {Freise}, {Frey}, {Frey}, {Fritschel}, {Frolov}, {Fulda},
  {Fyffe}, {Gabbard}, {Gair}, {Gammaitoni}, {Gaonkar}, {Garufi}, {Gaur},
  {Gehrels}, {Gemme}, {Geng}, {Genin}, {Gennai}, {George}, {Gergely},
  {Germain}, {Ghosh}, {Ghosh}, {Ghosh}, {Giaime}, {Giardina}, {Giazotto},
  {Gill}, {Glaefke}, {Goetz}, {Goetz}, {Gondan}, {Gonz{\'a}lez}, {Gonzalez
  Castro}, {Gopakumar}, {Gordon}, {Gorodetsky}, {Gossan}, {Gosselin}, {Gouaty},
  {Grado}, {Graef}, {Graff}, {Granata}, {Grant}, {Gras}, {Gray}, {Greco},
  {Green}, {Groot}, {Grote}, {Grunewald}, {Guidi}, {Guo}, {Gupta}, {Gupta},
  {Gushwa}, {Gustafson}, {Gustafson}, {Hacker}, {Hall}, {Hall}, {Hamilton},
  {Hammond}, {Haney}, {Hanke}, {Hanks}, {Hanna}, {Hannam}, {Hanson},
  {Hardwick}, {Harms}, {Harry}, {Harry}, {Hart}, {Hartman}, {Haster},
  {Haughian}, {Healy}, {Heidmann}, {Heintze}, {Heitmann}, {Hello}, {Hemming},
  {Hendry}, {Heng}, {Hennig}, {Henry}, {Heptonstall}, {Heurs}, {Hild}, {Hoak},
  {Hofman}, {Holt}, {Holz}, {Hopkins}, {Hough}, {Houston}, {Howell}, {Hu},
  {Huang}, {Huerta}, {Huet}, {Hughey}, {Husa}, {Huttner}, {Huynh-Dinh},
  {Indik}, {Ingram}, {Inta}, {Isa}, {Isac}, {Isi}, {Isogai}, {Iyer}, {Izumi},
  {Jacqmin}, {Jang}, {Jani}, {Jaranowski}, {Jawahar}, {Jian},
  {Jim{\'e}nez-Forteza}, {Johnson}, {Johnson-McDaniel}, {Jones}, {Jones},
  {Jonker}, {Ju}, {K}, {Kalaghatgi}, {Kalogera}, {Kandhasamy}, {Kang},
  {Kanner}, {Kapadia}, {Karki}, {Karvinen}, {Kasprzack}, {Katsavounidis},
  {Katzman}, {Kaufer}, {Kaur}, {Kawabe}, {K{\'e}f{\'e}lian}, {Kehl}, {Keitel},
  {Kelley}, {Kells}, {Kennedy}, {Key}, {Khalili}, {Khan}, {Khan}, {Khan},
  {Khazanov}, {Kijbunchoo}, {Kim}, {Kim}, {Kim}, {Kim}, {Kim}, {Kim}, {Kim},
  {Kimbrell}, {King}, {King}, {Kissel}, {Klein}, {Kleybolte}, {Klimenko},
  {Koehlenbeck}, {Koley}, {Kondrashov}, {Kontos}, {Korobko}, {Korth},
  {Kowalska}, {Kozak}, {Kringel}, {Krishnan}, {Kr{\'o}lak}, {Krueger}, {Kuehn},
  {Kumar}, {Kumar}, {Kuo}, {Kutynia}, {Lackey}, {Land ry}, {Lange}, {Lantz},
  {Lasky}, {Laxen}, {Lazzarini}, {Lazzaro}, {Leaci}, {Leavey}, {Lebigot},
  {Lee}, {Lee}, {Lee}, {Lee}, {Lenon}, {Leonardi}, {Leong}, {Leroy},
  {Letendre}, {Levin}, {Lewis}, {Li}, {Libson}, {Littenberg}, {Lockerbie},
  {Lombardi}, {London}, {Lord}, {Lorenzini}, {Loriette}, {Lormand}, {Losurdo},
  {Lough}, {Lousto}, {L{\"u}ck}, {Lundgren}, {Lynch}, {Ma}, {Machenschalk},
  {MacInnis}, {Macleod}, {Maga{\~n}a-Sandoval}, {Maga{\~n}a Zertuche}, {Magee},
  {Majorana}, {Maksimovic}, {Malvezzi}, {Man}, {Mandel}, {Mandic}, {Mangano},
  {Mansell}, {Manske}, {Mantovani}, {Marchesoni}, {Marion}, {M{\'a}rka},
  {M{\'a}rka}, {Markosyan}, {Maros}, {Martelli}, {Martellini}, {Martin},
  {Martynov}, {Marx}, {Mason}, {Masserot}, {Massinger}, {Masso-Reid},
  {Mastrogiovanni}, {Matichard}, {Matone}, {Mavalvala}, {Mazumder}, {McCarthy},
  {McClelland}, {McCormick}, {McGuire}, {McIntyre}, {McIver}, {McManus},
  {McRae}, {McWilliams}, {Meacher}, {Meadors}, {Meidam}, {Melatos}, {Mendell},
  {Mercer}, {Merilh}, {Merzougui}, {Meshkov}, {Messenger}, {Messick},
  {Metzdorff}, {Meyers}, {Mezzani}, {Miao}, {Michel}, {Middleton}, {Mikhailov},
  {Milano}, {Miller}, {Miller}, {Miller}, {Miller}, {Millhouse}, {Minenkov},
  {Ming}, {Mirshekari}, {Mishra}, {Mitra}, {Mitrofanov}, {Mitselmakher},
  {Mittleman}, {Moggi}, {Mohan}, {Mohapatra}, {Montani}, {Moore}, {Moore},
  {Moraru}, {Moreno}, {Morriss}, {Mossavi}, {Mours}, {Mow-Lowry}, {Mueller},
  {Muir}, {Mukherjee}, {Mukherjee}, {Mukherjee}, {Mukund}, {Mullavey}, {Munch},
  {Murphy}, {Murray}, {Mytidis}, {Nardecchia}, {Naticchioni}, {Nayak},
  {Nedkova}, {Nelemans}, {Nelson}, {Neri}, {Neunzert}, {Newton}, {Nguyen},
  {Nielsen}, {Nissanke}, {Nitz}, {Nocera}, {Nolting}, {Normandin}, {Nuttall},
  {Oberling}, {Ochsner}, {O'Dell}, {Oelker}, {Ogin}, {Oh}, {Oh}, {Ohme},
  {Oliver}, {Oppermann}, {Oram}, {O'Reilly}, {O'Shaughnessy}, {Ottaway},
  {Overmier}, {Owen}, {Pai}, {Pai}, {Palamos}, {Palashov}, {Palomba},
  {Pal-Singh}, {Pan}, {Pankow}, {Pannarale}, {Pant}, {Paoletti}, {Paoli},
  {Papa}, {Paris}, {Parker}, {Pascucci}, {Pasqualetti}, {Passaquieti},
  {Passuello}, {Patricelli}, {Patrick}, {Pearlstone}, {Pedraza}, {Pedurand},
  {Pekowsky}, {Pele}, {Penn}, {Perreca}, {Perri}, {Pfeiffer}, {Phelps},
  {Piccinni}, {Pichot}, {Piergiovanni}, {Pierro}, {Pillant}, {Pinard}, {Pinto},
  {Pitkin}, {Poe}, {Poggiani}, {Popolizio}, {Post}, {Powell}, {Prasad},
  {Predoi}, {Prestegard}, {Price}, {Prijatelj}, {Principe}, {Privitera},
  {Prix}, {Prodi}, {Prokhorov}, {Puncken}, {Punturo}, {Puppo}, {P{\"u}rrer},
  {Qi}, {Qin}, {Qiu}, {Quetschke}, {Quintero}, {Quitzow-James}, {Raab},
  {Rabeling}, {Radkins}, {Raffai}, {Raja}, {Rajan}, {Rakhmanov}, {Rapagnani},
  {Raymond}, {Razzano}, {Re}, {Read}, {Reed}, {Regimbau}, {Rei}, {Reid},
  {Reitze}, {Rew}, {Reyes}, {Ricci}, {Riles}, {Rizzo}, {Robertson}, {Robie},
  {Robinet}, {Rocchi}, {Rolland}, {Rollins}, {Roma}, {Romano}, {Romano},
  {Romanov}, {Romie}, {Rosi{\'n}ska}, {Rowan}, {R{\"u}diger}, {Ruggi}, {Ryan},
  {Sachdev}, {Sadecki}, {Sadeghian}, {Sakellariadou}, {Salconi}, {Saleem},
  {Salemi}, {Samajdar}, {Sammut}, {Sanchez}, {Sandberg}, {Sandeen}, {Sand ers},
  {Sassolas}, {Sathyaprakash}, {Saulson}, {Sauter}, {Savage}, {Sawadsky},
  {Schale}, {Schilling}, {Schmidt}, {Schmidt}, {Schnabel}, {Schofield},
  {Sch{\"o}nbeck}, {Schreiber}, {Schuette}, {Schutz}, {Scott}, {Scott},
  {Sellers}, {Sengupta}, {Sentenac}, {Sequino}, {Sergeev}, {Setyawati},
  {Shaddock}, {Shaffer}, {Shahriar}, {Shaltev}, {Shapiro}, {Shawhan},
  {Sheperd}, {Shoemaker}, {Shoemaker}, {Siellez}, {Siemens}, {Sieniawska},
  {Sigg}, {Silva}, {Singer}, {Singer}, {Singh}, {Singh}, {Singhal}, {Sintes},
  {Slagmolen}, {Smith}, {Smith}, {Smith}, {Son}, {Sorazu}, {Sorrentino},
  {Souradeep}, {Srivastava}, {Staley}, {Steinke}, {Steinlechner},
  {Steinlechner}, {Steinmeyer}, {Stephens}, {Stevenson}, {Stone}, {Strain},
  {Straniero}, {Stratta}, {Strauss}, {Strigin}, {Sturani}, {Stuver},
  {Summerscales}, {Sun}, {Sunil}, {Sutton}, {Swinkels}, {Szczepa{\'n}czyk},
  {Tacca}, {Talukder}, {Tanner}, {T{\'a}pai}, {Tarabrin}, {Taracchini},
  {Taylor}, {Theeg}, {Thirugnanasamband am}, {Thomas}, {Thomas}, {Thomas},
  {Thorne}, {Thrane}, {Tiwari}, {Tiwari}, {Tokmakov}, {Toland}, {Tomlinson},
  {Tonelli}, {Tornasi}, {Torres}, {Torrie}, {T{\"o}yr{\"a}}, {Travasso},
  {Traylor}, {Trifir{\`o}}, {Tringali}, {Trozzo}, {Tse}, {Turconi},
  {Tuyenbayev}, {Ugolini}, {Unnikrishnan}, {Urban}, {Usman}, {Vahlbruch},
  {Vajente}, {Valdes}, {Vallisneri}, {van Bakel}, {van Beuzekom}, {van den
  Brand}, {Van Den Broeck}, {Vand er-Hyde}, {van der Schaaf}, {van Heijningen},
  {van Veggel}, {Vardaro}, {Vass}, {Vas{\'u}th}, {Vaulin}, {Vecchio},
  {Vedovato}, {Veitch}, {Veitch}, {Venkateswara}, {Verkindt}, {Vetrano},
  {Vicer{\'e}}, {Vinciguerra}, {Vine}, {Vinet}, {Vitale}, {Vo}, {Vocca},
  {Vorvick}, {Voss}, {Vousden}, {Vyatchanin}, {Wade}, {Wade}, {Wade}, {Walker},
  {Wallace}, {Walsh}, {Wang}, {Wang}, {Wang}, {Wang}, {Wang}, {Ward}, {Warner},
  {Was}, {Weaver}, {Wei}, {Weinert}, {Weinstein}, {Weiss}, {Wen}, {We{\ss}els},
  {Westphal}, {Wette}, {Whelan}, {Whiting}, {Williams}, {Williamson}, {Willis},
  {Willke}, {Wimmer}, {Winkler}, {Wipf}, {Wittel}, {Woan}, {Woehler}, {Worden},
  {Wright}, {Wu}, {Wu}, {Yablon}, {Yam}, {Yamamoto}, {Yancey}, {Yu}, {Yvert},
  {Zadro{\.z}ny}, {Zangrando}, {Zanolin}, {Zendri}, {Zevin}, {Zhang}, {Zhang},
  {Zhang}, {Zhao}, {Zhou}, {Zhou}, {Zhu}, {Zucker}, {Zuraw}, {Zweizig},
  {Boyle}, {Hemberger}, {Kidder}, {Lovelace}, {Ossokine}, {Scheel}, {Szilagyi},
  {Teukolsky}, {LIGO Scientific Collaboration}, \& {VIRGO
  Collaboration}}]{2016PhRvL.116x1103A}
---. 2016{\natexlab{b}}, \prl, 116, 241103

\bibitem[{{Abbott}(2017{\natexlab{a}})}]{2017PhRvL.118v1101A}
{Abbott}, B.~P. e.~a. 2017{\natexlab{a}}, \prl, 118, 221101

\bibitem[{{Abbott}(2017{\natexlab{b}})}]{2017PhRvL.119p1101A}
---. 2017{\natexlab{b}}, \prl, 119, 161101

\bibitem[{{Abbott}(2017{\natexlab{c}})}]{2017ApJ...851L..35A}
---. 2017{\natexlab{c}}, \apj, 851, L35

\bibitem[{{Adams} {et~al.}(2017){Adams}, {Kochanek}, {Gerke}, \&
  {Stanek}}]{2017MNRAS.469.1445A}
{Adams}, S.~M., {Kochanek}, C.~S., {Gerke}, J.~R., \& {Stanek}, K.~Z. 2017,
  \mnras, 469, 1445

\bibitem[{{Andrews} \& {Smith}(2018)}]{2018MNRAS.477...74A}
{Andrews}, J.~E., \& {Smith}, N. 2018, \mnras, 477, 74

\bibitem[{{Andrews} {et~al.}(2015){Andrews}, {Farr}, {Kalogera}, \&
  {Willems}}]{2015ApJ...801...32A}
{Andrews}, J.~J., {Farr}, W.~M., {Kalogera}, V., \& {Willems}, B. 2015, \apj,
  801, 32

\bibitem[{{Armitage} \& {Livio}(2000)}]{2000ApJ...532..540A}
{Armitage}, P.~J., \& {Livio}, M. 2000, \apj, 532, 540

\bibitem[{{Astropy Collaboration} {et~al.}(2013){Astropy Collaboration},
  {Robitaille}, {Tollerud}, {Greenfield}, {Droettboom}, {Bray}, {Aldcroft},
  {Davis}, {Ginsburg}, {Price-Whelan}, {Kerzendorf}, {Conley}, {Crighton},
  {Barbary}, {Muna}, {Ferguson}, {Grollier}, {Parikh}, {Nair}, {Unther},
  {Deil}, {Woillez}, {Conseil}, {Kramer}, {Turner}, {Singer}, {Fox}, {Weaver},
  {Zabalza}, {Edwards}, {Azalee Bostroem}, {Burke}, {Casey}, {Crawford},
  {Dencheva}, {Ely}, {Jenness}, {Labrie}, {Lim}, {Pierfederici}, {Pontzen},
  {Ptak}, {Refsdal}, {Servillat}, \& {Streicher}}]{2013A&A...558A..33A}
{Astropy Collaboration}, {Robitaille}, T.~P., {Tollerud}, E.~J., {et~al.} 2013,
  \aap, 558, A33

\bibitem[{{Barkov} \& {Komissarov}(2008)}]{2008MNRAS.385L..28B}
{Barkov}, M.~V., \& {Komissarov}, S.~S. 2008, \mnras, 385, L28

\bibitem[{{Barkov} \& {Komissarov}(2011)}]{2011MNRAS.415..944B}
---. 2011, \mnras, 415, 944

\bibitem[{{Barrett} {et~al.}(2018){Barrett}, {Gaebel}, {Neijssel},
  {Vigna-G{\'o}mez}, {Stevenson}, {Berry}, {Farr}, \&
  {Mandel}}]{2018MNRAS.477.4685B}
{Barrett}, J.~W., {Gaebel}, S.~M., {Neijssel}, C.~J., {et~al.} 2018, \mnras,
  477, 4685

\bibitem[{{Bartel} \& {Bietenholz}(2003)}]{2003ApJ...591..301B}
{Bartel}, N., \& {Bietenholz}, M.~F. 2003, \apj, 591, 301

\bibitem[{{Bartel} \& {Bietenholz}(2008)}]{2008ApJ...682.1065B}
---. 2008, \apj, 682, 1065

\bibitem[{{Bartunov} \& {Blinnikov}(1992)}]{1992SvAL...18...43B}
{Bartunov}, O.~S., \& {Blinnikov}, S.~I. 1992, Soviet Astronomy Letters, 18, 43

\bibitem[{{Belczynski} {et~al.}(2002){Belczynski}, {Kalogera}, \&
  {Bulik}}]{2002ApJ...572..407B}
{Belczynski}, K., {Kalogera}, V., \& {Bulik}, T. 2002, \apj, 572, 407

\bibitem[{{Belczynski} {et~al.}(2008){Belczynski}, {Kalogera}, {Rasio}, {Taam},
  {Zezas}, {Bulik}, {Maccarone}, \& {Ivanova}}]{2008ApJS..174..223B}
{Belczynski}, K., {Kalogera}, V., {Rasio}, F.~A., {et~al.} 2008, The
  Astrophysical Journal Supplement Series, 174, 223

\bibitem[{{Bellm} \& {Kulkarni}(2017)}]{2017NatAs...1E..71B}
{Bellm}, E., \& {Kulkarni}, S. 2017, Nature Astronomy, 1, 0071

\bibitem[{{Blandford} \& {Znajek}(1977)}]{1977MNRAS.179..433B}
{Blandford}, R.~D., \& {Znajek}, R.~L. 1977, \mnras, 179, 433

\bibitem[{{Blondin} {et~al.}(1996){Blondin}, {Lundqvist}, \&
  {Chevalier}}]{1996ApJ...472..257B}
{Blondin}, J.~M., {Lundqvist}, P., \& {Chevalier}, R.~A. 1996, \apj, 472, 257

\bibitem[{{Chandra}(2018)}]{2018SSRv..214...27C}
{Chandra}, P. 2018, \ssr, 214, 27

\bibitem[{{Chen} \& {Beloborodov}(2007)}]{2007ApJ...657..383C}
{Chen}, W.-X., \& {Beloborodov}, A.~M. 2007, \apj, 657, 383

\bibitem[{{Chevalier}(1993)}]{1993ApJ...411L..33C}
{Chevalier}, R.~A. 1993, \apj, 411, L33

\bibitem[{{Chevalier}(1996)}]{1996ApJ...459..322C}
---. 1996, \apj, 459, 322

\bibitem[{{Chevalier}(2012)}]{2012ApJ...752L...2C}
---. 2012, \apj, 752, L2

\bibitem[{{Chevalier} \& {Irwin}(2011)}]{2011ApJ...729L...6C}
{Chevalier}, R.~A., \& {Irwin}, C.~M. 2011, \apj, 729, L6

\bibitem[{{Chevalier} \& {Irwin}(2012)}]{2012ApJ...747L..17C}
---. 2012, \apj, 747, L17

\bibitem[{{Chugai}(1997)}]{1997ARep...41..672C}
{Chugai}, N.~N. 1997, Astronomy Reports, 41, 672

\bibitem[{{Chugai}(2001)}]{2001MNRAS.326.1448C}
---. 2001, \mnras, 326, 1448

\bibitem[{{Chugai} \& {Danziger}(1994)}]{1994MNRAS.268..173C}
{Chugai}, N.~N., \& {Danziger}, I.~J. 1994, \mnras, 268, 173

\bibitem[{{Chugai} {et~al.}(2004){Chugai}, {Blinnikov}, {Cumming}, {Lundqvist},
  {Bragaglia}, {Filippenko}, {Leonard}, {Matheson}, \&
  {Sollerman}}]{2004MNRAS.352.1213C}
{Chugai}, N.~N., {Blinnikov}, S.~I., {Cumming}, R.~J., {et~al.} 2004, \mnras,
  352, 1213

\bibitem[{{Clayton} {et~al.}(2017){Clayton}, {Podsiadlowski}, {Ivanova}, \&
  {Justham}}]{2017MNRAS.470.1788C}
{Clayton}, M., {Podsiadlowski}, P., {Ivanova}, N., \& {Justham}, S. 2017,
  \mnras, 470, 1788

\bibitem[{{Das} \& {Ray}(2017)}]{2017ApJ...851..138D}
{Das}, S., \& {Ray}, A. 2017, \apj, 851, 138

\bibitem[{{de Kool}(1990)}]{de1990common}
{de Kool}, M. 1990, \apj, 358, 189

\bibitem[{{De Marco} \& {Izzard}(2017)}]{2017PASA...34....1D}
{De Marco}, O., \& {Izzard}, R.~G. 2017, Publications of the Astronomical
  Society of Australia, 34, e001

\bibitem[{{de Mink} {et~al.}(2014){de Mink}, {Sana}, {Langer}, {Izzard}, \&
  {Schneider}}]{2014ApJ...782....7D}
{de Mink}, S.~E., {Sana}, H., {Langer}, N., {Izzard}, R.~G., \& {Schneider},
  F.~R.~N. 2014, \apj, 782, 7

\bibitem[{{Dessart} {et~al.}(2015){Dessart}, {Audit}, \&
  {Hillier}}]{2015MNRAS.449.4304D}
{Dessart}, L., {Audit}, E., \& {Hillier}, D.~J. 2015, \mnras, 449, 4304

\bibitem[{{Dexter} \& {Kasen}(2013)}]{2013ApJ...772...30D}
{Dexter}, J., \& {Kasen}, D. 2013, \apj, 772, 30

\bibitem[{{Eggleton}(1983)}]{1983ApJ...268..368E}
{Eggleton}, P.~P. 1983, \apj, 268, 368

\bibitem[{{Eldridge} {et~al.}(2018){Eldridge}, {Xiao}, {Stanway}, {Rodrigues},
  \& {Guo}}]{2018PASA...35...49E}
{Eldridge}, J.~J., {Xiao}, L., {Stanway}, E.~R., {Rodrigues}, N., \& {Guo},
  N.~Y. 2018, \pasa, 35, 49

\bibitem[{{Ensman} \& {Burrows}(1992)}]{1992ApJ...393..742E}
{Ensman}, L., \& {Burrows}, A. 1992, \apj, 393, 742

\bibitem[{{Fassia} {et~al.}(2001){Fassia}, {Meikle}, {Chugai}, {Geballe},
  {Lundqvist}, {Walton}, {Pollacco}, {Veilleux}, {Wright}, \&
  {Pettini}}]{2001MNRAS.325..907F}
{Fassia}, A., {Meikle}, W.~P.~S., {Chugai}, N., {et~al.} 2001, \mnras, 325, 907

\bibitem[{{Feng} {et~al.}(2018){Feng}, {Shen}, \& {Lin}}]{2018ApJ...867..130F}
{Feng}, E.-H., {Shen}, R.-F., \& {Lin}, W.-P. 2018, \apj, 867, 130

\bibitem[{{Forster} {et~al.}(2018){Forster}, {Moriya}, {Maureira}, {Anderson},
  {Blinnikov}, {Bufano}, {Cabrera-Vives}, {Clocchiatti}, {de Jaeger},
  {Estevez}, {Galbany}, {Gonzalez-Gaitan}, {Grafener}, {Hamuy}, {Hsiao},
  {Huentelemu}, {Huijse}, {Kuncarayakti}, {Martinez}, {Medina}, {Olivares},
  {Pignata}, {Razza}, {Reyes}, {San}, {Smith}, {Vera}, {Vivas}, {de Ugarte
  Postigo}, {Yoon}, {Ashall}, {Fraser}, {Gal-Yam}, {Kankare}, {Le Guillou},
  {Mazzali}, {Walton}, \& {Young}}]{2018NatAs...2..808F}
{Forster}, F., {Moriya}, T.~J., {Maureira}, J.~C., {et~al.} 2018, Nature
  Astronomy, 2, 808

\bibitem[{{Frank} {et~al.}(2002){Frank}, {King}, \&
  {Raine}}]{2002apa..book.....F}
{Frank}, J., {King}, A., \& {Raine}, D.~J. 2002, {Accretion Power in
  Astrophysics: Third Edition}

\bibitem[{{Fryer} {et~al.}(1996){Fryer}, {Benz}, \&
  {Herant}}]{1996ApJ...460..801F}
{Fryer}, C.~L., {Benz}, W., \& {Herant}, M. 1996, \apj, 460, 801

\bibitem[{{Fryer} {et~al.}(2014){Fryer}, {Rueda}, \&
  {Ruffini}}]{2014ApJ...793L..36F}
{Fryer}, C.~L., {Rueda}, J.~A., \& {Ruffini}, R. 2014, \apj, 793, L36

\bibitem[{{Fryer} \& {Woosley}(1998)}]{1998ApJ...502L...9F}
{Fryer}, C.~L., \& {Woosley}, S.~E. 1998, \apj, 502, L9

\bibitem[{{Fryer} {et~al.}(1999){Fryer}, {Woosley}, \&
  {Hartmann}}]{1999ApJ...526..152F}
{Fryer}, C.~L., {Woosley}, S.~E., \& {Hartmann}, D.~H. 1999, \apj, 526, 152

\bibitem[{{Fuller}(2017)}]{2017MNRAS.470.1642F}
{Fuller}, J. 2017, \mnras, 470, 1642

\bibitem[{{Gilkis} {et~al.}(2019){Gilkis}, {Soker}, \&
  {Kashi}}]{2019MNRAS.482.4233G}
{Gilkis}, A., {Soker}, N., \& {Kashi}, A. 2019, \mnras, 482, 4233

\bibitem[{{Ginzburg} \& {Balberg}(2012)}]{2012ApJ...757..178G}
{Ginzburg}, S., \& {Balberg}, S. 2012, \apj, 757, 178

\bibitem[{{Ginzburg} \& {Balberg}(2014)}]{2014ApJ...780...18G}
---. 2014, \apj, 780, 18

\bibitem[{{Grichener} \& {Soker}(2018)}]{2018arXiv181003889G}
{Grichener}, A., \& {Soker}, N. 2018, arXiv e-prints, arXiv:1810.03889

\bibitem[{{Guillochon} {et~al.}(2017){Guillochon}, {Parrent}, {Kelley}, \&
  {Margutti}}]{2017ApJ...835...64G}
{Guillochon}, J., {Parrent}, J., {Kelley}, L.~Z., \& {Margutti}, R. 2017, \apj,
  835, 64

\bibitem[{{Holgado} {et~al.}(2018){Holgado}, {Ricker}, \&
  {Huerta}}]{2018ApJ...857...38H}
{Holgado}, A.~M., {Ricker}, P.~M., \& {Huerta}, E.~A. 2018, \apj, 857, 38

\bibitem[{{Holoien} {et~al.}(2019){Holoien}, {Brown}, {Vallely}, {Stanek},
  {Kochanek}, {Shappee}, {Prieto}, {Dong}, {Brimacombe}, {Bishop}, {Bose},
  {Beacom}, {Bersier}, {Chen}, {Chomiuk}, {Falco}, {Holmbo}, {Jayasinghe},
  {Morrell}, {Pojmanski}, {Shields}, {Strader}, {Stritzinger}, {Thompson},
  {Wo{\'z}niak}, {Bock}, {Cacella}, {Carballo}, {Cruz}, {Conseil}, {Farfan},
  {Fernandez}, {Kiyota}, {Koff}, {Krannich}, {Marples}, {Masi}, {Monard},
  {Mu{\~n}oz}, {Nicholls}, {Post}, {Stone}, {Trappett}, \&
  {Wiethoff}}]{2019MNRAS.484.1899H}
{Holoien}, T.~W.~S., {Brown}, J.~S., {Vallely}, P.~J., {et~al.} 2019, \mnras,
  484, 1899

\bibitem[{{Houck} \& {Chevalier}(1991)}]{1991ApJ...376..234H}
{Houck}, J.~C., \& {Chevalier}, R.~A. 1991, \apj, 376, 234

\bibitem[{{Huang}(1963)}]{1963ApJ...138..471H}
{Huang}, S.-S. 1963, \apj, 138, 471

\bibitem[{{Iben} \& {Livio}(1993)}]{1993PASP..105.1373I}
{Iben}, Icko, J., \& {Livio}, M. 1993, Publications of the Astronomical Society
  of the Pacific, 105, 1373

\bibitem[{{Ivanova} {et~al.}(2013){Ivanova}, {Justham}, {Chen}, {De Marco},
  {Fryer}, {Gaburov}, {Ge}, {Glebbeek}, {Han}, {Li}, {Lu}, {Marsh},
  {Podsiadlowski}, {Potter}, {Soker}, {Taam}, {Tauris}, {van den Heuvel}, \&
  {Webbink}}]{2013A&ARv..21...59I}
{Ivanova}, N., {Justham}, S., {Chen}, X., {et~al.} 2013, Astronomy and
  Astrophysics Review, 21, 59

\bibitem[{{Justham} {et~al.}(2014){Justham}, {Podsiadlowski}, \&
  {Vink}}]{2014ApJ...796..121J}
{Justham}, S., {Podsiadlowski}, P., \& {Vink}, J.~S. 2014, \apj, 796, 121

\bibitem[{{Kalogera} {et~al.}(2007){Kalogera}, {Belczynski}, {Kim},
  {O'Shaughnessy}, \& {Willems}}]{2007PhR...442...75K}
{Kalogera}, V., {Belczynski}, K., {Kim}, C., {O'Shaughnessy}, R., \& {Willems},
  B. 2007, \physrep, 442, 75

\bibitem[{{Kalogera} {et~al.}(2004){Kalogera}, {Kim}, {Lorimer}, {Burgay},
  {D'Amico}, {Possenti}, {Manchester}, {Lyne}, {Joshi}, {McLaughlin}, {Kramer},
  {Sarkissian}, \& {Camilo}}]{2004ApJ...601L.179K}
{Kalogera}, V., {Kim}, C., {Lorimer}, D.~R., {et~al.} 2004, \apj, 601, L179

\bibitem[{{Kangas} {et~al.}(2016){Kangas}, {Mattila}, {Kankare}, {Lundqvist},
  {V{\"a}is{\"a}nen}, {Childress}, {Pignata}, {McCully}, {Valenti}, \&
  {Vink{\'o}}}]{2016MNRAS.456..323K}
{Kangas}, T., {Mattila}, S., {Kankare}, E., {et~al.} 2016, \mnras, 456, 323

\bibitem[{{Kasen} \& {Bildsten}(2010)}]{2010ApJ...717..245K}
{Kasen}, D., \& {Bildsten}, L. 2010, \apj, 717, 245

\bibitem[{{Kasen} \& {Woosley}(2009)}]{2009ApJ...703.2205K}
{Kasen}, D., \& {Woosley}, S.~E. 2009, \apj, 703, 2205

\bibitem[{{Kiewe} {et~al.}(2012){Kiewe}, {Gal-Yam}, {Arcavi}, {Leonard},
  {Emilio Enriquez}, {Cenko}, {Fox}, {Moon}, {Sand }, {Soderberg}, \&
  {CCCP}}]{2012ApJ...744...10K}
{Kiewe}, M., {Gal-Yam}, A., {Arcavi}, I., {et~al.} 2012, \apj, 744, 10

\bibitem[{{Kleiser} {et~al.}(2018){Kleiser}, {Kasen}, \&
  {Duffell}}]{2018MNRAS.475.3152K}
{Kleiser}, I. K.~W., {Kasen}, D., \& {Duffell}, P.~C. 2018, \mnras, 475, 3152

\bibitem[{{Kochanek} {et~al.}(2008){Kochanek}, {Beacom}, {Kistler}, {Prieto},
  {Stanek}, {Thompson}, \& {Y{\"u}ksel}}]{2008ApJ...684.1336K}
{Kochanek}, C.~S., {Beacom}, J.~F., {Kistler}, M.~D., {et~al.} 2008, \apj, 684,
  1336

\bibitem[{{Kruckow} {et~al.}(2016){Kruckow}, {Tauris}, {Langer}, {Sz{\'e}csi},
  {Marchant}, \& {Podsiadlowski}}]{2016A&A...596A..58K}
{Kruckow}, M.~U., {Tauris}, T.~M., {Langer}, N., {et~al.} 2016, \aap, 596, A58

\bibitem[{{Lee} \& {Ramirez-Ruiz}(2006)}]{2006ApJ...641..961L}
{Lee}, W.~H., \& {Ramirez-Ruiz}, E. 2006, \apj, 641, 961

\bibitem[{{Levesque} {et~al.}(2014){Levesque}, {Massey}, {Zytkow}, \&
  {Morrell}}]{2014MNRAS.443L..94L}
{Levesque}, E.~M., {Massey}, P., {Zytkow}, A.~N., \& {Morrell}, N. 2014,
  \mnras, 443, L94

\bibitem[{{LSST Science Collaboration} {et~al.}(2009){LSST Science
  Collaboration}, {Abell}, {Allison}, {Anderson}, {Andrew}, {Angel}, {Armus},
  {Arnett}, {Asztalos}, {Axelrod}, {Bailey}, {Ballantyne}, {Bankert},
  {Barkhouse}, {Barr}, {Barrientos}, {Barth}, {Bartlett}, {Becker}, {Becla},
  {Beers}, {Bernstein}, {Biswas}, {Blanton}, {Bloom}, {Bochanski}, {Boeshaar},
  {Borne}, {Bradac}, {Brandt}, {Bridge}, {Brown}, {Brunner}, {Bullock},
  {Burgasser}, {Burge}, {Burke}, {Cargile}, {Chand rasekharan}, {Chartas},
  {Chesley}, {Chu}, {Cinabro}, {Claire}, {Claver}, {Clowe}, {Connolly}, {Cook},
  {Cooke}, {Cooray}, {Covey}, {Culliton}, {de Jong}, {de Vries}, {Debattista},
  {Delgado}, {Dell'Antonio}, {Dhital}, {Di Stefano}, {Dickinson}, {Dilday},
  {Djorgovski}, {Dobler}, {Donalek}, {Dubois-Felsmann}, {Durech},
  {Eliasdottir}, {Eracleous}, {Eyer}, {Falco}, {Fan}, {Fassnacht}, {Ferguson},
  {Fernandez}, {Fields}, {Finkbeiner}, {Figueroa}, {Fox}, {Francke}, {Frank},
  {Frieman}, {Fromenteau}, {Furqan}, {Galaz}, {Gal-Yam}, {Garnavich},
  {Gawiser}, {Geary}, {Gee}, {Gibson}, {Gilmore}, {Grace}, {Green}, {Gressler},
  {Grillmair}, {Habib}, {Haggerty}, {Hamuy}, {Harris}, {Hawley}, {Heavens},
  {Hebb}, {Henry}, {Hileman}, {Hilton}, {Hoadley}, {Holberg}, {Holman},
  {Howell}, {Infante}, {Ivezic}, {Jacoby}, {Jain}, {R}, {Jedicke}, {Jee},
  {Garrett Jernigan}, {Jha}, {Johnston}, {Jones}, {Juric}, {Kaasalainen},
  {Styliani}, {Kafka}, {Kahn}, {Kaib}, {Kalirai}, {Kantor}, {Kasliwal},
  {Keeton}, {Kessler}, {Knezevic}, {Kowalski}, {Krabbendam}, {Krughoff},
  {Kulkarni}, {Kuhlman}, {Lacy}, {Lepine}, {Liang}, {Lien}, {Lira}, {Long},
  {Lorenz}, {Lotz}, {Lupton}, {Lutz}, {Macri}, {Mahabal}, {Mandelbaum},
  {Marshall}, {May}, {McGehee}, {Meadows}, {Meert}, {Milani}, {Miller},
  {Miller}, {Mills}, {Minniti}, {Monet}, {Mukadam}, {Nakar}, {Neill}, {Newman},
  {Nikolaev}, {Nordby}, {O'Connor}, {Oguri}, {Oliver}, {Olivier}, {Olsen},
  {Olsen}, {Olszewski}, {Oluseyi}, {Padilla}, {Parker}, {Pepper}, {Peterson},
  {Petry}, {Pinto}, {Pizagno}, {Popescu}, {Prsa}, {Radcka}, {Raddick},
  {Rasmussen}, {Rau}, {Rho}, {Rhoads}, {Richards}, {Ridgway}, {Robertson},
  {Roskar}, {Saha}, {Sarajedini}, {Scannapieco}, {Schalk}, {Schindler},
  {Schmidt}, {Schmidt}, {Schneider}, {Schumacher}, {Scranton}, {Sebag},
  {Seppala}, {Shemmer}, {Simon}, {Sivertz}, {Smith}, {Allyn Smith}, {Smith},
  {Spitz}, {Stanford}, {Stassun}, {Strader}, {Strauss}, {Stubbs}, {Sweeney},
  {Szalay}, {Szkody}, {Takada}, {Thorman}, {Trilling}, {Trimble}, {Tyson}, {Van
  Berg}, {Vand en Berk}, {VanderPlas}, {Verde}, {Vrsnak}, {Walkowicz}, {Wand
  elt}, {Wang}, {Wang}, {Warner}, {Wechsler}, {West}, {Wiecha}, {Williams},
  {Willman}, {Wittman}, {Wolff}, {Wood-Vasey}, {Wozniak}, {Young}, {Zentner},
  \& {Zhan}}]{2009arXiv0912.0201L}
{LSST Science Collaboration}, {Abell}, P.~A., {Allison}, J., {et~al.} 2009,
  arXiv e-prints, arXiv:0912.0201

\bibitem[{{MacFadyen} \& {Woosley}(1999)}]{1999ApJ...524..262M}
{MacFadyen}, A.~I., \& {Woosley}, S.~E. 1999, \apj, 524, 262

\bibitem[{{MacFadyen} {et~al.}(2001){MacFadyen}, {Woosley}, \&
  {Heger}}]{2001ApJ...550..410M}
{MacFadyen}, A.~I., {Woosley}, S.~E., \& {Heger}, A. 2001, \apj, 550, 410

\bibitem[{{MacLeod} {et~al.}(2017{\natexlab{a}}){MacLeod}, {Antoni},
  {Murguia-Berthier}, {Macias}, \& {Ramirez-Ruiz}}]{2017ApJ...838...56M}
{MacLeod}, M., {Antoni}, A., {Murguia-Berthier}, A., {Macias}, P., \&
  {Ramirez-Ruiz}, E. 2017{\natexlab{a}}, \apj, 838, 56

\bibitem[{{MacLeod} {et~al.}(2017{\natexlab{b}}){MacLeod}, {Macias},
  {Ramirez-Ruiz}, {Grindlay}, {Batta}, \& {Montes}}]{2017ApJ...835..282M}
{MacLeod}, M., {Macias}, P., {Ramirez-Ruiz}, E., {et~al.} 2017{\natexlab{b}},
  \apj, 835, 282

\bibitem[{{MacLeod} {et~al.}(2018{\natexlab{a}}){MacLeod}, {Ostriker}, \&
  {Stone}}]{2018ApJ...863....5M}
{MacLeod}, M., {Ostriker}, E.~C., \& {Stone}, J.~M. 2018{\natexlab{a}}, \apj,
  863, 5

\bibitem[{{MacLeod} {et~al.}(2018{\natexlab{b}}){MacLeod}, {Ostriker}, \&
  {Stone}}]{2018ApJ...868..136M}
---. 2018{\natexlab{b}}, \apj, 868, 136

\bibitem[{{MacLeod} \& {Ramirez-Ruiz}(2015)}]{2015ApJ...798L..19M}
{MacLeod}, M., \& {Ramirez-Ruiz}, E. 2015, \apj, 798, L19

\bibitem[{{MacLeod} {et~al.}(2018{\natexlab{c}}){MacLeod}, {Vick}, {Lai}, \&
  {Stone}}]{2018arXiv181207594M}
{MacLeod}, M., {Vick}, M., {Lai}, D., \& {Stone}, J.~M. 2018{\natexlab{c}},
  arXiv e-prints, arXiv:1812.07594

\bibitem[{{Marcaide} {et~al.}(2009){Marcaide}, {Mart{\'\i}-Vidal},
  {Perez-Torres}, {Alberdi}, {Guirado}, {Ros}, \&
  {Weiler}}]{2009A&A...503..869M}
{Marcaide}, J.~M., {Mart{\'\i}-Vidal}, I., {Perez-Torres}, M.~A., {et~al.}
  2009, \aap, 503, 869

\bibitem[{{Margutti} {et~al.}(2014){Margutti}, {Milisavljevic}, {Soderberg},
  {Chornock}, {Zauderer}, {Murase}, {Guidorzi}, {Sanders}, {Kuin}, {Fransson},
  {Levesque}, {Chandra}, {Berger}, {Bianco}, {Brown}, {Challis},
  {Chatzopoulos}, {Cheung}, {Choi}, {Chomiuk}, {Chugai}, {Contreras}, {Drout},
  {Fesen}, {Foley}, {Fong}, {Friedman}, {Gall}, {Gehrels}, {Hjorth}, {Hsiao},
  {Kirshner}, {Im}, {Leloudas}, {Lunnan}, {Marion}, {Martin}, {Morrell},
  {Neugent}, {Omodei}, {Phillips}, {Rest}, {Silverman}, {Strader},
  {Stritzinger}, {Szalai}, {Utterback}, {Vinko}, {Wheeler}, {Arnett},
  {Campana}, {Chevalier}, {Ginsburg}, {Kamble}, {Roming}, {Pritchard}, \&
  {Stringfellow}}]{2014ApJ...780...21M}
{Margutti}, R., {Milisavljevic}, D., {Soderberg}, A.~M., {et~al.} 2014, \apj,
  780, 21

\bibitem[{{Margutti} {et~al.}(2017){Margutti}, {Kamble}, {Milisavljevic},
  {Zapartas}, {de Mink}, {Drout}, {Chornock}, {Risaliti}, {Zauderer},
  {Bietenholz}, {Cantiello}, {Chakraborti}, {Chomiuk}, {Fong}, {Grefenstette},
  {Guidorzi}, {Kirshner}, {Parrent}, {Patnaude}, {Soderberg}, {Gehrels}, \&
  {Harrison}}]{2017ApJ...835..140M}
{Margutti}, R., {Kamble}, A., {Milisavljevic}, D., {et~al.} 2017, \apj, 835,
  140

\bibitem[{{Margutti} {et~al.}(2019){Margutti}, {Metzger}, {Chornock}, {Vurm},
  {Roth}, {Grefenstette}, {Savchenko}, {Cartier}, {Steiner}, {Terreran},
  {Margalit}, {Migliori}, {Milisavljevic}, {Alexand er}, {Bietenholz},
  {Blanchard}, {Bozzo}, {Brethauer}, {Chilingarian}, {Coppejans}, {Ducci},
  {Ferrigno}, {Fong}, {G{\"o}tz}, {Guidorzi}, {Hajela}, {Hurley}, {Kuulkers},
  {Laurent}, {Mereghetti}, {Nicholl}, {Patnaude}, {Ubertini}, {Banovetz},
  {Bartel}, {Berger}, {Coughlin}, {Eftekhari}, {Frederiks}, {Kozlova},
  {Laskar}, {Svinkin}, {Drout}, {MacFadyen}, \&
  {Paterson}}]{2019ApJ...872...18M}
{Margutti}, R., {Metzger}, B.~D., {Chornock}, R., {et~al.} 2019, \apj, 872, 18

\bibitem[{{Mason} {et~al.}(2010){Mason}, {Diaz}, {Williams}, {Preston}, \&
  {Bensby}}]{2010A&A...516A.108M}
{Mason}, E., {Diaz}, M., {Williams}, R.~E., {Preston}, G., \& {Bensby}, T.
  2010, \aap, 516, A108

\bibitem[{{Mattei} {et~al.}(1979){Mattei}, {Johnson}, {Rosino}, {Rafanelli}, \&
  {Kirshner}}]{1979IAUC.3348....1M}
{Mattei}, J., {Johnson}, G.~E., {Rosino}, L., {Rafanelli}, P., \& {Kirshner},
  R. 1979, International Astronomical Union Circular, 3348, 1

\bibitem[{{Matzner}(2003)}]{2003MNRAS.345..575M}
{Matzner}, C.~D. 2003, \mnras, 345, 575

\bibitem[{{Mauerhan} {et~al.}(2014){Mauerhan}, {Williams}, {Smith}, {Smith},
  {Filippenko}, {Hoffman}, {Milne}, {Leonard}, {Clubb}, {Fox}, \&
  {Kelly}}]{2014MNRAS.442.1166M}
{Mauerhan}, J., {Williams}, G.~G., {Smith}, N., {et~al.} 2014, \mnras, 442,
  1166

\bibitem[{{McDowell} {et~al.}(2018){McDowell}, {Duffell}, \&
  {Kasen}}]{2018ApJ...856...29M}
{McDowell}, A.~T., {Duffell}, P.~C., \& {Kasen}, D. 2018, \apj, 856, 29

\bibitem[{{Metzger} \& {Pejcha}(2017)}]{2017MNRAS.471.3200M}
{Metzger}, B.~D., \& {Pejcha}, O. 2017, \mnras, 471, 3200

\bibitem[{{Moe} \& {Di Stefano}(2017)}]{2017ApJS..230...15M}
{Moe}, M., \& {Di Stefano}, R. 2017, The Astrophysical Journal Supplement
  Series, 230, 15

\bibitem[{{Montes} {et~al.}(2000){Montes}, {Weiler}, {Van Dyk}, {Panagia},
  {Lacey}, {Sramek}, \& {Park}}]{2000ApJ...532.1124M}
{Montes}, M.~J., {Weiler}, K.~W., {Van Dyk}, S.~D., {et~al.} 2000, \apj, 532,
  1124

\bibitem[{{Moriya}(2018)}]{2018MNRAS.475L..49M}
{Moriya}, T.~J. 2018, \mnras, 475, L49

\bibitem[{{Moriya} {et~al.}(2013{\natexlab{a}}){Moriya}, {Blinnikov},
  {Tominaga}, {Yoshida}, {Tanaka}, {Maeda}, \& {Nomoto}}]{2013MNRAS.428.1020M}
{Moriya}, T.~J., {Blinnikov}, S.~I., {Tominaga}, N., {et~al.}
  2013{\natexlab{a}}, \mnras, 428, 1020

\bibitem[{{Moriya} \& {Maeda}(2012)}]{2012ApJ...756L..22M}
{Moriya}, T.~J., \& {Maeda}, K. 2012, \apj, 756, L22

\bibitem[{{Moriya} {et~al.}(2013{\natexlab{b}}){Moriya}, {Maeda}, {Taddia},
  {Sollerman}, {Blinnikov}, \& {Sorokina}}]{2013MNRAS.435.1520M}
{Moriya}, T.~J., {Maeda}, K., {Taddia}, F., {et~al.} 2013{\natexlab{b}},
  \mnras, 435, 1520

\bibitem[{{Morozova} {et~al.}(2015{\natexlab{a}}){Morozova}, {Ott}, \&
  {Piro}}]{2015ascl.soft05033M}
{Morozova}, V., {Ott}, C.~D., \& {Piro}, A.~L. 2015{\natexlab{a}}, {SNEC:
  SuperNova Explosion Code}, , , ascl:1505.033

\bibitem[{{Morozova} {et~al.}(2015{\natexlab{b}}){Morozova}, {Piro}, {Renzo},
  {Ott}, {Clausen}, {Couch}, {Ellis}, \& {Roberts}}]{2015ApJ...814...63M}
{Morozova}, V., {Piro}, A.~L., {Renzo}, M., {et~al.} 2015{\natexlab{b}}, \apj,
  814, 63

\bibitem[{{Morozova} {et~al.}(2017){Morozova}, {Piro}, \&
  {Valenti}}]{2017ApJ...838...28M}
{Morozova}, V., {Piro}, A.~L., \& {Valenti}, S. 2017, \apj, 838, 28

\bibitem[{{Morozova} {et~al.}(2018){Morozova}, {Piro}, \&
  {Valenti}}]{2018ApJ...858...15M}
---. 2018, \apj, 858, 15

\bibitem[{{Morozova} \& {Stone}(2018)}]{2018ApJ...867....4M}
{Morozova}, V., \& {Stone}, J.~M. 2018, \apj, 867, 4

\bibitem[{{Murguia-Berthier} {et~al.}(2017){Murguia-Berthier}, {MacLeod},
  {Ramirez-Ruiz}, {Antoni}, \& {Macias}}]{2017ApJ...845..173M}
{Murguia-Berthier}, A., {MacLeod}, M., {Ramirez-Ruiz}, E., {Antoni}, A., \&
  {Macias}, P. 2017, \apj, 845, 173

\bibitem[{{Murguia-Berthier} {et~al.}(2014){Murguia-Berthier}, {Montes},
  {Ramirez-Ruiz}, {De Colle}, \& {Lee}}]{2014ApJ...788L...8M}
{Murguia-Berthier}, A., {Montes}, G., {Ramirez-Ruiz}, E., {De Colle}, F., \&
  {Lee}, W.~H. 2014, \apj, 788, L8

\bibitem[{{Ofek} {et~al.}(2013{\natexlab{a}}){Ofek}, {Lin}, {Kouveliotou},
  {Younes}, {G{\"o}{\v{g}}{\"u}{\c{s}}}, {Kasliwal}, \&
  {Cao}}]{2013ApJ...768...47O}
{Ofek}, E.~O., {Lin}, L., {Kouveliotou}, C., {et~al.} 2013{\natexlab{a}}, \apj,
  768, 47

\bibitem[{{Ofek} {et~al.}(2013{\natexlab{b}}){Ofek}, {Fox}, {Cenko},
  {Sullivan}, {Gnat}, {Frail}, {Horesh}, {Corsi}, {Quimby}, \&
  {Gehrels}}]{2013ApJ...763...42O}
{Ofek}, E.~O., {Fox}, D., {Cenko}, S.~B., {et~al.} 2013{\natexlab{b}}, \apj,
  763, 42

\bibitem[{{Ofek} {et~al.}(2013{\natexlab{c}}){Ofek}, {Sullivan}, {Cenko},
  {Kasliwal}, {Gal-Yam}, {Kulkarni}, {Arcavi}, {Bildsten}, {Bloom}, {Horesh},
  {Howell}, {Filippenko}, {Laher}, {Murray}, {Nakar}, {Nugent}, {Silverman},
  {Shaviv}, {Surace}, \& {Yaron}}]{2013Natur.494...65O}
{Ofek}, E.~O., {Sullivan}, M., {Cenko}, S.~B., {et~al.} 2013{\natexlab{c}},
  \nat, 494, 65

\bibitem[{{Ofek} {et~al.}(2014){Ofek}, {Arcavi}, {Tal}, {Sullivan}, {Gal-Yam},
  {Kulkarni}, {Nugent}, {Ben-Ami}, {Bersier}, {Cao}, {Cenko}, {De Cia},
  {Filippenko}, {Fransson}, {Kasliwal}, {Laher}, {Surace}, {Quimby}, \&
  {Yaron}}]{2014ApJ...788..154O}
{Ofek}, E.~O., {Arcavi}, I., {Tal}, D., {et~al.} 2014, \apj, 788, 154

\bibitem[{{{\"O}pik}(1924)}]{opik1924statistical}
{{\"O}pik}, E. 1924, Publications of the Tartu Astrofizica Observatory, 25

\bibitem[{{Paczynski}(1976)}]{1976IAUS...73...75P}
{Paczynski}, B. 1976, in IAU Symposium, Vol.~73, Structure and Evolution of
  Close Binary Systems, ed. P.~{Eggleton}, S.~{Mitton}, \& J.~{Whelan}, 75

\bibitem[{{Paczynski}(1983)}]{1983ApJ...267..315P}
{Paczynski}, B. 1983, \apj, 267, 315

\bibitem[{{Paczy{\'n}ski} \& {Sienkiewicz}(1972)}]{1972AcA....22...73P}
{Paczy{\'n}ski}, B., \& {Sienkiewicz}, R. 1972, \actaa, 22, 73

\bibitem[{{Pan} {et~al.}(2013){Pan}, {Patnaude}, \&
  {Loeb}}]{2013MNRAS.433..838P}
{Pan}, T., {Patnaude}, D., \& {Loeb}, A. 2013, \mnras, 433, 838

\bibitem[{{Patnaude} {et~al.}(2011){Patnaude}, {Loeb}, \&
  {Jones}}]{2011NewA...16..187P}
{Patnaude}, D.~J., {Loeb}, A., \& {Jones}, C. 2011, New Astronomy, 16, 187

\bibitem[{{Pejcha}(2014)}]{2014ApJ...788...22P}
{Pejcha}, O. 2014, \apj, 788, 22

\bibitem[{{Pejcha} {et~al.}(2016{\natexlab{a}}){Pejcha}, {Metzger}, \&
  {Tomida}}]{2016MNRAS.455.4351P}
{Pejcha}, O., {Metzger}, B.~D., \& {Tomida}, K. 2016{\natexlab{a}}, \mnras,
  455, 4351

\bibitem[{{Pejcha} {et~al.}(2016{\natexlab{b}}){Pejcha}, {Metzger}, \&
  {Tomida}}]{2016MNRAS.461.2527P}
---. 2016{\natexlab{b}}, \mnras, 461, 2527

\bibitem[{{Pejcha} {et~al.}(2017){Pejcha}, {Metzger}, {Tyles}, \&
  {Tomida}}]{2017ApJ...850...59P}
{Pejcha}, O., {Metzger}, B.~D., {Tyles}, J.~G., \& {Tomida}, K. 2017, \apj,
  850, 59

\bibitem[{{Pejcha} \& {Thompson}(2012)}]{2012ApJ...746..106P}
{Pejcha}, O., \& {Thompson}, T.~A. 2012, \apj, 746, 106

\bibitem[{{Podsiadlowski}(2007)}]{2007ASPC..367..541P}
{Podsiadlowski}, P. 2007, in Astronomical Society of the Pacific Conference
  Series, Vol. 367, Massive Stars in Interactive Binaries, ed. N.~{St. -Louis}
  \& A.~F.~J. {Moffat}, 541

\bibitem[{{Podsiadlowski} {et~al.}(1995){Podsiadlowski}, {Cannon}, \&
  {Rees}}]{1995MNRAS.274..485P}
{Podsiadlowski}, P., {Cannon}, R.~C., \& {Rees}, M.~J. 1995, \mnras, 274, 485

\bibitem[{{Popham} {et~al.}(1999){Popham}, {Woosley}, \&
  {Fryer}}]{1999ApJ...518..356P}
{Popham}, R., {Woosley}, S.~E., \& {Fryer}, C. 1999, \apj, 518, 356

\bibitem[{{Popov}(1993)}]{1993ApJ...414..712P}
{Popov}, D.~V. 1993, \apj, 414, 712

\bibitem[{{Prieto} {et~al.}(2013){Prieto}, {Brimacombe}, {Drake}, \&
  {Howerton}}]{2013ApJ...763L..27P}
{Prieto}, J.~L., {Brimacombe}, J., {Drake}, A.~J., \& {Howerton}, S. 2013,
  \apj, 763, L27

\bibitem[{{Quataert} {et~al.}(2016){Quataert}, {Fern{\'a}ndez}, {Kasen},
  {Klion}, \& {Paxton}}]{2016MNRAS.458.1214Q}
{Quataert}, E., {Fern{\'a}ndez}, R., {Kasen}, D., {Klion}, H., \& {Paxton}, B.
  2016, \mnras, 458, 1214

\bibitem[{{Quataert} \& {Kasen}(2012)}]{2012MNRAS.419L...1Q}
{Quataert}, E., \& {Kasen}, D. 2012, \mnras, 419, L1

\bibitem[{{Quataert} \& {Shiode}(2012)}]{2012MNRAS.423L..92Q}
{Quataert}, E., \& {Shiode}, J. 2012, \mnras, 423, L92

\bibitem[{{Salpeter}(1955)}]{salpeter1955luminosity}
{Salpeter}, E.~E. 1955, \apj, 121, 161

\bibitem[{{Sana} {et~al.}(2012){Sana}, {de Mink}, {de Koter}, {Langer},
  {Evans}, {Gieles}, {Gosset}, {Izzard}, {Le Bouquin}, \&
  {Schneider}}]{2012Sci...337..444S}
{Sana}, H., {de Mink}, S.~E., {de Koter}, A., {et~al.} 2012, Science, 337, 444

\bibitem[{{Senno} {et~al.}(2016){Senno}, {Murase}, \&
  {M{\'e}sz{\'a}ros}}]{2016PhRvD..93h3003S}
{Senno}, N., {Murase}, K., \& {M{\'e}sz{\'a}ros}, P. 2016, \prd, 93, 083003

\bibitem[{{Shiode} \& {Quataert}(2014)}]{2014ApJ...780...96S}
{Shiode}, J.~H., \& {Quataert}, E. 2014, \apj, 780, 96

\bibitem[{{Shivvers} {et~al.}(2015){Shivvers}, {Groh}, {Mauerhan}, {Fox},
  {Leonard}, \& {Filippenko}}]{2015ApJ...806..213S}
{Shivvers}, I., {Groh}, J.~H., {Mauerhan}, J.~C., {et~al.} 2015, \apj, 806, 213

\bibitem[{{Shu} {et~al.}(1979){Shu}, {Lubow}, \&
  {Anderson}}]{1979ApJ...229..223S}
{Shu}, F.~H., {Lubow}, S.~H., \& {Anderson}, L. 1979, \apj, 229, 223

\bibitem[{{Siegel} {et~al.}(2018){Siegel}, {Barnes}, \&
  {Metzger}}]{2018arXiv181000098S}
{Siegel}, D.~M., {Barnes}, J., \& {Metzger}, B.~D. 2018, arXiv e-prints,
  arXiv:1810.00098

\bibitem[{{Smith}(2017)}]{2017hsn..book..403S}
{Smith}, N. 2017, {Interacting Supernovae: Types IIn and Ibn}, 403

\bibitem[{{Smith} \& {Arnett}(2014)}]{2014ApJ...785...82S}
{Smith}, N., \& {Arnett}, W.~D. 2014, \apj, 785, 82

\bibitem[{{Smith} {et~al.}(2014){Smith}, {Mauerhan}, \&
  {Prieto}}]{2014MNRAS.438.1191S}
{Smith}, N., {Mauerhan}, J.~C., \& {Prieto}, J.~L. 2014, \mnras, 438, 1191

\bibitem[{{Smith} \& {McCray}(2007)}]{2007ApJ...671L..17S}
{Smith}, N., \& {McCray}, R. 2007, \apj, 671, L17

\bibitem[{{Smith} {et~al.}(2010){Smith}, {Miller}, {Li}, {Filippenko},
  {Silverman}, {Howard}, {Nugent}, {Marcy}, {Bloom}, {Ghez}, {Lu}, {Yelda},
  {Bernstein}, \& {Colucci}}]{2010AJ....139.1451S}
{Smith}, N., {Miller}, A., {Li}, W., {et~al.} 2010, \aj, 139, 1451

\bibitem[{{Soker}(2019)}]{2019arXiv190201187S}
{Soker}, N. 2019, arXiv e-prints, arXiv:1902.01187

\bibitem[{{Soker} \& {Gilkis}(2018)}]{2018MNRAS.475.1198S}
{Soker}, N., \& {Gilkis}, A. 2018, \mnras, 475, 1198

\bibitem[{{Soker} {et~al.}(2019){Soker}, {Grichener}, \&
  {Gilkis}}]{2019MNRAS.484.4972S}
{Soker}, N., {Grichener}, A., \& {Gilkis}, A. 2019, \mnras, 484, 4972

\bibitem[{{Song} \& {Liu}(2019)}]{2019ApJ...871..117S}
{Song}, C.-Y., \& {Liu}, T. 2019, \apj, 871, 117

\bibitem[{{Stevenson} {et~al.}(2017){Stevenson}, {Vigna-G{\'o}mez}, {Mandel},
  {Barrett}, {Neijssel}, {Perkins}, \& {de Mink}}]{2017NatCo...814906S}
{Stevenson}, S., {Vigna-G{\'o}mez}, A., {Mandel}, I., {et~al.} 2017, Nature
  Communications, 8, 14906

\bibitem[{{Sukhbold} {et~al.}(2018){Sukhbold}, {Woosley}, \&
  {Heger}}]{2018ApJ...860...93S}
{Sukhbold}, T., {Woosley}, S.~E., \& {Heger}, A. 2018, \apj, 860, 93

\bibitem[{{Taam} {et~al.}(1978){Taam}, {Bodenheimer}, \&
  {Ostriker}}]{1978ApJ...222..269T}
{Taam}, R.~E., {Bodenheimer}, P., \& {Ostriker}, J.~P. 1978, \apj, 222, 269

\bibitem[{{Taam} \& {Sandquist}(2000)}]{2000ARA&A..38..113T}
{Taam}, R.~E., \& {Sandquist}, E.~L. 2000, Annual Review of Astronomy and
  Astrophysics, 38, 113

\bibitem[{{Taddia} {et~al.}(2013){Taddia}, {Stritzinger}, {Sollerman},
  {Phillips}, {Anderson}, {Boldt}, {Campillay}, {Castell{\'o}n}, {Contreras},
  {Folatelli}, {Hamuy}, {Heinrich-Josties}, {Krzeminski}, {Morrell}, {Burns},
  {Freedman}, {Madore}, {Persson}, \& {Suntzeff}}]{2013A&A...555A..10T}
{Taddia}, F., {Stritzinger}, M.~D., {Sollerman}, J., {et~al.} 2013, \aap, 555,
  A10

\bibitem[{{Tauris} {et~al.}(2017){Tauris}, {Kramer}, {Freire}, {Wex}, {Janka},
  {Langer}, {Podsiadlowski}, {Bozzo}, {Chaty}, {Kruckow}, {van den Heuvel},
  {Antoniadis}, {Breton}, \& {Champion}}]{2017ApJ...846..170T}
{Tauris}, T.~M., {Kramer}, M., {Freire}, P.~C.~C., {et~al.} 2017, \apj, 846,
  170

\bibitem[{{Terman} {et~al.}(1995){Terman}, {Taam}, \&
  {Hernquist}}]{1995ApJ...445..367T}
{Terman}, J.~L., {Taam}, R.~E., \& {Hernquist}, L. 1995, \apj, 445, 367

\bibitem[{{The LIGO Scientific Collaboration} \& {the Virgo
  Collaboration}(2018{\natexlab{a}})}]{2018arXiv181112940T}
{The LIGO Scientific Collaboration}, \& {the Virgo Collaboration}.
  2018{\natexlab{a}}, arXiv e-prints, arXiv:1811.12940

\bibitem[{{The LIGO Scientific Collaboration} \& {the Virgo
  Collaboration}(2018{\natexlab{b}})}]{2018arXiv181112907T}
---. 2018{\natexlab{b}}, arXiv e-prints, arXiv:1811.12907

\bibitem[{{Thorne} \& {Zytkow}(1977)}]{1977ApJ...212..832T}
{Thorne}, K.~S., \& {Zytkow}, A.~N. 1977, \apj, 212, 832

\bibitem[{{Tylenda} {et~al.}(2011){Tylenda}, {Hajduk}, {Kami{\'n}ski},
  {Udalski}, {Soszy{\'n}ski}, {Szyma{\'n}ski}, {Kubiak}, {Pietrzy{\'n}ski},
  {Poleski}, {Wyrzykowski}, \& {Ulaczyk}}]{2011A&A...528A.114T}
{Tylenda}, R., {Hajduk}, M., {Kami{\'n}ski}, T., {et~al.} 2011, \aap, 528, A114

\bibitem[{{Vigna-G{\'o}mez} {et~al.}(2018){Vigna-G{\'o}mez}, {Neijssel},
  {Stevenson}, {Barrett}, {Belczynski}, {Justham}, {de Mink}, {M{\"u}ller},
  {Podsiadlowski}, {Renzo}, {Sz{\'e}csi}, \& {Mandel}}]{2018MNRAS.481.4009V}
{Vigna-G{\'o}mez}, A., {Neijssel}, C.~J., {Stevenson}, S., {et~al.} 2018,
  \mnras, 481, 4009

\bibitem[{{Webbink}(1984)}]{1984ApJ...277..355W}
{Webbink}, R.~F. 1984, \apj, 277, 355

\bibitem[{{Weiler} {et~al.}(1986){Weiler}, {Sramek}, {Panagia}, {van der
  Hulst}, \& {Salvati}}]{1986ApJ...301..790W}
{Weiler}, K.~W., {Sramek}, R.~A., {Panagia}, N., {van der Hulst}, J.~M., \&
  {Salvati}, M. 1986, \apj, 301, 790

\bibitem[{{Woosley} \& {Heger}(2012)}]{2012ApJ...752...32W}
{Woosley}, S.~E., \& {Heger}, A. 2012, \apj, 752, 32

\bibitem[{{Zhang} \& {Fryer}(2001)}]{2001ApJ...550..357Z}
{Zhang}, W., \& {Fryer}, C.~L. 2001, \apj, 550, 357

\end{thebibliography}

\end{document}